\documentclass[a4paper,fleqn,usenatbib]{mnras}

\usepackage{newtxtext,newtxmath}
\usepackage[T1]{fontenc}
\usepackage{ae,aecompl}

\usepackage{graphicx}	% Including figure files
\usepackage{amsmath}	% Advanced maths commands
\newcommand{\oversim}[2]{\protect{\mbox{\lower0.5ex\vbox{%
  \baselineskip=0pt\lineskip=0.2ex
  \ialign{$\mathsurround=0pt #1\hfil##\hfil$\crcr#2\crcr\sim\crcr}}}}}
\newcommand{\simgreat}{\mbox{$\,\mathrel{\mathpalette\oversim>}\,$}} % >~ sign
\newcommand{\simless} {\mbox{$\,\mathrel{\mathpalette\oversim<}\,$}} % <~ sign
\newcommand{\mstar}{m_{\rm BH}}

\newcommand{\ledd}{L_\mathrm{Edd}}
\newcommand{\lacc}{L_\mathrm{acc}}
\newcommand{\Mp}{m_\mathrm{p}}
\newcommand{\dragc}{f_\Lambda}
\newcommand{\virc}{f_\mathrm{v}}
\newcommand{\gasc}{f_\mathrm{g}}
\newcommand{\eddc}{f_\mathrm{Edd}}

\newcommand{\tedd}{\tau_\mathrm{Edd}}

\newcommand{\thompson}{\sigma_\mathrm{T}}

\newcommand{\myr}{\mathrm{Myr}}

\newcommand{\dd}[1]{\mathrm{d}{#1}}
\newcommand{\ff}[1]{\mathrm{#1}}

\newcommand{\tce}{t_{\mathrm{ce}}}
\newcommand{\tgw}{t_{\mathrm{gw}}}
\newcommand{\ma}{m_{\mathrm{1}}}
\newcommand{\mb}{m_{\mathrm{2}}}
\newcommand{\ab}{a_{\mathrm{h/s}}}
\newcommand{\m}{\langle m \rangle}

\title[Quasars and SMBHs]{Very high redshift quasars and the rapid emergence of super-massive black holes}

\author[Kroupa et al.]{Pavel Kroupa$^{1,2,3}$\thanks{pkroupa@uni-bonn.de; pavel.kroupa@mff.cuni.cz},
   Ladislav Subr$^{2}$\thanks{subr@sirrah.troja.mff.cuni.cz}, 
   Tereza Jerabkova$^{1,2,3,4,5,6}$\thanks{tereza.jerabkova@esa.int}\thanks{ESA Fellow}, Long  Wang$^{7}$\thanks{long.wang@astron.s.u-tokyo.ac.jp}
\\ 
$^1$Helmholtz-Institut f\"ur Strahlen- und Kernphysik, University
        of Bonn, Nussallee 14-16, D-53115 Bonn, Germany\\
$^2$Charles University in Prague, Faculty of Mathematics and 
        Physics, Astronomical Institute, V Hole\v{s}ovi\v{c}k\'ach 2, CZ-180
        00 Praha 8, Czech Republic\\
$^3$Astronomical Institute, Czech Academy of Sciences, Fri\v{c}ova 298,
25165, Ond\v{r}ejov, Czech Republic\\
$^4$Instituto de Astrofisica de Canarias, E-38205 La Laguna, Tenerife, Spain\\
$^5$GRANTECAN, Cuesta de San Jose s/n, 38712 Brena Baja, La Palma, Spain\\
$^6$European Space Research and Technology Centre (ESA ESTEC),
Keplerlaan 1, 2201 AZ Nordwijk, Netherlands\\
$^7$Department of Astronomy, School of Science, The University of
Tokyo 7-3-1 Hongo, Bunkyo-ku, Tokyo 113-0033, JAPAN\\
}

\pubyear{20xx}

\begin{document}
\label{firstpage}
\pagerange{\pageref{firstpage}--\pageref{lastpage}}
\maketitle

\begin{abstract}
  The observation of quasars at very high redshift such as
    P{\={o}}niu{\={a}}'ena is a challenge for models of super-massive
  black hole (SMBH) formation.  This work presents a study of SMBH
  formation via known physical processes in star-burst clusters formed
  at the onset of the formation of their hosting galaxy.  While at the
  early stages hyper-massive star-burst clusters reach the
  luminosities of quasars, once their massive stars die, the ensuing
  gas accretion from the still forming host galaxy compresses its
  stellar black hole (BH) component to a compact state overcoming
  heating from the BH--BH binaries such that the cluster collapses,
  forming a massive SMBH-seed within about a hundred~Myr.  Within this
  scenario the SMBH--spheroid correlation emerges near-to-exactly. The
  highest-redshift quasars may thus be hyper-massive star-burst
  clusters or young ultra-compact dwarf galaxies (UCDs), being
  the precursors of the SMBHs that form therein within about 200~Myr
  of the first stars. For spheroid masses $\simless 10^{9.6}\,M_\odot$
  a SMBH cannot form and instead only the accumulated nuclear cluster
  remains. The number evolution of the quasar phases with redshift is
  calculated and the possible problem of missing quasars at very
    high redshift is raised. SMBH-bearing UCDs and the formation of
    spheroids are discussed critically in view of the high redshift
    observations. A possible tension is found between the high
    star-formation rates (SFRs) implied by downsizing and the observed
    SFRs, which may be alleviated within the IGIMF theory and if the
    downsizing times are somewhat longer.
\end{abstract}


\section{Introduction}
\label{sec:introd}

The existence of super massive black holes (SMBHs, masses
$M_{\rm SMBH} \simgreat 10^6\,M_\odot$) at the centers of galaxies has
been surmised since a long time \citep{Salpeter64, WB70,
  LyndenBellRees71} and appears now well secured through the
observation of the orbit of the S2 star near the centre of the Milky
Way with the GRAVITY instrument \citep{Gravity+18}, astrometric
measurements of the motion of Sgr~A* \citep{ReidBrunthaler20} and the
observation of the shadow of the SMBH in M87 with the Event Horizon
Telescope \citep{EventHorizon+19}.  The observed rapid appearance of
SMBHs in the early Universe, their correlation with their host galaxy
properties and the presence of recently discovered SMBHs in
ultra-compact dwarf galaxies (UCDs) remain, however, unexplained
\citep{FM02, KormendyHo13, Seth14, HeckmanBest14, Janz15a, DiMatteo17, Mezcua17, Banados+17, Ahn17,  Ahn18, Afanasiev+18}, with UCDs with mass $<10^7\,M_\odot$ apparently not hosting SMBHs \citep{Voggel+18}.  SMBHs with masses $M_{\rm SMBH} \approx 10^{10}\,M_\odot$ have been argued to exist a few hundred~Myr after the Big Bang \citep{Banados+16,DiMatteo17,
  Banados+17, Mezcua17}.  This poses a challenge for theories of the
formation of SMBHs because they are observed at high redshift to be
accreting at \citep{Willott+10, Banados+17} or below
\citep{Mazzuchelli+17,Kim+18} the Eddington limit (Eq.~\ref{eq:mt} in
Sec.~\ref{sec:BHaccr}).  Assuming super-Eddington accretion with only
10~per cent of the infalling matter being radiated away
($\epsilon_{\rm r} = 0.1$ in Eq.~\ref{eq:mt} in Sec.~\ref{sec:BHaccr}), a
stellar-mass black hole (BH) seed, $M_{\rm seed} =10\,M_\odot$, would
still need to continuously accrete for $t_{\rm accr} \approx 1 \,$Gyr
to reach $10^{10}\,M_\odot$. This poses a problem if the
hosting spheroid galaxy\footnote{
Many disk galaxies have central old spheroidal (classical) bulges
which have largely indistinguishable properties from those of elliptical galaxies
\citep{GadottiKauffmann09} and we refer to elliptical galaxies and classical bulges as spheroids or spheroid galaxies.} 
forms on a much shorter downsizing
time-scale (Eq.~\ref{eq:downs} below) after which gas infall shuts
off.  Quasars have life-times of about 10--100~Myr
\citep{MartiniWeinberg01, YuTremaine02} and the first stars would have formed
after~200~Myr after the Big Bang due to the time required for
metal-free gas to cool \citep{Yoshida+04}.  Indeed, a star-forming,
about $10^9\,M_\odot$ galaxy has been discovered at a redshift $z=9.1$
which is about 250~Myr after the birth of the Universe
\citep{Hashimoto+18}. These authors find the bulk of its
formation to have occurred at $z\approx15$, with a suppression of its star-formation rate SFR\footnote{SFR is the acronym for "star-formation rate", while $SFR$ is the corresponding physical quantity.} and a rejuvenated gas inflow leading to the 
detected star formation at $z\approx 9.1$. It remains a challenge to understand
  the observed rapid appearance of the SMBH--host galaxy correlations
  in view of the theoretically predicted hierarchical formation of
  galaxies in standard cosmology (e.g. \citealt{Kotilainen+09,
    Portinari+12, LatifFerrara16}), noting that the SMBH--host galaxy
  correlation extends from elliptical galaxies to the bulge component
  of disk galaxies \citep{Sanghvi+14}. In the recent review of high redshift ($z \simgreat 5$) quasars and their host galaxies, \cite{Trakhtenbrot20} writes
  that 
  the multi-wavelength data show these SMBHs to be consistent with Eddington-limited, radiatively efficient accretion which had to proceed near-to continuously since very early epochs. \cite{Trakhtenbrot20} also writes that ALMA observations of the host's inter-stellar medium uncover gas-rich, well developed galaxies, with SFRs that may exceed $\approx 1000\,M_\odot$/yr.
 % quotation:
 % "The available multi-wavelength data show that these SMBHs are consistent with Eddington-limited, radiatively efficient accretion that had to proceed almost continuously since very early epochs. ALMA observations of the host's ISM reveal gas-rich, well developed galaxies, with a wide range of SFRs that may exceed $\approx 1000\,M_\odot$/yr."
  
  Possible theoretical explanations (see \citealt{LatifFerrara16} for
  a review) for much larger $M_{\rm seed}$ or more rapid mass growth
  are primordial black holes \citep{CarrHawking74,DV17},
  hyper-Eddington or supra-exponential accretion \citep{AN14, BV17},
  the very massive first metal-free population~III stars leaving
  massive BH seeds or runaway stellar mergers in massive star clusters
  which would have had very large masses leaving direct-collapse
    SMBH seeds \citep{Portegies02, Rasio04, Portegies04,
    Goswami+12,Mapelli16}.  These suggestions rely on the uncertain
  stellar evolution of hyper-massive stars (see e.g. \citealt{Diaz+19}
  for a discussion of the physical barriers concerning direct cloud
  collapse to form SMBH-seeds, \citealt{Seguel+19} for a discussion of
  mass loss during mergers of population~III stars, \citealt{Woods+20}
  for an exploration of the theoretical limiting one-dimensional
  evolution of zero-metallicity, non-rotating stars).  Also studied is
  the possible direct collapse of the first baryonic structures
  \citep{LatifSchleicher15,MayerBonoli18, Ardaneh+18, Latif+19}, which
  rely on early structure formation in standard dark matter models
  (see \citealt{Haemmerle+20} for a review).

With this contribution an explanation for the rapid emergence of SMBHs
of all masses is sought which may operate by itself or in conjunction
with the above suggestions, but does not need to invoke additional
non-standard physics. Growing a black hole by accretion of gas is
inefficient because a significant fraction of the rest-mass energy of the 
infalling matter is radiated as photons and ejected via jets. The key idea followed here 
is based on the high efficiency of mass-growth of black holes when they
merge. According to the LIGO/VIRGO observations of gravitational wave signals
from merging binary black holes, the merged black hole
mass is only less than 5~per cent lighter than the sum of the
pre-merger black holes \citep{Ligo+16,Ligo+19}.  A rapidly functioning
physical mechanism ensuring the merger of many black holes may thus be
a necessary key ingredient for understanding the observed rapid
appearance of SMBHs in the Universe within a conservative framework.

To this end, we consider here the stellar population formed from
extremely low metallicity gas in the first highest-density peaks of
star formation, relying on previously independently obtained knowledge
on the properties of such a population.  The evolution is formulated
of the stellar black hole (BH) sub-cluster with initial mass
$M_{\rm BH,0}$ which is left after the massive stars
have evolved. The cluster most likely forms mass-segregated
(e.g. \citealt{Pavlik+19}) and if not it will dynamically
mass-segregate to a half mass radius which depends on the
mass-fraction of BHs \citep{BH13b}. \cite{Giersz15} show that such a
cluster can form an intermediate mass black hole (IMBH, mass
$10^3 \simless M_{\rm IMBH}/M_\odot <10^6$) through runaway merging of
its BHs, but the time-scale for this, being of the order of
10--15~median two-body relaxation times, $t_{\rm rlx}$, is too long
($>\,{\rm few}\,$Gyr) to account for the rapid ($\simless 300\,$Myr)
appearance of SMBHs after the birth of the Universe.  For the BH
sub-cluster to form a SMBH-seed mass comprising a significant fraction
of $M_{\rm BH,0}$, it needs to reach a velocity dispersion of about
1~per cent of the speed of light \citep{Lee93, Lee95, Kupi06}, because
then the energy dissipation through the radiation of gravitational
waves due to BH--BH encounters makes the relativistic collapse
inevitable on the core-collapse time-scale of the BH sub-cluster,
being as above $10< t_{\rm collapse}/t_{\rm rlx} < 15$. This
relativistic state can only be reached if the BH sub-cluster has a
half-mass radius less than $0.02\,$pc (for
$M_{\rm BH,0}=10^6\,M_\odot$).  In this case $t_{\rm collapse}$ is of
the order of a~Myr (assuming the BHs have masses
$m_{\rm BH}=50\,M_\odot$). The problem is that BH-binaries which form
in the core of the BH sub-cluster heat it, leading to long-term
balanced evolution of the cluster \citep{BH13} such that such small
radii cannot be reached.  But if gas from the assembling spheroid accretes onto
the central BH sub-cluster it may squeeze it into the relativistic state. 
This idea is placed into the
context of the formation of spheroids and their stellar population.
An essential part of the here presented theory is to calculate the radius shrinkage of the BH
sub-cluster within the massive star-burst cluster, which, according to the IGIMF theory (Sec.~\ref{sec:gwIMF}), is expected to have formed during the formation of a  spheroid.

The observed correlations between the galaxy-wide
SFR and the population of forming star clusters
and the stellar initial mass function (IMF, the number of stars per
mass interval) are discussed in Sec.~\ref{sec:galaxy_stcl} and in
Sec.~\ref{sec:stpop}, respectively.  Following \cite{Jerab18} we distinguish
between the composite or galaxy-wide IMF, gwIMF, and the IMF which
constitutes the population of stars formed in one embedded star
cluster, see also \cite{Kroupa13, Hopkins18}. 
These set the stage for the
calculation of the formation of the SMBH (Sec.~\ref{sec:model}, 
Sec.~\ref{sec:sm}). The resulting galaxy--SMBH-mass correlation and the 
expected number of quasars at high redshift are addressed, respectively, in Sec.~\ref{sec:corrs} and~\ref{sec:ndens}. The latter section includes a discussion of the newly discovered P{\={o}}niu{\={a}}'ena.
Sec.~\ref{sec:disc} deals with caveats and compares the model with observations of star-forming objects at high redshift.
Sec.~\ref{sec:concs} contains the conclusions.

\section{The galaxy--star-cluster correlation}
\label{sec:galaxy_stcl}

The correlation between galaxy mass and its star-cluster population is
discussed in this section, before continuing with a discussion of the
initial stellar population in massive star-burst clusters in
Sec.~\ref{sec:stpop}. In the following we first discuss the downsizing
problem (Sec.~\ref{sec:downs}), then populating a forming spheroid
with star clusters (Sec.~\ref{sec:selfreg}).

\subsection{The formation time scale for spheroids}
\label{sec:downs} 

Observations of the chemical abundances suggest that the
most massive spheroids, which typically harbour the most-massive SMBHs
\citep{FM02}, may have formed  
very early and on a time scale, $\Delta \tau$, of
less than a Gyr. This downsizing result was deduced originally by \cite{Matteucci94, Thomas+99, Thomas05, Recchi09, Yan+2019, SR+19} through the following general argument: Core-collapse supernovae (e.g. SN~type~II) enrich the inter-stellar medium (ISM) with $\alpha$ elements and iron, Fe, while the type~Ia supernovae contribute most of the Fe. Stars formed from the ISM which was being enriched by SN~II will thus have similar [$\alpha$/Fe] abundance ratios.
When SN~Ia begin to explode with a delay after the SN~II cease
the [$\alpha$/Fe] abundance ratio of the ISM
begins to decrease.  The observed super-solar [$\alpha$/Fe] abundance
of the stellar population in spheroids requires $\Delta \tau$ to
be shorter than the time needed for SN~Ia to contribute significant Fe
to the gas from which the population formed (for a discussion on the uncertainty of the delay-time distribution see \citealt{YJK20}). 
By implication, the
so-constrained $\Delta \tau$ leads to a high average SFR ($=M_{\rm
  igal}/\Delta \tau$) and \cite{Matteucci94} additionally finds the
gwIMF to need to be top-heavy in order to explain the high observed
(near-to solar or super-solar) metallicity, $Z$, of the massive
spheroids.  

From fig.~18 in \cite{Recchi09} the downsizing relation
follows (see also Fig.~\ref{fig:SFR} below),
\begin{equation} \Delta \tau/{\rm Gyr} = 8.16 \, e^{ \left( -0.556 \,
      {\rm log}_{10}\left( M_{\rm igal}/M_\odot \right) + 3.401
    \right)} + 0.027, 
\label{eq:downs} 
\end{equation}
%\label{eq:downsizing}
% TJ: 21.03.18:   (need to multiply by 1000 to bet the time in Myr
% rather than Gyr which is the fit)
%f(x) = a*exp(b*x+c)+d
%[a,b,c,d] = [8.16174963 0.5561232  3.40105768 0.02682047]
%for value 10^12 you should get delta t = 0.336
%
where $10^6 \simless M_{\rm igal}/M_\odot \simless 10^{12}$ is the
initial stellar mass of the galaxy ("initial" referring in this context to 
the total stellar mass assembled without stellar-evolutionary mass loss, 
but compare with Fig.~\ref{fig:ipgal} below).  
For example, a giant spheroid
with an initial stellar mass $M_{\rm igal}=10^{12}\,M_\odot$ would have
formed over a time-scale
of $\Delta \tau = 0.34\,$Gyr implying a star-formation rate
$SFR = 2941\,M_\odot$/yr (the formation of spheroids and these model vs observed SFRs are discussed in more detail in Sec.~\ref{sec:SFRs}).
Such SFRs are indeed observed at high
redshift ($z>4$: \citealt{Glazebrook17, Pavesi+18, Miller+18,
  Fan+19, Nguyen+20}). The REQUIEM Survey~I finds high redshift ($z\approx 6$)
quasars to be surrounded by massive circum-galactic media on scales of
dozens of~kpc typical of massive galaxies \citep{Farina+19},
suggesting a significant cool gas component to exist from which the
spheroid hosting the quasar accretes. Stellar-population synthesis of spheroids confirms 
the downsizing relation but suggests a systematically longer $\Delta
\tau$ by a factor of about two (\citealt{McDermid+15}, see also \citealt{delaRosa+11}).

The implied short formation times thus suggest a monolithic formation of spheroids:
Following \cite{Comeron+16}, monolithic formation or collapse of a stellar system is understood to be its rapid formation where several generations of stars form in rapid succession, essentially on a free-fall time-scale.
Assume a spherical post-Big-Bang gas cloud of mass
  $M_{\rm cloud}$ in units of $M_\odot$ with initial radius
  $R_{\rm cloud}$ in units of~pc collapses on a free-fall time scale
  and forms a spheroid. Thus, if the formation time
  $\Delta \tau_{\rm Myr}$ in units of Myr is assumed to be the
  free-fall time (Eq.~\ref{eq:tff}) it follows that
  $R_{\rm cloud} \approx (\Delta \tau_{\rm Myr} /16.6)^{2/3} \,M_{\rm
    cloud}^{1/3}$. This means that for a spheroid which forms
  $10^{12}\,M_\odot$ in stars in $\Delta \tau = 0.34\,$Gyr
  (Eq.~\ref{eq:downs}), $R_{\rm cloud}\approx 100\,$kpc assuming a
  star-formation efficiency of $1/3$.  This rough estimate is smaller
  in reality due to cosmic expansion (needing a higher density and
  thus smaller radius for gravitational decoupling from the initial
  Hubble flow), but suggests (in this model) that the gas which forms the bulk of the
  final spheroid accretes from a region with a radius
  of~$\simless 100\,$kpc. Since no fully-fledged independent galaxy
  would have been able to develop within the flow within $0.34\,$Gyr,
  the so-formed spheroid can be referred to as having formed
  monolithically, i.e., in one collapse. However, it is likely that the
  in-falling gas would not be homogeneous such that sub-regions
  collapse faster forming stars in sub-galactic objects such that the
  "monolithic" formation may well be associated with the rapid merging
  of developing sub-galactic building blocks. This monolithic formation model would be
  replaced by a merging model if $\Delta \tau$ would be longer, but the 
  essence of the model is the 
natural physical correlation between the central gas density, where the most massive (and first) star cluster forms, 
and the overall proto-spheroid gas mass which ultimately leads to the observed SMBH-mass–spheroid-mass correlation. This natural physical correlation is at the heart of the IGIMF theory discussed next.

\subsection{The most massive embedded cluster in the forming
  spheroid}
\label{sec:selfreg}

Extragalactic observations \citep{LarsenRichtler00, Larsen02a,
  Larsen02b, Rand13, Whitmore14} have shown that the stellar mass,
$M_{\rm ecl, max}$, of the most-massive star cluster forming in a
galaxy follows the empirical WKL correlation with the galaxy-wide SFR
\citep{Weidner04b},
%$ M_{\rm ecl, max}/M_\odot = 0.014 \times \left(SFR /
%  \left(M_\odot/{\rm yr}\right) \right)^{0.75} \times 10^{6.77}$
\begin{equation}
M_{\rm ecl, max}/M_\odot = 8.24 \times 10^4 \times 
\left(
SFR / \left(M_\odot/{\rm yr}\right) 
\right)^{0.75}.
\label{eq:WKL}
\end{equation}
\cite{Rand13} find the dispersion of the masses of the most-massive clusters at a
given SFR to be smaller than expected from random sampling, and
\cite{Larsen02a} emphasises that the mass of the most massive young
cluster depends on the SFR of the galaxy, and thus the pressure
\citep{EE97}, such that globular (GC) and open star clusters are part
of the same family, rather than requiring distinct formation
conditions. In the example above ($SFR=2491\,M_\odot$/yr),
$M_{\rm ecl,max}\approx 2.9 \times 10^7\,M_\odot$. 

The gaseous
inter-stellar medium transforms to a new population of stars on a
time-scale of about $\delta t'=10\,$Myr. This time scale is observed
to be the life-time of molecular clouds \citep{Fukui99, Yamaguchi01,
  Tamburro08, FK10, Meidt15} and is also deduced from the offset of
dusty spiral patterns from the freshly hatched stellar population in
spiral galaxies \citep{Egusa04, Egusa09}. The mass of the stars formed
in the galaxy over the time interval $\delta t$ is
\begin{equation}
M_{\delta t}(t) = SFR(t) \times \delta t \, .  
\label{eq:mstformed}
\end{equation}
According to observations, most stars form clustered in the density
peaks in molecular clouds \citep{Lada2003a, Lada10, Megeath16,
  Joncour+18}, a conclusion also reached from binary-star dynamical
population synthesis \citep{Kroupa95, Marks+11}.  The distribution
function of stellar masses, $M_{\rm ecl}$, of these embedded clusters
has been observed to be a power-law \citep{ZhangFall99, Larsen02b,
  Lada2003a, Hunter+03, Weidner04b,McCradyGraham07},
\begin{equation}
\xi_{\rm ecl}(M_{\rm ecl}) = k_{\rm ecl} M_{\rm ecl}^{-\beta}, \quad M_{\rm
  ecl} \ge  M_{\rm ecl, min}, 
  \label{eq:ECMF}
\end{equation}
where $M_{\rm ecl, min}\approx 5\,M_\odot$ is the minimum embedded
stellar group or cluster mass (as observed in e.g. the Taurus-Auriga
star-forming region, \citealt{Joncour+18}), 
$dN_{\rm ecl} = \xi_{\rm ecl}\,dM_{\rm ecl}$ being the number of embedded
clusters with birth masses (in stars) in the mass range $M_{\rm ecl}$
to $M_{\rm ecl}+dM_{\rm ecl}$. 

The power-law index of the embedded cluster mass function is suggested by \cite{Weidner04b} to be
$\beta=2.4$ based on $M_{\rm ecl, max}-SFR$ data for late-type galaxies.  \cite{Murray09} summarises that $1.5< \beta < 2$. According to \cite{Weidner13}
$\beta \le 2.0$ is possible with  $\xi_{\rm ecl}(M_{\rm ecl})$ becoming more top-heavy with increasing galaxy-wide $SFR$ and decreasing metallicity.
We will discuss both possibilities ($\beta=2$ and $2.4$, for a discussion and references see \citealt{Schulz15}) in Sec.~\ref{sec:corrs} but take $\beta=2$ as the default canonical value.  

Solving for $M_{\rm ecl,max}$ the following two equations,
\begin{equation}
\begin{aligned}
M_{\delta t}(t) = \int_{M_{\rm ecl, min}}^{M_{\rm ecl, max}(SFR)} M_{\rm
  ecl}\, \xi_{\rm ecl}(M_{\rm ecl})\,dM_{\rm ecl}, \\
1 = \int_{M_{\rm ecl, max}(SFR)}^{M_{\rm ecl,max*}} \xi_{\rm ecl}(M_{\rm ecl})\,dM_{\rm ecl}, 
\label{eq:Mdeltat}
\end{aligned}
\end{equation}
where $M_{\rm ecl,max*}$ is the physical allowed maximum cluster mass (we adopt $M_{\rm ecl,max*}=10^{9.7}\,M_\odot$, but the results are not sensitive to this value as long as $M_{\rm ecl,max*}\ge10^{9}\,M_\odot$, see \citealt{Weidner04b})
the calculated $M_{\rm ecl,max}(SFR)$ relation follows the above WKL
correlation (Eq.~\ref{eq:WKL}) very closely if
$\delta t \approx 10\,$Myr in general, and with $\beta=2.4$. 
The values of $M_{\rm ecl,max}$ are significantly larger for $\beta=2$ than for $\beta=2.4$. 
For example, if $SFR=3 \times 10^3\,M_\odot$/yr then $M_{\rm ecl,max} \approx 3.3 \times 10^7\,M_\odot (\beta=2.4)$ and $M_{\rm ecl,max}\approx 3\times 10^9\,M_\odot$ ($\beta=2.0$; see fig.2 in \citealt{YJK17}). This has an important bearing on the mass of the SMBH-seed which can form in the central most-massive star-burst cluster within a few hundred~Myr as discussed in Sec.~\ref{sec:corrs} below).
Note the consistency of this model with
the data, because $\delta t \approx \delta t'$.

Assuming
$SFR \approx M_{\rm igal}/\Delta \tau$ remains about constant over the
time of formation of the galaxy (i.e. we assume a box-shaped star-formation history, SFH), 
%the cluster-galaxy %relation results,
%\begin{equation}
%M_{\rm ecl, max}/M_\odot = 8.24\times 10^4 \, \left( (M_{\rm
%    igal}/M_\odot) / (\Delta \tau/{\rm yr}) \right)^{0.75}.
%\label{eq:Meclmax_deltat}
%\end{equation}
\begin{equation}
N_{\rm gen}=\Delta \tau / \delta t
\label{eq:Ngen}
\end{equation}
generations of embedded cluster populations form, each following the
above relations.  This is supported by observational evidence in that subsequently-populated $\xi_{\rm ecl}(M_{\rm ecl})$ can be extracted from observed luminosity functions of GCs and UCDs in galaxy clusters
\citep{Schulz16}.  Therefore, a $M_{\rm
  igal}=10^{12}\,M_\odot$ elliptical galaxy will have formed
$N_{\rm gen} \approx 34$ most-massive clusters of mass
$M_{\rm ecl,max} \approx 3.3 \times 10^7 \,
M_\odot$ (if $\beta=2.4$) up to $M_{\rm ecl,max} \approx 3\times 10^9 \,
M_\odot$ (for $\beta=2.0$)
during its assembly.  Only the first-formed clusters will be
extremely metal poor though \citep{Yan+2019}, this being important for
the shape of the IMF and thus the content of BHs \citep{Jerab17}.  The
true star-formation history is likely to have been more complex with
peak star formation rates that may surpass the average
$SFR$. 

Spheroids would thus have formed many massive star-burst clusters in
the densest gas clouds usually observationally
\citep{Joseph85,Wright88, McCradyGraham07, Stolte14, Ando+17,
  Leroy+18, Rand19} and theoretically \citep{Norman87, LiGnedin16}
found near their centres where the highest gas pressures reign
\citep{EE97}.  Cases in point for the ongoing formation of the
most-massive clusters in the central region are the galaxies Arp~220
\citep{Lonsdale+06}, M33 \citep{Pflamm13} and Henize~2-10
\citep{Nguyen+14}.  Evidence for higher and earlier past innermost
star-forming intensities find support from metallicity and age tracers
\citep{Martin+18, Martin+19} and morphological studies at high
redshift \citep{Pavesi+18} as well as measured radial metallicity and
age gradients in spheroids \citep{Zibetti+19}. The observed
compactness ($\approx 40\,$pc) of $6 \simless z \simless 8$ galaxies
with stellar masses $10^6-10^7\,M_\odot$ may thus be naturally
explained in this way \citep{Ploeckinger+19}.

\subsection{Summary}

Tthe masses of the most-massive young clusters in
star-forming galaxies correlate positively with the SFR and with
decreasing galactocentric distance.

\section{The stellar population in massive star-burst clusters and in 
ultra compact dwarf galaxies}
\label{sec:stpop}

In this section the initial stellar population in massive star-burst
clusters is discussed.  First (Sec.~\ref{sec:canIMF}) the canonical
stellar IMF is introduced.  In Sec.~\ref{sec:varIMF} the
observational evidence on the molecular cloud-core scale for a
systematically varying IMF is elucidated. Consistency with the
galaxy-wide initial stellar populations is tested for thereafter
(Sec.~\ref{sec:gwIMF}). The content of stellar black holes (BHs) in
individual clusters is then formulated (Sec.~\ref{sec:BHcontent}), in
preparation for Sec.~\ref{sec:model}.

\subsection{The canonical stellar IMF}
\label{sec:canIMF}

One conservative assumption is the stellar population to be always
described by an invariant IMF. Thus, for the invariant canonical IMF
\citep{Kroupa2001, Kroupa02, Chabrier03, Bastian10, Kroupa13,
  Offner+14}, the number of freshly formed stars with masses in the
range $m$ to $m+dm$ in an embedded cluster is $dN=\xi(m)\,dm$, where
the IMF, $\xi(m)=k\,m^{-\alpha_i}$, with $\alpha_1=1.3$ for
$0.08<m/M_\odot\le 0.5$ and $\alpha_2=2.3$ for $0.5<m/M_\odot<1$.  The
canonical IMF, which is typical for star formation in the present-day
Galaxy, has the Salpeter-Massey index $\alpha_3=\alpha_2=2.3$
\citep{Salpeter55, Massey95, Massey03} for
$1\,M_\odot\le m \le m_{\rm max}(M_{\rm ecl})  \le m_{\rm max*}$, with
$m_{\rm max*} \approx 150\,M_\odot$ being the empirically found
maximum stellar mass \citep{weidner2004a,Figer05,OC05, Koen06, Maiz08,
  BKO12}, and $m_{\rm max}(M_{\rm ecl})$ being the most-massive-star--$M_{\rm ecl}$ relation \citep{OK18}.

\subsection{The variable stellar IMF in dependence of the physical environment}
\label{sec:varIMF}

Observation and stellar-dynamical analysis has
suggested that low-metallicity star-burst clusters with
$M_{\rm ecl}\approx 10^4-10^5\,M_\odot$, ages$\,\simless 10\,$Myr,
which can be resolved into individual stars, have an over-surplus of
massive stars relative to their low-mass star content when compared to
the canonical IMF \citep{BK12,Schneider+18,Kalari+18}. That is, the
IMF appears to be top-heavy in massive and/or low-metallicity
star-burst clusters compared to the canonical IMF.  Evidence for an
IMF which becomes increasingly top-heavy with decreasing metallicity
and increasing star-formation rate per volume on a pc-scale
consistent with these results has emerged from an analysis of the
present-day properties of very old Milky Way GCs, which had birth
masses $\approx 10^5-10^7\,M_\odot$ and that have different but low
metal abundances \citep{Marks12}. The star clusters in the Andromeda
galaxy likewise suggest a similar metallicity dependence of their IMFs
\citep{Zonoozi16, Haghi+17}. Present-day ultra compact dwarf galaxies
(UCDs) have properties which resemble ultra-massive GCs with masses in
the range $10^6-10^9\,M_\odot$ \citep{Hilker07,Mieske08,Brodie11} and
many have been shown to comply to this dependency of the IMF
\citep{Dabring09, Murray09,Dabring12, Marks12}. 
The high rate of core-collapse supernovae per year
observed in the central region of the star-bursting galaxy Arp~220
suggests the stars to be forming with a top-heavy IMF. These observed
massive central star-bursts appear to occur in individual UCD-type
objects \citep{Lonsdale+06, Dabring12}.  The extreme star-formation
conditions within 0.04 to $0.4\,$pc of the Galactic centre lead,
apparently, to a very top-heavy ($\alpha_3\approx 0.45$, see
Eq.~\ref{eq:IMF}) IMF significantly depleted in low mass stars
(\citealt{Bartko+2010}, see also discussion in \citealt{Kroupa13}),
despite the high metallicity.

Theoretical work on star formation has also been suggesting that a
top-heavy IMF ought to emerge in low metallicity and high density gas
peaks in gas clouds. A shift to higher average stellar 
masses is needed for gravitational instability due to the lower efficiency of cooling in low-metallicity environments 
and environments with higher gas temperatures \citep{Larson98, Murray09}, as induced for example from cosmic
ray heating produced by supernovae \citep{Papadopoulos10}, due to the
increased rate of accretion resulting from a lower photon pressure in low-metallicity regions \citep{AF96,
  AL96}, higher ambient temperature \citep{Riaz+20} and due to the coagulation in very dense proto-clusters of
cloud cores to more massive ones before their individual collapse to
proto stars can occur \citep{Dib07}.

Given the above empirical evidence and the need for compliance with
the on-going star-forming activity in the MW, the variation of the IMF
with metallicity and density of the star-forming gas on the scale of
an embedded cluster ($\approx 1\,$pc) has been formulated as a
dependency of the power-law indices describing the distribution of
stellar masses \citep{Dabring09, Dabring12, Marks12, Recchi15, YJK17,
  Jerab18}. Thus,
\begin{equation}
\alpha_3= -0.41\,x + 1.94; x\ge -0.87,
\label{eq:IMF}
\end{equation}
and $\alpha_3= 2.3; x < -0.87$, where
\begin{equation}
x=-0.14\,{\rm [M/H]} + 0.99\,{\rm log}_{10} \left(\rho_{\rm cl} /
  (10^6\,M_\odot\,{\rm pc}^{-3})\right),
\end{equation}
with [M/H] being the metal (all nuclei heavier than He providing
cooling through electronic transitions) abundance and $\rho_{\rm cl}$
the density of the cluster-forming gas cloud core (\citealt{Marks12},
as obtained from the embedded-cluster radius-$M_{\rm ecl}$ relation of
\citealt{Marks12b}; see also \citealt{Recchi15, Jerab18}). Since the star-formation history
building up an individual embedded cluster from the monolithic
collapse of a molecular cloud core takes about 1~Myr
(e.g. \citealt{Beccari+17, Kroupa+18}), the star-formation rate
density during the formation of an embedded cluster is
$\approx \epsilon \rho_{\rm cl} / 1\,{\rm Myr}$, where $\epsilon$ is the star-formation efficiency 
(stellar mass formed per total initial mass). Caveats concerning
the variation of the IMF as defined by Eq.~\ref{eq:IMF} are discussed
in Sec.~\ref{sec:other}.

\subsection{Consistency Check: the galaxy-wide IMF}
\label{sec:gwIMF}

A consistency check as to whether the variation of the IMF on the
scales of star-forming events detailed in Sec.~\ref{sec:varIMF}
captures the broadly correct physical behaviour is provided by
comparing with the galaxy-wide IMF (gwIMF) calculated by summing all
star-formation events in a galaxy as detailed via the IGIMF theory
\citep{KW03, Pflamm11b, Kroupa13, Recchi15, YJK17, Jerab18, Dabring19}.  Calculating the
gwIMF by summing all IMFs in the newly formed embedded clusters to
obtain the composite or integrated IMF (i.e. the IGIMF) 
leads to galaxies with low/high SFRs having a
top-light/top-heavy gwIMF, respectively.

This has indeed been found to be the case observationally
\citep{Lonsdale+06,HG08,Lee+09,Meurer+09,Gun11,Rowlands14,Romano17,ZRI+18}.
Elliptical galaxies, having formed with $SFR\,>10^3\,M_\odot$/yr,
have been found to have had early top-heavy gwIMFs in order to account
for their high metallicity, despite their brief formation time-scales
and a possibly bottom-heavy gwIMF during the final metal-rich
formation phase \citep{Matteucci94,GM97,Vazdekis+97, Weidner13,
  Weidner13c, Martin+16, Fontanot17,DeMasi+18, Jerab18}.  The
top-heavy gwIMFs in galaxies which have or had high SFRs comes, within
the model used here, from such galaxies forming very massive clusters
(Sec.~\ref{sec:galaxy_stcl}, Eq.~\ref{eq:Mdeltat}), which have top-heavy
IMFs. The observed dynamical mass-to-light ratios of early type
galaxies are well consistent with the IGIMF being top-light at small
galaxy masses ($\simless 10^9\,M_\odot$) becoming increasingly
top-heavy for massive spheroids leaving up to about halve their
present-day dynamical masses in the form of stellar remnants
\citep{Dabring19}.  This observational and theoretical agreement of
how the gwIMF changes with a changing SFR and metallicity is an
important positive consistency-test for the IMF variation on the
scale of individual star clusters (Sec.~\ref{sec:varIMF}).

\subsection{The BH content of high redshift star burst clusters, their
appearance as quasars and later as UCDs}
\label{sec:BHcontent}

Assuming the canonical IMF above (Sec.~\ref{sec:canIMF}), a few-Myr
old massive star-burst cluster/UCD with mass $M_{\rm ecl}$ formed
through a monolithic gas-cloud collapse contains
\begin{equation}
N_{\rm can}= 6.3\times 10^{-3} \, M_{\rm ecl} /M_\odot
\label{eq:canIMF_BHs}
\end{equation}
stars more massive than~$20\,M_\odot$ (see also \citealt{Kroupa13},
their table~4-1, average stellar mass $\overline{m}=0.55\,M_\odot$).
Thus, a $M_{\rm ecl}=10^7\,M_\odot$ cluster would contain $1.8\times10^7$ stars
and $6.3\times 10^4$ stars more massive than~$20\,M_\odot$, each of
these becoming, after about 50~Myr and ignoring stellar mergers and
ejections, a BH.

According to the systematically varying IMF (Sec.~\ref{sec:varIMF}),
such a monolithically-formed cluster (MC) would, if formed with very
low metallicity, be dominated in mass by massive stars and would
have  a bolometric luminosity of 
$L_{\rm bol} \approx 4\times 10^{10}\,L_{{\rm bol} \odot}$ for
dozens of Myr \citep{Jerab17}. That is, such an object would 
appear quasar-like (see also an example below in this section) for a time comparable to 
quasar-life-times (Sec.~\ref{sec:introd}). 
The above work shows the stellar populations would reach the luminosities of bright quasars and have comparable photometrical colors. The computed SEDs represent only stellar populations. In reality the massive star-burst clusters are complex systems of stars (and stellar BHs early on), gas and dust. In order to be able to compare detailed differences between the spectral properties of actual quasars (accreting SMBHs) and the hyper-massive clusters, detailed realistic hydro-dynamical simulations with radiative transfer in hot dense gas with dust would be necessary. While this certainly represents a fruitful pathway for future investigations, in this work "quasar-like" describes objects with photometric properties of a quasar in terms of brightness and color.
The first such formed
hyper-massive cluster may inhibit the formation of further clusters
due to its strong feedback \citep{Ploeckinger+19}. Once its massive
stars have died, the supernovae, which lead to a variability of
  $L_{\rm bol}$ on a monthly time-scale by dozens of~per cent, will
have enriched the surrounding gas with metals such that obscuration by dust may be significant and the IMF in the cluster
forming from the enriched gas may rapidly evolve to a more canonical
form \citep{Jerab18}. It is thus the first such massive cluster which
may play a decisive role in the rapid emergence of SMBH-seeds.

On the other hand, some UCDs can form from the coagulation of massive
star-cluster complexes (SCCs) \citep{Kroupa98,FK02, Bruens11} observed to have formed in strongly
interacting star-bursting galaxies and also in the expelled gas-rich
tidal arms (e.g. in the Tadpole galaxy, fig.~10 in
\citealt{Kroupa15}). The composite IMF
of such composite SCC UCDs is calculated assuming the UCD is made of a
full ensemble of embedded clusters as above
(Sec.~\ref{sec:galaxy_stcl}). The composite IMF
  (cIMF) of a UCD is the sum over all its embedded clusters and their
  individual IMFs as given by the IGIMF theory (Sec.~\ref{sec:gwIMF}, 
  see also \citealt{Schulz15}), and is
top-light relative to that of the MC UCD type, therewith containing a
smaller number of remnants. Given that the assembly of spheroids will
have involved mergers, it is likely that some of the very massive
young clusters may be ejected from the inner regions or some may form
in gas-rich tidal tails.  Thus, at a given present-day UCD mass, a
range of objects with different dynamical mass to light ratios is
expected to exist in the vicinity of present-day spheroids in this
theory.

\begin{figure}
\includegraphics[scale=0.95]{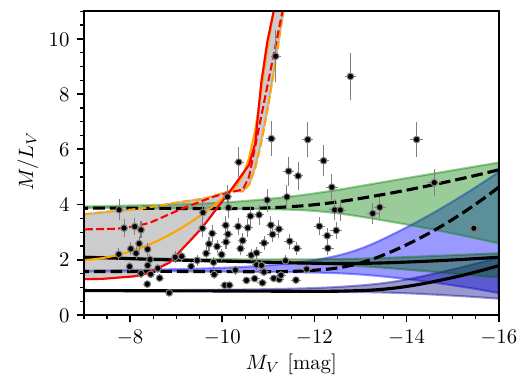}
\caption{The dynamical mass to V-band light ($M/L_V$) ratio of the MC
  (red and orange) and SCC (green and blue) models for different
 metallicities.  The MC models which assume the varying IMF
 (Eq.~\ref{eq:IMF}, \citealt{Marks12}) are shown for an age of
 10~Gyr.  The dashed red line is for [Fe/H]$=-2$ assuming all BHs
  are retained.  The dashed orange line is for [Fe/H]$=0$ assuming all
 BHs are retained.  The full red line is for [Fe/H]$=-2$ assuming the BHs
 are retained dynamically, i.e. in dependence of the distribution of
 kick velocities of neutron stars and BHs (assuming a kick velocity dispersion of $190\,$km/s) and the depth of the potential well 
  (assuming the initial radius-mass relation of \citealt{Marks12b}, 
  see \citealt{Jerab17,Pavlik+18} for details).  The full orange line is for [Fe/H]$=0$
 assuming the BHs are retained dynamically.  The SCC models assume
  the composite IMF to be given by the sum of all varying IMFs  (Eq.~\ref{eq:IMF})
  in all star clusters which form the final UCD following the IGIMF
  theory (Sec.~\ref{sec:gwIMF}, see \citealt{Jerab18} for the details, the IGIMF3
  formulation being applied here).  The lower boundaries of the
  coloured regions show models which retain no BHs, while the upper
 boundaries assume all BHs are retained. The black lines show the
 results if BHs are retained dynamically.  Note the down trend of the
 lower boundaries for the brighter (more massive) UCDs. It is due to
  removing BHs in increasingly top-heavy composite IMFs for which the BHs contribute an increasing fraction of the mass.  The black
  line is dashed for an age of 13~Gyr and full for an age of 5~Gyr, both for [Fe/H]$=-2$. The
  upper blue region is for an age of 13~Gyr and the lower dark blue
  region assumes models at an age of 5~Gyr, both having [Fe/H]$=-2$.
  The upper green models assume an age of 13~Gyr, and the lower dark
  green models assume an age of 5~Gyr, both being for [Fe/H]$=0$. 
  The luminosities of the models are calculated with the second
  release of the spectral evolution code PEGASE2 \citep{Fioc97}.  The
  data points are observed UCDs with their uncertainties
  (\citealt{Voggel+19}, not their normalised data). The stellar-population and dynamical properties of UCDs in terms of their formation process is studied in much more detail by Mahani et al. (inp prep.).}
\label{fig:UCDML}
\end{figure}

A test of whether the model developed here is consistent with
observational data on the pc-scale is provided by comparing the
calculated photometric V-band mass-to-light ($M/L_V$) ratios with
observational data of UCDs. The calculated values are dynamical
$M/L_V$ ratios because they include the mass from remnants and are
thus directly comparable to the observed ratios, which are derived
from the stellar velocity dispersions in the UCDs (see
\citealt{Dabring08} for an in-depth discussion of these issues).  The
comparison is done in Fig.~\ref{fig:UCDML} which demonstrates that the
model well accounts for the data, given that most real UCDs are likely
a mixture of the MC and SCC types (e.g., a certain fraction of the
proto-UCD may form monolithically and may merge with many other
massive clusters which formed in the same star cluster
complex). Interesting is that the upper and lower boundaries of the
models for plausible metallicities and ages encompass the observed
data nearly completely.  An invariant canonical IMF would lead to a
horizintal line at a level depending on the age and metallicity. The
brighter UCDs do not appear to be consistent with an invariant IMF 
\citep{Dabring09, Dabring12}.

\begin{figure}
\includegraphics[scale=0.9]{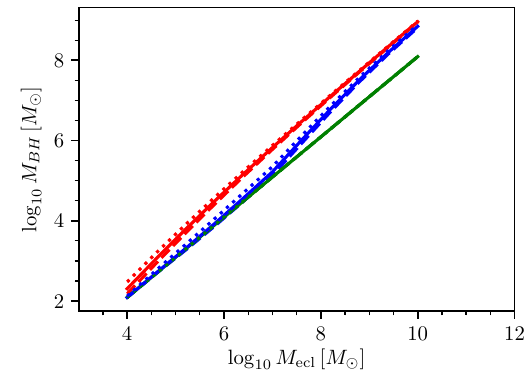}
\includegraphics[scale=0.9]{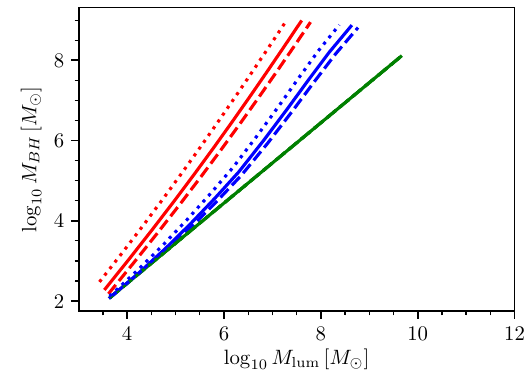}
\caption{\small The model mass in BHs ($M_{\rm BH}=M_{\rm BH, 0}$) 
in UCDs in dependence of the (upper
  panel) UCD birth mass in stars, $M_{\rm ecl}$, and of the (lower panel) UCD
  luminous mass ($M_{\rm lum}$, defined as the mass in all stars less
  massive than $1\,M_\odot$; see also Fig.~\ref{fig:UCDlum}). The
  models assume all BHs are retained and that a BH comprises 10~per
  cent of the birth mass of the star and that only stars more massive
 than $20\,M_\odot$ leave BHs. The red lines are for MC models and
  assume the varying IMF (Eq.~\ref{eq:IMF}), while the
  blue lines are for SCC models and assume the UCDs are made from a
  fully sampled population of star clusters (following the IGIMF3
  formulation in \citealt{Jerab17}). The green line is for the
  invariant canonical IMF. The dotted lines are for [Fe/H]$ = -5$, the
  full lines are for [Fe/H]$=-2$ and the dashed lines are for
  [Fe/H]$= 0$.  
}
\label{fig:UCDBH}
\end{figure}

\begin{figure}
\includegraphics[scale=0.9]{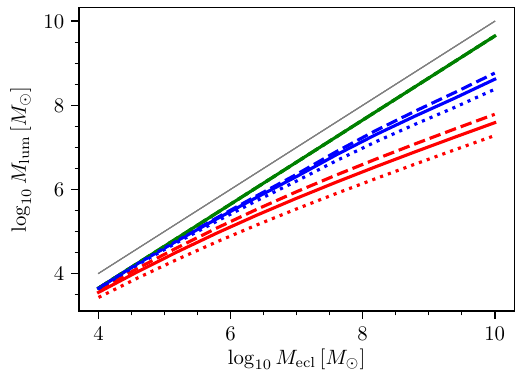}
\caption{\small The luminous mass $M_{\rm lum}$ in dependence of the initial
  MC (red) and SCC (blue)  stellar mass, $M_{\rm ecl}$, of the UCD models as in
  Fig.~\ref{fig:UCDBH}. The thin gray line is the 1:1 relation. 
}
\label{fig:UCDlum}
\end{figure}

The likely mass of all BHs (stemming from stars more massive than $20\,M_\odot$) as
a function of $M_{\rm ecl}$ and of the luminous mass for the typical
UCD ages for the MC and SCC types is shown in
Fig.~\ref{fig:UCDBH}. The luminous mass, $M_{\rm lum}$, is
approximated here to be the mass in all stars less massive than
$1\,M_\odot$.  Fig.~\ref{fig:UCDlum} depicts $M_{\rm lum}$ vs the
initial or birth stellar mass of the MC and SCC types of UCDs.

Thus, according to the model developed here, a present-day UCD 
with a mass in shining
stars of $M_{\rm lum} \approx 10^8\,M_\odot$ would have had the following properties:
\begin{itemize}

\item If born monolithically (MC type), a total birth mass in stars
comprising $M_{\rm ecl}\simgreat 10^{10}\,M_\odot$ and a mass in BHs of
$M_{\rm BH,0}\simgreat 10^9\,M_\odot$. This type of object would
have $L_{\rm bol} \approx 10^{13}\,L_{{\rm bol} \odot}$ which varies by
10~per cent on a monthly time-scale due to the large number of
core-collapse supernovae. It would drive a metal-rich, massive-star- and 
supernova-driven outflow of $\approx 200\,M_\odot/$yr with an outflow speed of probably $>1000\,$km/s
(outflow velocities $\approx 1000\,$km/s are computed for $10^5 < M_{\rm ecl}/M_\odot \le 10^6$ by \citealt{ChevalierClegg85, Tenorio-Tagle+07,Tenorio-Tagle+10, 
Silich+11}). \cite{Trakhtenbrot20} discusses similar observed outflows from high redshift quasar-hosting systems.
 An observer may interpret the luminosity to be the
accretion luminosity onto a SMBH of Eddington mass 
$\approx 3 \times 10^7\,M_\odot$ (Eq.~\ref{eq:Ledd1}).
This similarity in spatially unresolved observational data between very young super-star-clusters and accreting SMBHs, which 
extends also to the spectral energy distribution, has been discussed in the past 
(\citealt{TerlevichMelnick85}, for a historical review see \citealt{Shields99}).
Such an MC type cluster may thus appear, to a certain degree, as a very high redshift
quasar (see \citealt{Jerab17}, who however only calculate the stellar SED of such objects). The masses of SMBHs at high redshift are estimated from the 
width of their spectral lines as well as their accretion luminosity, as successfully 
obtained for the first time for a $z=5.9$ quasar by \cite{Eilers+18}.
    
\item If born from a full distribution of star clusters (SCC type), a
  total birth mass in stars comprising
  $M_{\rm ecl} \approx 10^{9.5}\,M_\odot$ with
  $M_{\rm BH,0} \approx 10^{8.5}\,M_\odot$.

\item If born with a canonical IMF, a
  total birth mass in stars comprising
  $M_{\rm ecl} \approx 10^{8.1}\,M_\odot$ with
  $M_{\rm BH,0} \approx 10^{6.5}\,M_\odot$. 
  
\end{itemize}
  
Note that the apparent quasar-likeness extends also to this (canonical) and the previous SCC case, with the MC~case  being the most extreme in the present models (subject to the possibility that the IMF may have been even more top-heavy at very low metallicity).
Whether such objects do in fact resemble the observed properties of high redshift quasars needs to be ascertained and tested by calculating the spectrum of the metal-rich $>1000\,$km/s fast outflow in combination with that of the stellar SED and dust self-obscuration. 
\cite{Trakhtenbrot20} writes
that the (rest-frame) UV spectra of the highest- and lower-redshift quasars are remarkably similar to each other (when matched in luminosity), referring in particular to the broad emission lines of C$_{\rm IV} \lambda1549$, C$_{\rm III]} \lambda1909$, and Mg$_{\rm II} \lambda2798$, and that such comparisons need to account for the tendency of highly accreting quasars to show blue-shifted (UV) broad lines.
%quotation:
%"The (rest-frame) UV spectra of the highest-redshift quasars are remarkably similar to those of lower-redshift ones (matched in luminosity), including in particular the broad emission lines of C$_{\rm IV} \lambda1549$, C$_{\rm III]} \lambda1909$, and Mg$_{\rm II} \lambda2798$ [...]. Such comparisons have to account for the tendency of highly accreting quasars to show blue-shifted (UV) broad lines". 
It is beyond the scope of this work to perform the necessary radiation transfer calculations within the dynamical winds. The similarities and differences of these to the outflows powered by accreting SMBHs at low to moderate redshift (e.g. \citealt{Arav+20}) will need to be studied. This may thus be seen as an interesting topic for future research.

\subsection{Summary}

The top-heaviness of the IMF in star-burst clusters
correlates positively with decreasing metallicity and increasing
star-formation rate density and thus birth mass of the cluster. The
observationally constrained galaxy-wide IMFs are consistent with this
variation. The observed properties of present-day UCDs are also
consistent with this variation. The massive star-burst clusters thus
contain populations of stellar BHs that can be estimated using this
variation of the IMF. The case can be raised that the first (very metal-poor)
hyper-massive star-burst clusters which most likely formed at the centres of the
spheroids resemble, for some dozens of~Myr, quasars in terms of spatial compactness,
luminosity and outflow speeds.

\section{The cluster of stellar mass black holes and its collapse}
\label{sec:model}

In this section the theoretical framework is developed, and
Sec.~\ref{sec:sm} presents the results of the calculations.

To introduce the model, assume for now that the central region of the
collapsing gas cloud, which will become the spheroid with (for
example) a peak global $SFR\simgreat 3 \times 10^3\,M_\odot$/yr, forms within about
a~Myr at least one massive ($M_{\rm ecl,max} \simgreat 3\times 10^7\,M_\odot$)
star-burst cluster with a birth half-mass radius
$r_{\rm h} \approx 1\,$pc \citep{Marks12b} and of very low metallicity
([M/H]$\,\approx-6$) and thus with a top-heavy IMF
($\alpha_3\approx1.05$, Eq.~\ref{eq:IMF}).
Note that the {\it first} stars will be formed as a population at the centre of the forming spheroid before the spheroid exists and before it has self-enriched with metals. The IMF of this population of stars in this first hyper-massive cluster will be very top-heavy, assuming \cite{Marks12} holds. After this population has formed its massive stars will begin to enrich the gas in their surrounding such that further star formation will be associated with a less top-heavy IMF. At the same time the star-burst will enshroud itself by dust produced by the evolving stars.
Such a cluster forms during the first few~Myr with an average SFR of only 
$\approx 100\,M_\odot$/yr with the SFR picking up over time as the rest of the spheroid forms.

During the first
about~50~Myr the star-burst cluster would be quasar-like
(\citealt{Jerab17}, Sec.~\ref{sec:BHcontent}) 
and may suppress further star formation within its vicinity \citep{Ploeckinger+19}.
One extreme assumption would be that 
it blows off a large fraction of
its mass ($\simgreat 75$~per cent, table~3 in \citealt{Dabring10}) in
the form of stellar winds and combined supernova ejecta thus probably
driving a massive metal-rich outflow
($\dot{M}_{\rm ofl} \simgreat 1\,M_\odot$/yr) and would, in this case,
expand by a factor of more than ten \citep{Dabring10}. This object
does not dissolve despite loosing more than 75~per cent of its initial
mass (in the form of winds and supernova ejecta) because it is confined
at the centre of the potential well of the forming~spheroid. 
It is also likely that such a central cluster will not be able to blow out
its stellar ejecta and residual gas due to the large escape speed from
its centre (in the above example, $v_{\rm esc} \simgreat 400\,$km/s). The binding energy of the central cluster $\propto M^2_{\rm cluster}$ while the feedback energy scales almost linearly with $M_{\rm cluster}$ such that for $M_{\rm cluster}\simgreat 10^7\,M_\odot$ the feedback energy will not suffice to remove the residual gas (fig.~3 in \citealt{Baumgardt+08} and sec.~4.6 in
\citealt{Wang+19b}). An example of a low-redshift star-burst where the high rate of central core-collapse supernovae per year, detected at radio wavelengths, is unable to blow out the gas is evident in Arp~220 \citep{Lonsdale+06, Dabring12}. 

The star-burst cluster is most likely to form mass-segregated
(\citealt{Pavlik+19} and references therein). If not, it will
mass-segregate; idealised N-body simulations having shown that the BH
sub-system shrinks to about 10~per cent of the original radius of the
cluster but on a Gyr time-scale \citep{BH13b}. During the explosion a BH may
receive a substantial kick if the explosion is not symmetric but the envelope
fall back fraction limits the velocities (e.g. \citealt{Fryer+12, Belczynski+16}).
Given the deep potential well of the case in study here, it is safe to assume
all BHs are retained \citep{Pavlik+18, Jerab17, Wang20}. Independently of
primordial mass-segregation, the whole BH-containing cluster will shrink due
to gas inflow, as discussed below (but neglecting the stars; the stars would aid in the shrinkage due to the deeper potential and would lead to a faster mass growth of the merging BHs so neglecting them here leads to a conservative estimate of the shrinking time).

When the supernova explosions cease in the central hyper-massive cluster, it is unavoidable that the gas from which the rest of the spheroid is still forming falls into it \citep{Pflamm-A09}. The evolving star-burst which forms the spheroid will not blow out all the gas as it is confined by the potential of the baryonic component and its particle or phantom dark matter halo \citep{FamaeyMcGaugh12, Lueghausen+13}. The survey of~1957 massive galaxies at 1 < z < 3 by \cite{Ramasawmy+19} shows the feedback from their accreting SMBHs to not measurably affect their star formation and thus their gas content. 

The time scale of gas in-fall is given by the free-fall or dynamical time, 
\begin{equation}
t_{\rm dyn} \approx \left( {3\,\pi \over  32\,{\rm G}\, \rho}  \right)^{1/2}, 
\label{eq:tff}
\end{equation}
with the density being (assuming a constant-density sphere) 
$\rho = 3\,M / (4\,\pi\,R^3)$ and 
$G\approx 0.0045\,{\rm pc}\,M_\odot^{-1}/$(pc/Myr)$^2$ is the
gravitational constant.  Assuming for the spheroid a mass $M=10^{11}\,M_\odot$ and a characteristic radius of $R = 2000\,$pc \citep{Fujimoto+20} and for the cluster $M=10^8\,M_\odot, R=10\,$pc \citep{Dabring08}, $t_{\rm dyn} = 4.7\,$Myr and $0.05\,$Myr, respectively. The in-falling gas is likely to form stars. Assuming a  star-formation efficiency of 10-30~per cent as is observed in the embedded clusters in molecular clouds of the Milky Way (e.g. \citealt{Megeath16})
(in actuality it may be smaller since the star-formation efficiency probably
increases with the cooling capability of the gas and thus with its metallicity),
most of the in-falling mass will remain gaseous, especially so since gas will
continue falling towards the centre as stars form from it.

As $t_{\rm dyn}$ is less than about 10~Myr, gas
inflow can be assumed to be essentially
instantaneous. The potential of
  a non-ionising cluster can draw-in a cooling flow \citep{Pflamm-A09, BM09}.  
  Observational evidence for gas
  accretion onto clusters triggering new star formation has been found
  \citep{FB17}.  For a global $SFR= 3\times 10^3 \,M_\odot$/yr about
$3 \times 10^{12}\,M_\odot$ of gas collapses to the~spheroid in
340~Myr assuming the star-formation efficiency
$\epsilon \approx \, $30~per cent. 
The innermost 1~per cent mass radius is
thus likely to have an inflow rate of
$\dot{M}_{\rm ifl} \approx 10^8\,M_\odot$/Myr. An example of
an observed significant mass inflow into the central region hosting an
active galactic nucleus is the galaxy~M81 at a distance of about
$3.6\,$Mpc \citep{Devereux2019}. The interested reader is referred to \cite{Boco+20} who discuss further observational evidence for gas-rich central galactic regions. The main conclusion here is that the central cluster of stellar-mass BHs will be in a significant star-forming gaseous environment during the down-sizing time-scale, Eq.~\ref{eq:downs}, which is the formation time-scale of the spheroid.

This leads to three connected effects: (i) the cluster shrinks due
to friction of the BHs on the gas, (ii)~the cluster shrinks due to the
inflowing gas (this is the inverse of the expansion of a star cluster
when it expels its residual gas), and (iii)~the stellar-mass black holes
(BHs) in the nuclear cluster undergo mass growth through accretion.

We assume the stars and other remnants in the cluster can be ignored
as their individual masses are much smaller than  those of the stellar BHs 
and thus they are either
absorbed by the BHs (enhancing their mass growth) or they are pushed
outside of $R$ via energy equipartition \citep{BK11, BH13b}. The
$\simgreat 50\,$Myr old cluster is assumed to consist of $N$ BHs, each
of mass $\mstar$. The example given below Eq.~\ref{eq:mstformed} has, for
$\mstar \approx 100\,M_\odot$ (ultra-low metallicity,
\citealt{Banerjee+2019}),
$N\approx 10^{4.3}\;(\beta = 2.4)$ to $N \approx 10^{6.1}\;(\beta=2.0)$.
A gravitational N-body system in a tidal field looses its 
members through energy-equipartition driven evaporation as
    ejections through binary encounters are rare \citep{Heggie2003,
      BM03}.  In the present context, in 200~Myr this cluster will
    have lost a negligible fraction of its BHs to galactocentric
    distances larger than $R$. Some of these may return due to
    dynamical friction on the dense stellar component in the inner
    region of the forming spheroid (see \citealt{BH13b} for a further 
    discussion). Assuming no BHs are lost is thus a reasonable
assumption in the present context.

This section discusses the processes acting as such a cluster of BHs
evolves.  The aim here is to understand if, at least in principle, the evolution
of a cluster of BHs subject to the presence and accretion of gas onto
it from the forming host galaxy may lead to the formation of an
SMBH-seed on an astrophysically relevant (within a few hundred~Myr)
time-scale.  The role of binary BHs which form near the core of the BH
cluster is discussed in Sec.~\ref{sec:heat}.  In
Sec.~\ref{sec:relcoll} it is shown that the BH sub-cluster will
core-collapse once a critical density is reached because the heating
from binary BHs ceases to be an energy source.  The evolution of the
BH sub-cluster taking into account gas drag is calculated in
Sec.~\ref{sec:DF}.  We emphasise that we only touch upon the relevant
physical processes. Detailed treatment awaits significant future
research involving radiation transport and relativistic stellar
dynamics.

%==============================================================================
\subsection{Heating from dynamically formed BH--BH binaries}
\label{sec:heat}

The physical situation we are studying is as follows: the central
region of the forming spheroid contains, after the first
$\approx 50\,$Myr, a cluster of $N$ BHs. This cluster will tend to
core collapse due to the energy-equipartition process.  As the density
of the BH sub-cluster increases the BHs form binaries through triple
encounters. These triple and subsequent encounters involving the BH
binaries stabilise the cluster against core collapse as the binaries
are a heating source (e.g. \citealt{Hills1975,
  Heggie1975,MH02,BBK10,MD12, Strader12, FP14, Giersz15}). That
  the energy-equipartition-driven core collapse of a star cluster is
  halted through energy production by binary encounters leading to
  gravothermal oscillations rather than runaway collapse of the core
  has been shown for the first time by means of numerical models by \cite{GierszSpurzem03}.

Thus a BH sub-cluster evolves such that the BH population
self-depletes through the dynamical formation of BH binaries in triple
encounters which, after their formation, exchange energy with a third
BH, leading to expulsion of the binary and single BH to distances $>R$
\citep{Banerjee17}.  Due to the deep potential well, the massive
nuclear cluster will retain most of its BHs (for a more detailed discussion on BH
retainment see \citealt{Jerab17, Pavlik+18}), but the binary activity will halt
core collapse of the BH sub-cluster. At the same time the cluster may
shrink as gas keeps falling into it (Sec.~\ref{sec:DF}), enhancing
heating through binary activity.

According to Henon’s principle \citep{Henon61, Henon75}, the energy
generating rate of the core from encounters between singles/binaries
with hard binaries is regulated by the bulk of the system.  Such
encounters transform internal binding energy of the binary into
kinetic energy of the centre of mass motions (eq. 1 in
\citealt{Antonini19}, where the numerical coefficient (see also below) $\zeta\approx 0.0926$ for isolated
systems, \citealt{Henon65}, and $\zeta\approx 0.0743$ for tidally
limited clusters, \citealt{Henon61}) which supports the cluster
against core collapse and thus leads to energy balance of the cluster.
Based on this Henon’s principle, \cite{Gieles+11} study the life cycle
of GCs in the Milky Way and suggest that almost all GCs are in energy
balance.  This balanced evolution theory is also the basis for
  Monte-Carlo simulation codes for GCs (e.g. \citealt{HypkiGiersz13,
    Joshi+2000}) having been confirmed to work well via comparison
  with direct $N$-body simulations (e.g. \citealt{Rodriguez+16}).
  Using direct N-body simulations, \cite{BH13} apply the balanced
  evolution theory to a system composed of a star cluster and a BH
  sub-cluster which generates the energy through binary BH encounters
  and \cite{Wang20} studies the implied dissolution rates of star
  clusters assuming top-heavy IMFs and thus enhanced BH contents.

The hard/soft binary boundary is
\citep{Heggie1975,Hills1975}
\begin{equation}
  \ab = \frac{G \ma \mb}{\m \sigma^2} \;.
\end{equation}
A binary with component masses $m_1, m_2$ and with a semi-major axis
$a>\ab$ is a soft binary and will be disrupted due to further
encounters (e.g. \citealt{Kroupa95}).  Thus only hard binaries with
$a<\ab$ can survive.  When the velocity dispersion of the BH sub-cluster,
$\sigma$, increases, $\ab$ decreases.  The heating process of the
cluster through the binaries is via the energy exchange between the
hard binaries encountering with other BHs and the cluster achieves
energy balance. The binary-single encounter timescale is
\citep{Heggie2003}
\begin{equation}
  \tce = \frac{\sigma }{8 \pi G \rho_{\rm BH} a} \;,
\end{equation}
where $\rho_{\rm BH}$ is the total BH mass density in the cluster and $a$ is the typical semi-major axis of a BH binary.
$\tce$ becomes longer when $\sigma$ increases or $\rho_{\rm BH}$ decreases.

The energy production rate through binary-BHs,
  $\dot{E}_{\rm bin, heat}$, can be written for equal-mass BHs by
  adopting eq.~1 and~2 in \cite{Antonini19} together with their
  equation for the total energy of the cluster, $E$ (we do not use
  their eq.~3). For the cluster mass we assume
  $M_{\rm cl}=M_{\rm g} + M_{\rm BH,0}$, which is the gas mass plus mass
  in $N$ equal-mass BHs,
  $M_{\rm g}=\eta_g\,M_{\rm BH,0}, M_{\rm BH,0} = N\,m_{\rm BH}$ for some
  number $\eta_g$.  Their $r_{\rm h}$ we assume to be the
  characteristic radius of the BH sub-cluster, $R$, and following the assumption by \cite{Gieles+11} we adopt for the numerical coefficient $\zeta = 0.1$ which relates the energy production to the total energy and the two-body relaxation time. The constant $\psi$ depends on the mass spectrum (for a discussion see \citealt{Antonini19}) and here $\psi=1$ for equal-mass BHs. The Coulomb logarithm,
  ln$\Lambda = {\rm ln}(N/2)$). Thus
\begin{equation}
\dot{E}_{\rm bin, heat} = 0.145 \frac{G^{1.5}(1+\eta_{\rm g})^{1.5}
  m_{\rm BH}^{2.5} N^{1.5}}{R^{2.5}} \, {\rm ln} \left( {N \over 2}
\right) \; .
\label{eq:binaryheating}
\end{equation}
Note that we ignore the rest of the stellar cluster in order to
simplify the problem. That is, we assume the forming spheroid
  contains at its centre a cluster of $N$ equal-mass BHs which is
  self-gravitating and in which binaries are formed dynamically which
  heat the cluster. Further assumptions are discussed as caveats in
  Sec.~\ref{sec:disc} and in particular in Sec.~\ref{sec:other}. 

In a BH sub-cluster with a velocity dispersion larger than a critical
value (Sec.~\ref{sec:relcoll}), the dynamically formed BH binaries
merge due to the radiation of gravitational waves before a further
dynamical encounter with another BH because these binaries have a
small encounter cross section. Three-body encounters which expel the
BHs from the cluster thus cease to be an important physical process.
Radiation of gravitational waves during the occurring BH mergers may
also impart sufficient recoils to allow the merged BHs to escape to
distances $>R$. Calculations which include gravitational redshift and
space-time curvature scattering of gravitational waves suggest that
recoil velocities are much less than $500\,$km/s \citep{Favata04}. In
a BH sub-cluster with $\sigma_{\rm h, BH}>100\,$km/s a sufficient
number of BHs are thus most probably retained despite the
gravitational-wave recoils (see also \citealt{Jerab17,
  Pavlik+18}). Runaway merging may proceed in this regime because
further mergers between the retained growing merged BH and a stellar
BH receive insignificant recoils. If the cluster has a velocity
dispersion larger than a critical value ($\simgreat 500\,$km/s,
Sec.~\ref{sec:relcoll}) the BH binaries typically merge through the
radiation of gravitational waves before the next encounter and the
binary BHs cease to be a heating source. When this occurs, the BH
cluster starts to shrink without the core being able to generate
sufficient energy to sustain balanced evolution.

%\begin{equation}
%\dot{E}_{\rm bin,heat} \approx 0.1 \, {|E_{\rm bind}| \over t_{\rm relax}},
%\label{eq:binaryheating}
%\end{equation}
%is significant relative to the binding energy,
%$E_{\rm bind}\approx -f_{\rm E}\,G\,(N\,m_{\rm BH})^2 / R$, of the
%cluster, where $f_{\rm E} \approx 0.2$ takes into account the
%distribution of mass within the cluster. The average two-body
%relaxation time of the cluster can be approximated as
%$t_{\rm relax} \approx {N \over 2} \, {\rm ln}\Lambda$, where
%${\rm ln}\Lambda$ is the Coulomb logarithm (see Eq.~\ref{eq:Coulombln}
%below). It is the time-scale over which the cluster reconfigures
%itself through evolving towards energy-equipartion and the form used
%here in the context of binary heating is (assuming a cluster of
%equal-mass BHs, eq.~2 in \citealt{Antonini19})
%\begin{equation}
%t_{\rm relax} = 0.138\,\sqrt{  {M_{\rm BH,0} \, R^3 \over G} } \; {1
% \over m_{\rm BH}\, {\rm ln}\Lambda}.
%\end{equation}

%==============================================================================

\subsection{Loss of support from BH binaries}
\label{sec:relcoll}

Once the cluster of BHs reaches a critical density and velocity
dispersion when the dynamically formed binaries are sufficiently tight
to merge through radiation of gravitational waves on a timescale
comparable to or faster than subsequent further encounters, the BH
binaries cease to be the energy source opposing core collapse (in
analogy to the final endothermic nuclear reactions in stars leading to
catastrophic stellar collapse and therefore core-collapse
supernovae). Balanced evolution breaks down and the cluster then 
collapses on a core-collapse time-scale,
forming a singularity, i.e. the SMBH-seed containing a significant
fraction of the BH sub-cluster mass (5~per cent in pure post-Newtonian
N-body computations, \citealt{Lee93,Kupi06}).

A hard BH binary can merge via gravitational wave (GW) radiation.  The
decay timescale of a BH binary \citep{Peters1964} is
\begin{equation}
  \tgw = \frac{a^4}{4 \beta}, \quad\quad {\rm where} \quad
  \beta= \frac{64}{5} \frac{G^3 \ma \mb (\ma + \mb )} {c^5} \;.
\label{eq:decaygw}
\end{equation}
Eq.~\ref{eq:decaygw} estimates the decay timescale for a circular
orbit for a given $a$.  For eccentric orbits, the merging timescale is
shorter.  Thus Eq.~\ref{eq:decaygw} provides a conservative, upper
limit.

When $a$ is sufficiently small, $\tce>\tgw$ obtains.  Such a binary
merges before any additional encounter happens.  This suggests that
when $\sigma$ is large enough in the core of the BH sub-cluster, $\ab$ can
be sufficiently small such that $\tce>\tgw$.  In this case, once hard
binaries form dynamically, they merge via GW radiation before they
experience a sufficient number of encounters to transfer their binding
energy to the kinetic energy of the field, i.e., the BH binaries stop
being stellar-dynamical heating sources for the BH sub-cluster
\citep{Lee95,Kupi06}.  When this occurs, it is expected that the
energy balance cannot be established and collapse happens to a
merged SMBH-seed.

\begin{figure}
  \includegraphics[scale=0.6]{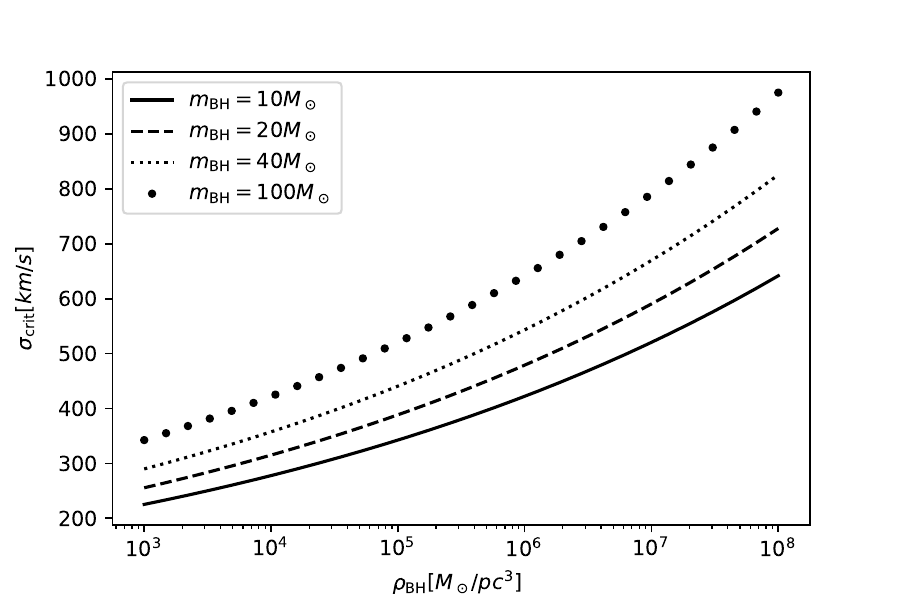}
  \caption{\small The velocity dispersion--density criterion
    (Eq.~\ref{eq:sigmacrit}) for the condition when the BH binaries
    cease to be heating sources (above the curves) such that the BH
    cluster can continue to shrink due to gas drag w/o opposition by
    BH binary encounters. Different line styles indicate different BH
    masses, $\mstar$. }
  \label{fig:sigma}
\end{figure}

The criterion $\tce>\tgw$ results in 
\begin{equation}
  a < \left (\frac{\beta \sigma}{2 \pi G \rho_{\rm BH}} \right )^{1/5} \;.
\end{equation}
By setting $a=\ab$, we can obtain a relation between $\sigma$ and
$\rho_{\rm BH}$,
\begin{equation}
  \sigma^{11} > \frac{5}{32} \pi \rho_{\rm BH} \frac{G^3 (\ma \mb )^4 c^5}{(\ma
    + \mb) \m^5} \;.
\label{eq:mergcondgw}
\end{equation}
In a system with equal-mass BHs ($m_1=m_2=\mstar$), the relation can
be simplified,
\begin{equation}
  \sigma_{\rm crit}^{11} = \frac{5}{64} \pi \rho_{\rm BH}  G^3 \mstar^2 c^5 \;,
\label{eq:sigmacrit}
\end{equation}
such that for $\sigma$ (Eq.~\ref{eq:veldispBH}) larger than
$\sigma_{\rm crit}$ the heating through binaries
(Eq.~\ref{eq:binaryheating}) ceases to be relevant.  With
Eq.~\ref{eq:sigmacrit} we can obtain the criterion on
$\sigma_{\rm crit}$ in dependence of $\rho_{\rm BH}$ when this occurs,
as shown in Fig.~\ref{fig:sigma}.

According to \cite{Guerkan+04} and \cite{FP14}, a cluster with $N$ BHs
and an average BH mass of $\mstar$ undergoes core collapse
after a time $t_{\rm cc} \approx 0.15 \, t_{\rm rc}$, where the
core two-body-relaxation time-scale is
\begin{equation}
t_{\rm rc} \approx  {0.065 \over G^2\, \mstar \, {\rm ln} \Lambda }
{\sigma_{\rm c}^3  \over \rho_{\rm c} } \;.
\label{eq:tcorecoll}
\end{equation}
As a rough approximation, the core-values for the velocity dispersion
and density, $\sigma_{\rm c}, \rho_{\rm c}$, can be replaced by
$\sigma_{\rm crit}, \rho_{\rm BH}$, respectively.  A cluster of $10^5$
BHs with an average mass of $50\,M_\odot$ will, once it reaches a
density of $\rho_{\rm BH} \approx 10^8\,M_\odot/$pc$^3$ and
$\sigma \approx 700\,$km/s, collapse within
$t_{\rm cc} \approx 3\,$Myr.

Inserting Eq.~\ref{eq:sigmacrit} into  Eq.~\ref{eq:tcorecoll},
\begin{equation}
{t_{\rm cc} \over {\rm Myr}} \approx 1.2 \times 10^8
 {1 \over {\rm ln}\Lambda } \; 
\left( { \mstar \over M_\odot} \right)^{-{5 \over 11}}  \; \left( 
{\rho_{\rm BH} \over (M_\odot / {\rm pc}^3)}
\right)^{-{8 \over 11}} \; .
\label{eq:finalgwcollapse}
\end{equation}

The BH sub-cluster collapses faster though, because gas drag
(Sec.~\ref{sec:DF}) shrinks it in the absence of binary heating. Once
$\sigma$ reaches 1~per cent of the speed of light, 
\begin{equation}
\sigma_{\rm rel} \equiv 3000\,{\rm km/s},
\label{eq:sigmarel}
\end{equation}
the BH sub-cluster is referred to throughout this text as being in the
relativistic state. This critical value is adopted here in view of the N-body work studying the relativistic collapse of BH clusters starting with comparable velocity dispersions \citep{Lee93, Lee95, Kupi06}.

%==============================================================================
\subsection{Shrinkage of BH sub-cluster through gas drag}
\label{sec:DF}

We consider a self-gravitating system consisting of BHs of total mass
$M_{\rm BH}(t) = N\mstar(t)$ and a gas mass of
\begin{equation}
M_{\rm g}(t) = (4/3)\,\pi\,R^3(t)\,\rho_{\rm g}(t) \; ,
\label{eq:gdensity}
\end{equation}
where the time dependence is explicitly written since the BHs may
accrete from the gas and the cluster may accrete additional gas,
$R(t)$ is the characteristic radius of the BH sub-cluster, and
$\rho_{\rm g}(t)$ is the gas mass volume density. The gas density
becomes 
\begin{equation}
\rho_{\rm g}(t) = \gasc \frac{3\, M_{\rm g}(t)}{4\pi R^3(t)}\;,
\label{eq:gasvoldens}
\end{equation}
where $\gasc\approx 1$ covers the density actually not being constant
throughout the cluster and that the characteristic radius $R$ does not
cover the whole cluster. 

Assuming the BH sub-cluster to be close to virial equilibrium, i.e., the
velocity dispersion $\sigma$ being related to, $R$, the absolute value
of the potential energy, $E$, is
%\begin{equation}
%-E \approx \virc\frac{(\etag + 1)^2 N^2 G\mstar^2}{R} \approx
%N\mstar\sigma^2\; .
%\label{eq:virial}
%\end{equation}
\begin{equation}
E(t) = \virc\frac{G\, \left(M_{\rm g}(t)  + N\,\mstar(t) \right)^2}{R(t)} =
N\,\mstar(t) \, \sigma^2(t)\; .
\label{eq:virial}
\end{equation}
The dimensionless factor $\virc \approx 1$ and covers, e.g., a
departure from virial equilibrium or a particular shape of the
potential well. From this follows the velocity dispersion of the BHs
(dropping from hereon the explicit time dependence),
\begin{equation}
\sigma = \left(
\virc \, {G\,\left(M_{\rm g}+N\,\mstar \right)^2 \over
 R\,N\,\mstar }
\right)^{1/2}.
\label{eq:veldispBH}
\end{equation}

We further assume that the BHs dissipate their kinetic energy as they
move through the gas medium. Dynamical friction will lead to a drag
force on each BH of
\begin{equation}
F = \dot{p} = -4\pi \dragc \ln\Lambda \,  \rho_{\rm g} \, \frac{G^2 \mstar^2}{v^2}\; ,
\label{eq:pdot}
\end{equation}
which points in the direction opposite to the BH's motion. 
This first order approximation
  is sufficient for the current context; a similar formula can be
  derived within the framework of Bondi-Hoyle accretion onto a moving
  object (Sec.~\ref{sec:BHaccr}). In
Eq.~\ref{eq:pdot}, $v$ is the speed of the BH with respect to the gas,
$\dragc \approx 1$, the exact value of which depends on the particular
derivation (e.g., in the case of Chandrasekhar dynamical friction,
$\dragc \approx \pi / 3$) and $\ln\Lambda = \ln(R/b_{\rm min})$ is the
Coulomb logarithm with $b_{\rm min}$ being the physically relevant
minimum gravitational encounter distance. The Coulomb logarithm can be
estimated by equating the kinetic energy of a particle with its
binding energy at minumum distance, 
\begin{equation}
\ln\Lambda =  \ln\frac{Rv^2}{2\,G\mstar}\;.
\label{eq:Coulombln}
\end{equation}
For $v^2 \approx \sigma^2$ and for a self-gravitating sytem near
virial equilibrium, $\ln\Lambda = \ln (N/2)$ is a good approximation.
The drag force (Eq.~\ref{eq:pdot}) will lead do a decrease
(dissipation) of the BH's (kinetic) energy,
\begin{equation}
\dot{E}_{\rm BH} = v\dot{p} = -4\pi\dragc \ln (N/2) \,
\rho_{\rm g} \, \frac{G^2 \mstar^2}{v}\;.
\label{eq:ediss_star}
\end{equation}
Considering a cluster of $N$ bodies of mass $\mstar$ and since $v
\approx \sigma$, the net energy dissipation per
unit time is estimated as
\begin{equation}
\dot{E}_{\rm diss} = -4\pi\dragc \ln{(N/2)} \, \rho_{\rm g} \, \frac{N G^2 \mstar^2}{\sigma}\;.
\label{eq:edisstot}
\end{equation}
The energy generation rate due to binary heating is given by
Eq.~\ref{eq:binaryheating} such that the overall rate of change of
binding energy of the BH sub-cluster becomes
\begin{equation}
\dot{E} = \dot{E}_{\rm diss} + \dot{E}_{\rm bin, heat}.
\label{eq:overalenergychange}
\end{equation}

From Eq.~\ref{eq:virial} follows (assuming $N=\,$constant)
%\begin{equation}
%\dot{E} = -f_{\rm v} \, G \, \left(
%{\left( M_{\rm g} + N\,\mstar  \right)^2 \over R^2} \, \dot{R}- 
%{2\,(M_{\rm g} + N\,\mstar)\,(\dot{M_{\rm g}} + N\,\dot{\mstar})}
%\right)
%\label{eq:master1}
%\end{equation}
%or
\begin{equation}
\dot{R} = 
\left( 
{\dot{E} \over f_{\rm v}\,G} - {2 \over R}\, (M_{\rm g} +
  N\,\mstar)\,(\dot{M_{\rm g}} + N\,\dot{\mstar})
\right) 
{R^2 \over \left( M_{\rm g} + N\,\mstar \right)^2} \;.
\label{eq:master}
\end{equation}
Note that if the cluster of BHs accretes gas then $\dot{M}_{\rm g}>0$
and the cluster shrinks (the inverse of gas expulsion). It also
shrinks due to gas drag, since $\dot{E}_{\rm diss} < 0$, but will
expand if binary heating dominates. This is the case in the absence of
gas ($E_{\rm diss}=0$, neglecting other energy dissipation mechanisms
for now).

Note also that if the binary heating just balances the energy
dissipation through gas drag,
$\dot{E}_{\rm diss}  + \dot{E}_{\rm bin, heat} = 0$, it follows, with
Eq.~\ref{eq:gasvoldens} and~\ref{eq:veldispBH} and assuming the gas
mass within $R$ is a constant 
fraction or multiple, $\eta_{\rm g}$, of the BH
cluster mass, $M_{\rm g}=\eta_{\rm g} \, N \, m_{\rm BH}$, that
\begin{equation}
\frac{\eta_g}{(1+\eta_g)^{2.5}} = 0.0483.
\label{eq:energybalance}
\end{equation}
Thus, for $\eta_{\rm g} < 0.055$ and for $\eta_{\rm g} > 5.78$ the BH
cluster expands because binary heating dominates over gas drag, while
for values in between the cluster contracts. At large $\eta_{\rm g}$
the velocity dispersion, $\sigma$, becomes too large for gas drag to
be significant (Eq.~\ref{eq:edisstot}).  For large values of
$\eta_{\rm g}$ the BH sub-cluster may also be in the relativistic regime
(Eq.~\ref{eq:sigmarel}) initially, experiencing core-collapse and thus
runaway BH--BH merging within 10--15 median two body relaxation times
as a consequence of energy loss through the emission of gravitational
waves \citep{Kupi06}. The collapse may be much faster though if gas
falls in from the forming and evolving spheroid.

Eq.~\ref{eq:master} will be used to investigate the shrinkage of the
BH sub-cluster.

\subsection{Summary}

The equation governing the radius of the BH sub-cluster
(Eq.~\ref{eq:master}) has been derived taking into account the heating
through dynamically formed BH--BH binaries
(Eq.~\ref{eq:binaryheating}) and thus the strive towards expansion of
$R$, loss of their support when the velocity dispersion of the BH
sub-cluster surpasses a critical value (Eq.~\ref{eq:sigmacrit}) and
the effect of gas on shrinking the cluster (Eq.~\ref{eq:edisstot}). If
a fraction $\eta_{\rm g}$ of the BH sub-cluster mass is in gas, then
for $\eta_{\rm g} < 0.06$ and $\eta_{\rm g}>6$ binary BH--BH heating
drives the cluster towards expansion. Nevertheless, for
$\eta_{\rm g}>6$ the BH sub-cluster may be in the relativistic state
(Eq.~\ref{eq:sigmarel}) initially. When the BH sub-cluster reaches, or
is, in this state, it will collapse within 10-15~median two-body
relaxation times, forming a runaway merger through gravitational-wave
emitting BH--BH encounters comprising about 5~per cent of the BH
sub-cluster mass \citep{Lee93, Kupi06}. This collapse time is likely
shorter and the SMBH-seed mass will be likely larger if gas from the
forming spheroid continues to fall onto the BH sub-cluster. This phase
of the evolution remains to be understood though.

%==============================================================================
\section{The shrinking BH sub-cluster}
\label{sec:sm}

The evolution of the BH sub-cluster through binary heating and gas
drag is calculated in Sec.~\ref{sec:sols}. The results are discussed in the context of other, independent, work in Sec.~\ref{sec:relatedwork}, and Sec.~\ref{sec:sm_summary} contains a brief summary of this section.  

\subsection{Solutions}
\label{sec:sols}
The aim is to compute the radius of the BH sub-cluster
in order to find the time, $t_{\rm rel}$,
in initial BH cluster radius and mass ($R_0, M_{\rm BH,0}$) space
when the BH cluster reaches the relativistic state
(Eq.~\ref{eq:sigmarel}). During the evolution we also note the time,
$t_{\rm crit}$, when the condition given by Eq.~\ref{eq:sigmacrit} on
its velocity dispersion (Eq.~\ref{eq:veldispBH}) is reached, i.e. when
the BH binaries cease to be a heating source. The mass of the 
BH sub-cluster is assumed to be constant, 
$M_{\rm BH}(t)=M_{\rm BH,0}=N\,m_{\rm BH}$ (see Sec.~\ref{sec:BHaccr}).

We assume the BH sub-cluster is embedded in gas of constant total 
mass $M_{\rm g}=\eta_{\rm g}\, N \, m_{\rm BH}$ within $R(t)$ for 
different
values of $\eta_{\rm g}$ and that $\dot{m}_{\rm BH}=0$. The gas mass
is assumed to be constant within $R(t)$ in a rough approximation of
the gas accretion from the forming spheroid being balanced by feedback
from the BH sub-cluster (see Sec.~\ref{sec:disc} for a further
discussion).

The differential equation for the evolution of $R(t)$
(Eq.~\ref{eq:master}) is integrated with the python scipy library
function called {\sc odeint} \citep{Virtanen19} for solving systems of the form
$dy/dt = f(y)$.  The library is based on the {\sc odepack} fortran77
ordinary differential equation solver and uses the {\sc lsoda} package
\citep{Hindmarsch83, Radh93, Brown89}. It uses the dense or banded
Jacobian for stiff problems and automatically evaluates if the system
is stiff or non-stiff.

In order to map-out the parameter space we solve the equation
$\dot{R}(t)$ (Eq.~\ref{eq:master}) for $R(t)$ for an equidistant  grid of the
parameters $N=M_{\rm BH,0}/m_{\rm BH}$ (from $10^3$ to $10^7\,$
with 100 
points in log$_{10}$ space), $R_0$ (between $0.5$ and
$10\,$pc in linear space with~100 points) and for several values of
$m_{\rm BH}$ and $\eta_{\rm g}$ (Fig.~\ref{fig:solutions_m1_1}).  To
solve the equation efficiently the following procedure is applied for
each set of parameters
($M_{\rm BH,0}, R_0, m_{\rm BH}, \eta_{\rm g}$):

\begin{enumerate}

\item If the system is initially relativistic it is noted and we move
  to the next set of parameters. This means its initial condition
  (Eq.~\ref{eq:veldispBH}) fulfils Eq.~\ref{eq:sigmarel},
i.e. $\sigma \ge \sigma_{\rm rel}$.

\item In the case that the system is not initially relativistic, at
  first $R(t)$ is found for the array ranging in time from
  $t = 10^3\,$Myr to $t = 10^4\,$Myr in $10^6$ log-steps to evaluate
  the nature of the given solution. We can thus estimate the
time-scale on which the radius evolves. In case the
  system is collapsing too slowly, that is, if this characteristic
  time scale is longer than $1\,$Gyr, then this is noted and we
  continue with the next set of parameters.

\item In the case that the characteristic time scale is shorter than
  $1\,$Gyr we compute a finer grid for $R(t)$ and identify the time,
  $t_{\rm crit}$, when the BH binaries cease to heat the cluster
  (Eq.~\ref{eq:sigmacrit}) and the time, $t_{\rm rel}$, when the BH
  cluster reaches the relativistic state (Eq.~\ref{eq:sigmarel}). The
  former time is noted for interest, but it is the latter time that we
  are mostly interested in.
\end{enumerate}

The computation of a whole grid (as seen in one of the panels in
Fig.~\ref{fig:solutions_m1_1}) takes about 2-6 hours of CPU time on a
modern PC with a 2.8~GHz QC~i7 processor.

\begin{figure}
\includegraphics[scale=0.9]{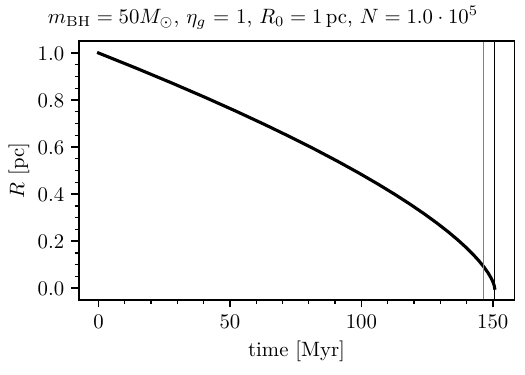}
\caption{\small The evolution of the cluster radius, $R(t)$, for
  $\eta_{\rm g}=1.0$, $m_{\rm BH}=50\,M_\odot$,
  $M_{\rm BH,0}=5\times 10^6\,M_\odot$ and $R_0=1\,$pc.  The cluster
  of BHs contracts because gas drag outweighs binary heating. See
  Fig.~\ref{fig:Erat} for the ratio of the rates of energy change due
  to gas drag and binary heating. The vertical thin grey line
  represents the time, $t_{\rm crit}$, when binary heating ceases to be relevant
  (Eq.~\ref{eq:sigmacrit}) and the vertical thick line shows the time,
  $t_{\rm rel}$, when the BH cluster reaches the relativstic state
  (Eq.~\ref{eq:sigmarel}). }
\label{fig:R}
\end{figure}

\begin{figure}
\includegraphics[scale=0.9]{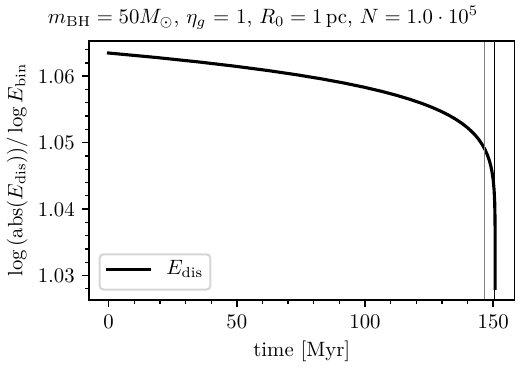}
\caption{\small The ratio of the rates of energy change due to gas
  drag and binary heating for the parameters as in
  Fig.~\ref{fig:R}. The ratio of the log$_{10}$ values of
  $|\dot{E}_{\rm diss}|$ and $\dot{E}_{\rm bin, heat}$ is larger
  than~one because gas drag overweighs binary heating. The ratio
  decreases because binary heating becomes increasingly significant as
  the cluster shrinks forming more energetic binaries.  The vertical
  lines have the same meaning as in Fig.~\ref{fig:R}.}
\label{fig:Erat}
\end{figure}

\begin{figure}
\includegraphics[scale=0.9]{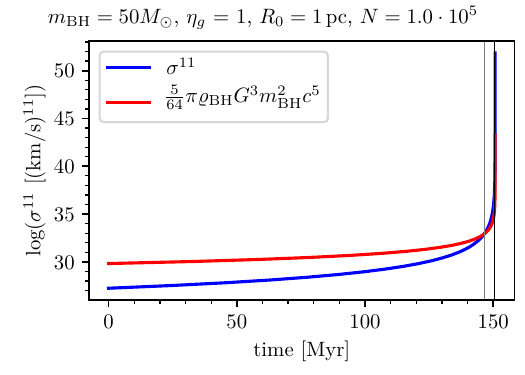}
\caption{\small The evolution of the velocity dispersion
  (Eq.~\ref{eq:veldispBH}, blue line) and of the condition when
  binaries cease to be a heating source (Eq.~\ref{eq:sigmacrit}, red
  line) for the parameters as in Fig.~\ref{fig:R}. The velocity
  dispersion increases as the cluster plus gas cloud shrinks. When the
  two lines intersect (the left vertical line) binary heating ceases
  to be active.  The vertical lines have the same meaning as in
  Fig.~\ref{fig:R}. Note that log$=$log$_{10}$. }
\label{fig:sigmarel}
\end{figure}

\begin{figure*}
\includegraphics[width=0.49\textwidth]{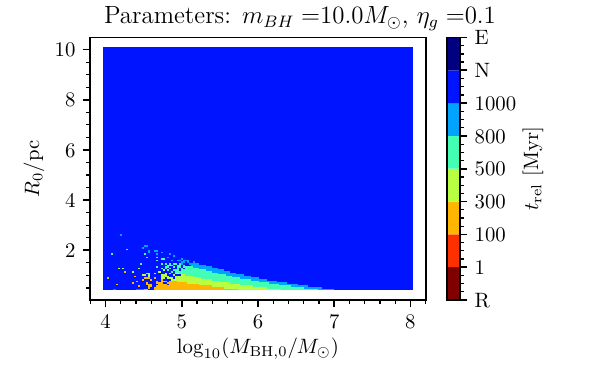}
\includegraphics[width=0.49\textwidth]{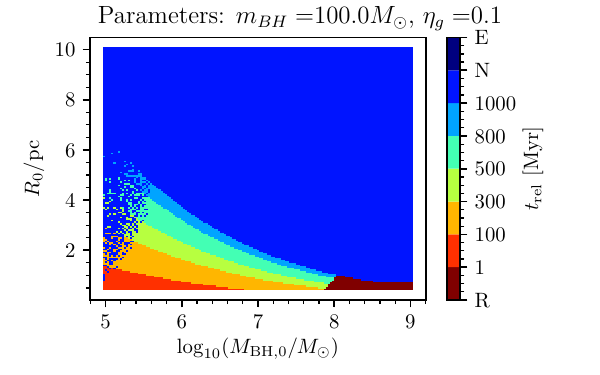}
\includegraphics[width=0.49\textwidth]{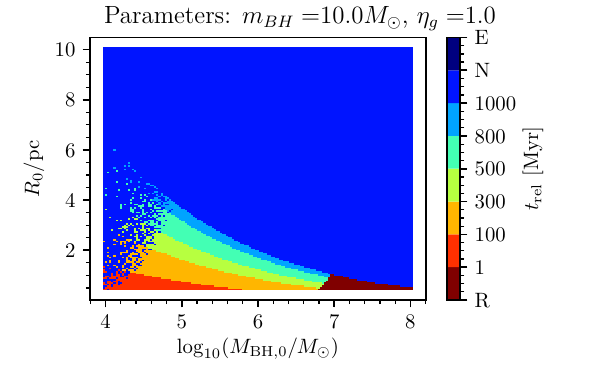}
\includegraphics[width=0.49\textwidth]{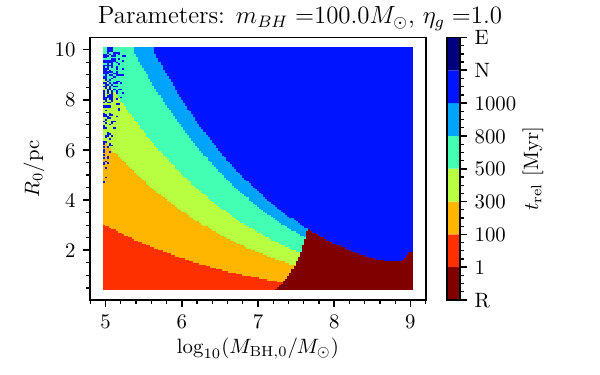}
\includegraphics[width=0.49\textwidth]{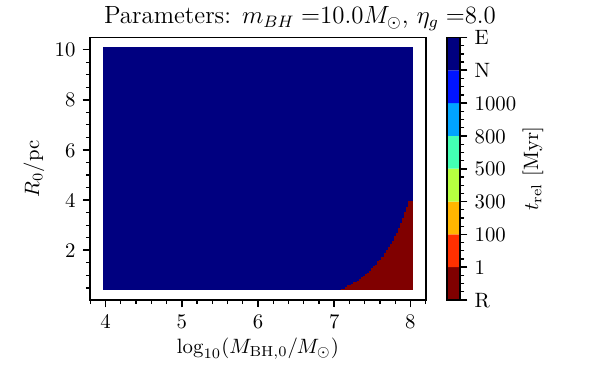}
\includegraphics[width=0.49\textwidth]{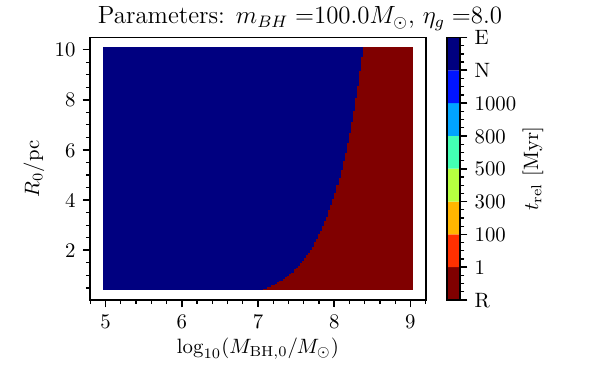}
\caption{\small The time, $t_{\rm rel}$, in
  $R_0, M_{\rm BH, 0}=N\,m_{\rm BH}$ space. $t_{\rm rel}$ is in Myr
  and is coloured according to the key: the middle- and dark-blue
  regions indicate, respectively, a shrinkage time longer than~1~Gyr
  and expansion, the brown region indicates the cluster to reach the
  relativistic state within~1~Myr.  Note that for $\eta_{\rm g}<0.055$
  the BH sub-cluster expands. For $\eta_{\rm g}>5.78$ (see
  Eq.~\ref{eq:energybalance}) it may also formally expand, but for 
  $M_{\rm BH, 0} > 10^7\,M_\odot$ and for a
  large fraction of the parameter space (shown here in brown in the
  bottom two panels) the BH sub-cluster is already initially in the
  relativistic state and will collapse and form a runaway BH merger
  within 10--15~median two-body relaxation times. For this to occur gas is therefore not needed, but the gas drag
  shortens this time. Note also that the $R_0 - M_{\rm BH,0}$ range for collapee to a SMBH seed becomes smaller for larger $M_{\rm BH,0}$ because the larger velocity dispersion leads to less efficient gas drag of the BHs.
  }
\label{fig:solutions_m1_1}
\end{figure*}

An example of the evolution of the radius of the BH sub-cluster, $R(t)$,
is visualised in Fig.~\ref{fig:R} which demonstrates a contracting
case. For the latter, the ratio of the energy dissipation through gas
drag to the heating of the cluster through binary BH activity is
plotted in Fig.~\ref{fig:Erat}. It demonstrates that, while the
dissipation dominates initially, binary heating becomes more
significant as the radius shrinks. The corresponding evolution of the
velocity dispersion of the BH sub-cluster in comparison to the critical
condition when binary heating ceases are shown in
Fig.~\ref{fig:sigmarel}.

Fig.~\ref{fig:solutions_m1_1} delineates the different behaviour in
$R_0, M_{\rm BH,0}=M_{\rm BH}$ space. For the parameters of the upper
four panels ($\eta_{\rm g}=0.1$ and~1), the whole shown space leads to
contracting solutions, i.e., heating through dynamically formed BH--BH
binaries is smaller than the gas drag which the BHs experience in the
BH sub-cluster. In the light blue region the shrinking time is longer
than a~Gyr.  But even with a moderate gas fraction
($0.1 < \eta_{\rm g} < 1$), BH sub-clusters with
$R_0 \simless \,1.5-4$pc and $M_{\rm BH,0}\simgreat 10^4\,M_\odot$
reach the relativstic state within much less than a~Gyr. As
$\eta_{\rm g}$ or $m_{\rm BH}$ increases, the collapsing solution
space increases significantly. For the cosmologically relevant cases
($m_{\rm BH} = 100\,M_\odot$ and $\eta_{\rm g}\approx {\rm few}$), BH
clusters with $R_0 \simless 6\,$pc and
$M_{\rm BH, 0} \approx 10^5\,M_\odot$ have
$t_{\rm rel} < 300\,$Myr.

Once in the relativistic state and if gas plays no role, the BH
sub-cluster experiences core-collapse within 10--15 median two-body
relaxation times \citep{Lee93, Lee95, Kupi06}, i.e.,
$t_{\rm collapse}/t_{\rm rlx} \approx 10-15$ with
$t_{\rm rlx} \approx t_{\rm cross} \, N/{\rm ln}(N/2)$ and the
crossing time $t_{\rm cross}=2\,R/\sigma$.  Taking a conservative
estimate,
\begin{equation}
t_{\rm collapse} /{\rm Myr} \approx {15 \over  {\rm ln}(0.5\,M_{\rm BH}/m_{\rm BH})} \,  
{M_{\rm BH} \over m_{\rm BH}} \, {2\,R_{\rm rel}/{\rm pc} \over
  \sigma_{\rm rel} / ({\rm km/s})}.
\label{eq:trelax}
\end{equation}
The radius of the BH cluster in this state is, 
by Eq.~\ref{eq:veldispBH},
\begin{equation}
R_{\rm rel}/{\rm pc} \approx 
{M_{\rm BH}\over 2\times 10^9\,M_\odot} \, \left( 1+\eta_g \right)^2.
\label{eq:Rrel}
\end{equation}
Thus, for example, for $M_{\rm BH,0}=10^7\,M_\odot, m_{\rm
  BH}=10\,M_\odot, \eta_g=1$, $R_{\rm rel}\approx 2\times 10^{-2}\,$pc and $t_{\rm
  collapse}\approx 15\,$Myr. In reality, the collapse time is likely
to be shorter still since the BHs will have different masses and gas
drag, accretion and heating is likely to contribute to cluster shrinkage. 

Thus, in this state the BH cluster core collapses through the
radiation of gravitational waves in $\simless 15\,$Myr.  An
independent estimate for the speed of the final relativistic collapse
can be obtained as follows: eq.~15 in \cite{Lee93} implies a
collapse time of about $10^5\,$yr for $10^6\,$BHs,
$m_{\rm BH}=10\,M_\odot$ and $\sigma=3000\,$km/s. A comparable
relativistic collapse time scale of about $600\,$crossing times from
that state is obtained with N-body simulations involving higher-order
post-Newtonian terms (\citealt{Kupi06}; e.g. within $0.3 \times 10^5\,$yr for
$R=1\,$pc).  Thus, once the relativistic state is reached, the BH
sub-cluster can be assumed to essentially collapse near to instantly
to a SMBH-seed (i.e.  $t_{\rm collapse} \ll t_{\rm rel}$).  About
5~per cent of the BH sub-cluster mass merges to the SMBH-seed in this
state \citep{Lee93, Kupi06}, but the inflow of gas and consequently
the additional orbital shrinkage of the BHs is likely to increase this
to a significant fraction of $M_{\rm BH,0}$.

The upper two panels of Fig.~\ref{fig:solutions_m1_1} also demonstrate
that as $M_{\rm BH,0}$ increases, the $R_0$ values decrease for which
$t_{\rm rel} < 300\,$Myr. This results from gas drag becoming less
efficient for larger $\sigma$ (Eq.~\ref{eq:edisstot}).

The lowest two panels (with $\eta_{\rm g}=8$) demonstrate that none of
the solutions lead to a shrinking $R(t)$ (most of the
$R_0, M_{\rm BH,0}$ space being dark blue, i.e. expanding, by
Eq.~\ref{eq:energybalance}).  For
$M_{\rm BH, 0}\simgreat 10^7\,M_\odot$ and certain small to large
values of $R_0$ (shown as the brown region) the BH sub-clusters are
initially in the relativistic regime.  For these, collapse to a
SMBH-seed is also inevitable and will occur faster than the
core-collapse time scale of $10-15\,$median two-body relaxation times
since the gas accretion onto the BHs speeds up the core collapse
\citep{Leigh2013b}.  For sufficiently small $R_0$ this BH--BH runaway
merging time may be shorter than a few~Myr. The extreme (brown)
regions in Fig.~\ref{fig:solutions_m1_1} are not accessible to the
present modelling though.

\subsection{Related work}
\label{sec:relatedwork}

Sec.~\ref{sec:sols} has shown that the BH sub-cluster which forms
within the first massive star-burst cluster at the centre of the later
spheroid can shrink within about 200~Myr to the relativistic state due
to the gas which falls into it from the forming spheroid.

The results obtained here are in line with the work
of \cite{Leigh2013b} who discussed that 
gas-accretion onto a BH sub-cluster
speeds-up its core collapse due to the mass-growth of the BHs. 
But the mass-growth is Eddington-limited (see Sec.~\ref{sec:BHaccr}), 
limiting the mass-growth within the time $\Delta \tau$.
Our results are also in line with the suggestion by
\cite{Davies+11} and \cite{Boco+20}\footnote{The work presented here was conducted with
  neither prior knowledge of the work of \cite{Davies+11, Leigh2013b} nor of that by \cite{Boco+20}. These were discovered during the
  writing-up stage of this manuscript.} who point out 
  that the shrinkage of a cluster with BHs through the
accretion of gas onto it may enhance the formation of a SMBH-seed. The
difference to their work is that here much more massive star burst
clusters are considered (consistent with observations, given the SFRs
of galaxies) and that this work suggests a direct link between the formation
of the spheroid to the SMBH-seed via the stellar population, as
derived from published scalings of the stellar populations and the SFR
of the forming spheroid via the IGIMF theory (Sec.~\ref{sec:gwIMF}). 
Also, the here presented work relaxes the need for $\eta_{\rm g}>1$ to reach
sufficient compression for SMBH-seed formation \citep{Davies+11} to
values where the gas mass needed is smaller than that of the BH
sub-cluster ($\eta_{\rm g}<1$). Also, here, for the first time,
explicit time-scales are associated with the process in view of the
formation times of the massive spheroids and the populations of stars
they ought to contain if star formation in the early Universe follows
similar rules as inferred from local populations, including
low-metallicity ones.

The results obtained here are thus well consistent
with the work of \cite{Davies+11, Boco+20} and of \cite{Leigh2013b}, and this
work allows for the first time a quantification of the expected
scaling of the SMBH-seed mass with the mass of the forming spheroid,
as shown in Fig.~\ref{fig:SMBHseed} below.

\subsection{Summary}
\label{sec:sm_summary}

Solutions for the evolution of the radius of the BH sub-cluster are
obtained by taking into account heating through dynamically formed
BH--BH binaries and dissipation of BH orbital energy through gas which
presumably is in the BH sub-cluster from the forming spheroid. The gas
mass is assumed to be proportional to the BH sub-cluster mass within
$R$, mimicking very roughly the self-regulation through feedback and
accretion from the forming spheroid.  The BH sub-cluster can shrink to
the relativistic state when its equation of state changes form an
incompressible one to a pressureless one. At this point it collapses
within $10-15$~two body relaxation times by emitting gravitational
waves.  The calculations indicate that for viable initial values,
$R_0, M_{\rm BH, 0}$, the total time of $R$ shrinking to a SMBH-seed
mass occurs on an astrophysically interesting ($<200\,$Myr) time
scale.

%===============================================================================

\section{The galaxy--SMBH correlation}
\label{sec:corrs}

Given the correlations between the most massive forming star cluster
($M_{\rm ecl,max}$), its stellar population (IMF) and the mass of the
forming spheroid ($M_{\rm igal}$) discussed in
Secs~\ref{sec:galaxy_stcl} and~\ref{sec:stpop}, it is now possible to
quantify the expected relation between the present-day luminous mass,
$M_{\rm pgal}$, of the spheroid and the mass of the SMBH-seed which
forms in the most massive cluster near the centre of an emerging
spheroid through the relativistic collapse of its BH sub-cluster
(Sec.~\ref{sec:model}).

\begin{figure}
\includegraphics[scale=0.95]{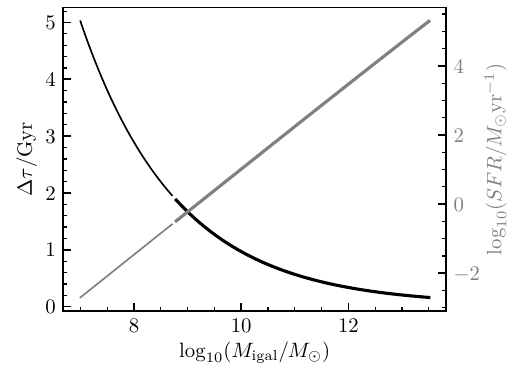}
\caption{\small The downsizing time, $\Delta \tau$
  (Eq.~\ref{eq:downs}, left axis) and the SFR (right axis) of a
  spheroid with a total mass in stars of $M_{\rm igal}$. The thick and
  thin lines with gap in between are explained in
  Fig.~\ref{fig:SMBHseed}.}
\label{fig:SFR} 
\end{figure}

\begin{figure}
\includegraphics[scale=0.9]{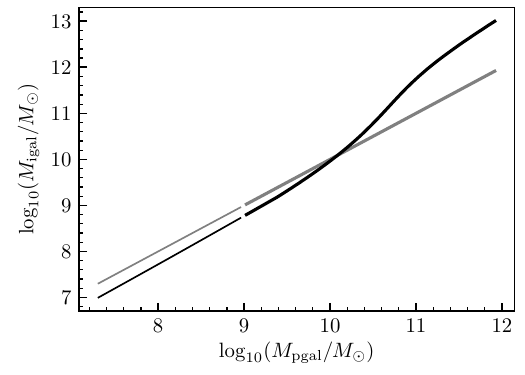}
\caption{\small The total stellar mass formed in the spheroid,
  $M_{\rm igal}$, as a function of its present-day luminous mass,
  $M_{\rm pgal}$, assuming the systematically varying IMF
  in each embedded cluster as given by Eq.~\ref{eq:IMF}
  (\citealt{Marks12}; see also \citealt{YJK17, Jerab18} and
  \citealt{Dabring19} for more detailed models).  An invariant
  canonical IMF (Sec.~\ref{sec:canIMF}) leads to a constant ratio of
  $M_{\rm pgal}/M_{\rm igal} \approx 0.7$ \citep{BM03}. The thick and
  thin lines with gap in between are explained in
  Fig.~\ref{fig:SMBHseed}. Note the upwards turn of the thick line
  towards larger masses. This is due to the increasingly top-heavy
  gwIMF with increasing SFR and thus a decreasing
  $M_{\rm pgal}/M_{\rm igal}$ ratio. 
  The gray thin and thick line is the 1:1 relation, the gap 
  having merely an orientative meaning). }
\label{fig:ipgal} 
\end{figure}

We proceed as follows: For a value of the total (i.e. initial) stellar
mass of the spheroid, $M_{\rm igal}$, the time-scale, $\Delta \tau$,
of its formation is calculated (Eq.~\ref{eq:downs}). Assuming the SFR
remains constant over this time-scale, $SFR=M_{\rm igal}/\Delta
\tau$ (Fig.~\ref{fig:SFR}).
The total stellar mass of the spheroid which forms over the downsizing
time, $M_{\rm igal}$, is plotted in Fig.~\ref{fig:ipgal} as a function
of the final luminous mass after about 12~Gyr of evolution,
$M_{\rm pgal}$, approximated here as the total mass in stars less
massive than $1\,M_\odot$.

Given a SFR, $M_{\rm ecl,max}$ follows from the WKL relation (Eq.~\ref{eq:WKL}) or more generally from Eq.~\ref{eq:Mdeltat}.
Under the assumptions made in Sec.~\ref{sec:galaxy_stcl}, as the 
spheroid assembles, a number of $N_{\rm gen}$ such
clusters form (Eq.~\ref{eq:Ngen}). But only the first-formed star-burst cluster 
is likely to be the most
relevant \citep{Ploeckinger+19}, as only it will have a very top-heavy
IMF as subsequently formed star-burst clusters will already be
metal-enriched and will therefore probably have an IMF which is closer
to the canonical one. From the expected number of stars more massive
than $20\,M_\odot$, the mass of the BH sub-cluster within
$M_{\rm ecl, max}$ follows (Fig.~\ref{fig:UCDBH}). This BH sub-cluster
is expected to be present within about 50~Myr of the formation of the
$M_{\rm ecl,max}$ cluster (see also Sec.~\ref{sec:ndens}).  We also assume, as a point of reference,
an invariant canonical IMF (Sec.~\ref{sec:canIMF}).

\begin{figure*}
\includegraphics[scale=0.9]{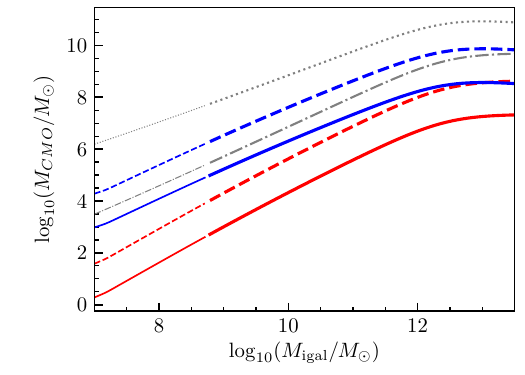}
\includegraphics[scale=0.9]{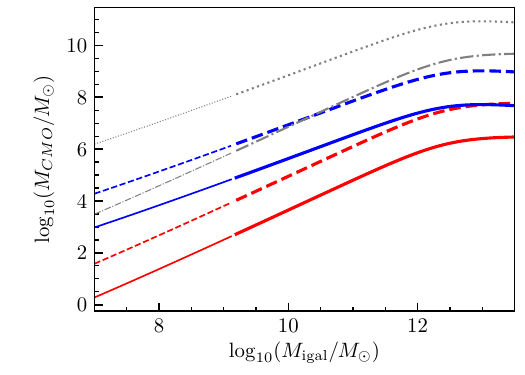}
\includegraphics[scale=0.9]{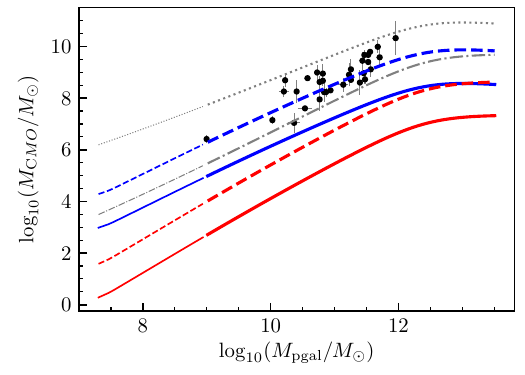}
\includegraphics[scale=0.9]{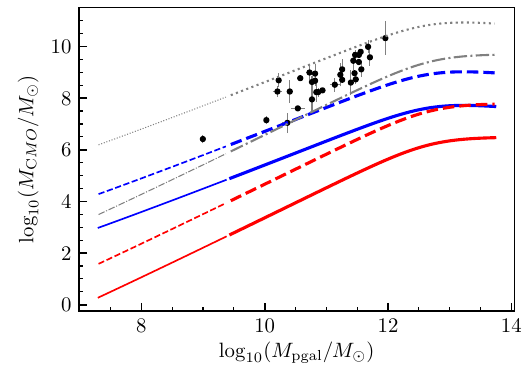}
\caption{\small The mass of the central massive object, $M_{\rm CMO}$,
  observed in a galaxy in dependence of the initial (total) stellar
  mass (upper panels) and the present-day luminous stellar mass (lower
  panels) of the hosting spheroid and assuming $\beta=2.0$ (see Fig.~\ref{fig:AppSMBHseed} for the case $\beta=2.4$). 
  The left panels are for the variable IMF (Sec.~\ref{sec:varIMF}) 
  while the right panels are for the invariant canonical IMF (Sec.~\ref{sec:canIMF}).
  The observational SMBH-mass data are
  the black dots with uncertainties (table~3 in \citealt{McConnell13},
  see also \citealt{Sanghvi+14, HeckmanBest14}).  Assuming that 5~per cent of the BH
  sub-cluster of mass $M_{\rm BH,0}$ in the first-formed cluster with a
  mass of $M_{\rm ecl,max}$ merges to a SMBH-seed within 100--300~Myr
  (Fig.~\ref{fig:solutions_m1_1}) yields the solid red line.  The
  dashed red line assumes all of $M_{\rm BH,0}$ merges to the SMBH-seed.
  The solid blue line assumes that the SMBH-seed mass is
  $0.05\,N_{\rm gen}\,M_{\rm BH,0}$, i.e., that 5~per cent of all BH
  sub-clusters of all formed $M_{\rm ecl,max}$-clusters merge to one
  SMBH-seed. The blue dashed line assumes $N_{\rm gen}\,M_{\rm BH,0}$ is
  the SMBH-seed mass, i.e. that all of the BH sub-clusters merge to an
  SMBH-seed due to gas infall onto the BH sub-clusters over the whole
  formation period (the downsizing time) of the spheroid.  The
  divergence between the respective blue and red lines with decreasing
  $M_{\rm igal}$ is a result of downsizing ($N_{\rm gen}$ increasing
  for decreasing $M_{\rm igal})$.  The lower grey dash-dotted line
  depicts the SMBH-seed mass if all of $M_{\rm ecl,max}$ collapses to
  the SMBH-seed. This scenario would only be valid if the gas-infall
  from the forming spheroid forces the whole cluster to coalesce to
  one massive object and is deemed to be merely an indicative possible
  upper mass limit. The case that all $N_{\rm gen}$ most massive 
  consecutively-formed central clusters collapse
  in this way is shown as the upper grey dotted line.  The gap and
  thinner lines for $M_{\rm igal, pgal} < 10^{9}\,M_\odot$ indicate
  the formal relationships but SMBH-seeds are unlikely to form for these
  spheroid masses as the BH clusters have $M_{\rm BH,0}<10^4\,M_\odot$
  which cannot coalesce to any reasonably-massive SMBH-seed
  (Fig.~\ref{fig:solutions_m1_1}).}
\label{fig:SMBHseed}
\end{figure*}

As gas from the forming spheroid falls into the central cluster, its
BH sub-cluster (and most realistically the whole cluster) shrinks.
For those cases where the BH sub-cluster undergoes a relativistic
collapse within 0.1--0.3~ Gyr (Fig.~\ref{fig:solutions_m1_1}),
Fig.~\ref{fig:SMBHseed} shows the expected SMBH-seed--spheroid mass
relation. Impressively, the slope of the analytical 
relation is very comparable to the observational data. Assuming that
only 5~per cent of the BH sub-cluster in the first cluster with a mass
$M_{\rm ecl, max}$ becomes the SMBH-seed, all such seeds need to grow
by three orders of magnitude in mass ($f_{\rm growth}= 10^3$) to reach
the observed SMBH masses shown as the data points in
Fig.~\ref{fig:SMBHseed}. This can be achieved by Eddington-limited constant 
super-Eddington accretion ($\epsilon_{\rm r}=0.1$, Sec.~\ref{sec:BHaccr})
by the SMBH-seeds within 347~Myr, since, from
Eq.~\ref{eq:mt}, the accretion time is
\begin{equation}
  t_{\rm accr} = (452/9)\; {\rm ln}(f_{\rm growth}).
\label{eq:factor}
\end{equation}
We emphasise that $t_{\rm accr} < \Delta \tau$, i.e., {\it the model
leads to the self-consistent result that the required accretion time
to reach the observed SMBH masses at a given $M_{\rm pgal}$ is
shorter than or comparable to the downsizing, i.e. spheroid formation, time scale
(Eq.~\ref{eq:downs}). This is a rather remarkable result of this model needing to be emphasised.}
In their review, \cite{MayerBonoli18} point out the lack of
direct radiation hydrodynamical calculations on nuclear or sub-nuclear scales that show that this growth mode can be sustained up to the extremely large BH masses needed to explain the bright high-z quasars. In this case the only advantage of the present BH merging model over that of population III remnant black holes would be that the present model leads to larger SMBH seed masses and thus shorter required accretion sustainment times and that the physical pathway of explaining SMBH seeds via gas-promoted merging BHs might, arguably, be better understood than the more uncertain physics of population III stars. However, other than that, both solution-ansatzae suffer from the same remaining problem of SMBH-seed growth to the SMBH masses observed, the present model to a significantly lesser extend though.
For larger SMBH-seed masses (the other coloured curves in
Fig.~\ref{fig:SMBHseed}) the required accretion time becomes shorter
or the accretion rate can be sub-Eddington.  The above holds true also
if the IMF does not vary, as shown by Fig.~\ref{fig:SMBHseed}. But in
this case the slope of the SMBH-seed--spheroid mass is somewhat
flatter, and the required growth factors, $f_{\rm growth}$, are
larger, but still requiring growth times of less than a Gyr to match
the observed relation.

Noteworthy is that if all centrally formed $M_{\rm ecl,max}$ clusters
merge to one cluster, then their combined mass continues the same
relation as that of the SMBH masses to
$M_{\rm pgal}\simless 10^9\,M_\odot$ for $\beta=2.0$
(Fig.~\ref{fig:SMBHseed}) and to
$M_{\rm pgal}\simless 10^{9.6}\,M_\odot$ for $\beta=2.4$
(Fig.~\ref{fig:AppSMBHseed}).  This is reminiscent of the
central-massive-object(CMO)--spheroid-mass correlation found by
\cite{WehnerHarris06} and further discussed by \cite{Capuzzo+17}. This
single continuous relation between the CMO mass and the luminosity of
their hosting spheroid \citep{WehnerHarris06} would, in the present
context, be consistent with the assembly of the spheroid: as the
spheroid forms over the downsizing time $\Delta \tau$, the inner,
nuclear-cluster-hosting region acquires a baryonic mass amounting to
$N_{\rm gen}\,M_{\rm ecl,max}$ (Eq.~\ref{eq:Ngen}).  The correlation
for dwarf elliptical galaxies and their nuclear cluster (the upper
grey dashed line in Figs~\ref{fig:SMBHseed} and~\ref{fig:AppSMBHseed})
may thus be explained. This amount of mass infall into the central
region would also lead to the compression and growth of the SMBH-seed
mass needed to reach the data points shown in both figures. While the
nuclear regions of spheroids form within the downsizing time scale,
late-type galaxies continue to grow in mass (e.g. \citealt{Speagle+14,
  Kroupa+20}) such that their nuclear star cluster is likely to grow
through in-situ star formation and mergers with freshly formed massive
clusters beyond the mass of that assembled during the initial
formation of its spheroidal component. Additional slow growth of the
SMBH through the disruption of stars continues until the present, most
efficiently though for low-mass SMBHs \citep{Brockamp+11}. This would
lead to a flatter $M_{\rm CMO}-M_{\rm pgal}$ relation for late-type
galaxies as found to be the case by \cite{Capuzzo+17}.

%
%%% Notes from TJ on how the figures were made:
%{\bf 
%Notes for figure from TJ: lover thick curves are assuming 5 per cent
%of MBH, upper ones for 100 per cent MBH The upper two sets are the
%same but for
%$M_{\rm BHtot} = (\Delta \tau / \delta 10\,{\rm Myr}) M_{\rm BH}$
%which is an upper limits assuming all most massive clusters and their
%BH contents forming near the centre of the developing spheroid merge.
%Thin upper dotted lines are for Meclmax and Meclmaxtot.
%$Mpgal = L_Vpgal*3$  for $M/L_V=3$ for 10Gyr metal rich pop w/o remnants
%(i.e. $m<1Msun$ stars) (Dab08).
%Migal is the mass in all stars formed in the galaxy, Mpgal is the present day
%luminous mass.
%$MBH<10^4Msun$  => no SMBH

The upper dotted lines in Figs~\ref{fig:SMBHseed} and~\ref{fig:AppSMBHseed} are obtained by adding all mass assembled in the nuclear region over the down-sizing time, i.e., it is $M_{\rm CMO}=N_{\rm gen}\,M_{\rm ecl,max}$. This means that the nuclear clusters, in this mathematical description of the highly complex central processes, form one after another and on top of each other, the second one forming when the first one is finished after $\delta t = 10\,$Myr. This is not entirely unphysical as the notion is that star-formation proceeds also in the central nuclear region as long as the spheroid assembles over the down sizing time scale. These $N_{\rm gen}$ clusters therefore do not need to first spiral towards the central region through dynamical friction.  The nuclear region is likely to grow further as non-nuclear massive clusters merge with the central region due to dynamical friction on the spheroid and due to gas drag (cf. \citealt{Bekki10}). The time-scale needed for a BH cluster with (for example) a mass $M_{\rm BH,0}=10^8\,M_\odot$ to sink to the center of its hosting spheroid due to dynamical friction on the stellar population can be estimated from the usual mass-segregation times scale, 
\begin{equation}
t_{\rm ms} \approx \left({ m_{\rm star} \over M_{\rm BH,0} }\right)\, t_{\rm sph,rlx},
\label{eq:t_msegr}
\end{equation}
where $m_{\rm star}\approx 0.5\,M_\odot$ is the mass of a typical star in the spheroid and $t_{\rm sph, rlx}\approx \left( N/8\,{\rm ln}\left(0.5\,N\right) \right)\, t_{\rm sph,cr} \approx 6\times 10^6\,$Gyr is the median two-body relaxation time 
for the spheroid with radius of 2~kpc and containing $N\approx 10^{11}\,$stars. The crossing time is $t_{\rm sph,cr}\approx 12\,$Myr and $t_{\rm ms}\approx 30\,$Myr. 
Thus, over less than $0.1\,$Gyr most such clusters will arrive at the centre, adding to the SMBH masses in some cases. However, how and if such BH clusters do merge with the SMBH is a major open question and goes beyond the scope of this work. This discussion is merely meant to indicate the great complexity of the variety of processes that will be acting in the nuclear regions of spheroids and cannot be treated in any detail in the present contribution.

%=============================================================================
\section{The redshift-dependent number of quasars}
\label{sec:ndens}

In this section the expected number of quasars at high redshift is assessed. Indeed, observations of the distribution of quasars on the sky and with redshift are being applied to constrain structure formation (e.g. \citealt{Song+16}).
From local observations it is known that most major galaxies host a SMBH and that the correlation between the SMBH mass and spheroid properties  exists but that the overall situation is complex \citep{KormendyHo13}. Taking the Local Group  of galaxies (which has a zero-velocity radius of about $1.5\,$Mpc) as an example, there are two SMBHs in two major galaxies, the MW and M31, which leads to an equivalent of about $1.4 \times 10^8$ galaxies hosting SMBHs per Gpc$^3$. The Catalogue of Nearby Galaxies \citep{Karachentsev+04, Karachentsev+13} contains about 1200 galaxies in total and 140~galaxies more massive than $10^{9}\,M_\odot$ within a sphere with a radius of about $11\,$Mpc. This is equivalent to $2.5 \times 10^7$ such galaxies per Gpc$^3$. Since it is to be expected that each of these galaxies hosted a SMBH in the early Universe, we would expect to observe of the order of $10^7-10^8$ quasars in a co-moving Gpc$^3$ volume at, e.g., a redshift of $z=6$ {\it if} all these SMBHs were in fact accreting at the same time and would not be obscured by dust. Observational surveys of the number density of quasars at a high redshift are extremely difficult 
(e.g. \citealt{Wu+11, Manti+17, Yang+18, PacucciLoeb19}) and prone to major observational bias towards detecting only the brightest quasars and missing those that are obscured by intervening dust or missing quasars if they have an anisotropic emission. 
For example, \cite{Jiang+2016} detect a few quasars per Gpc$^3$ at $z\approx 6$ with a decreasing number density with increasing redshift. There may thus be a problem in that, given the number of SMBHs per unit volume found in the local Universe, the expected number of quasars at a high redshift is not readily evident. 

\begin{figure}
\includegraphics[scale=0.33]{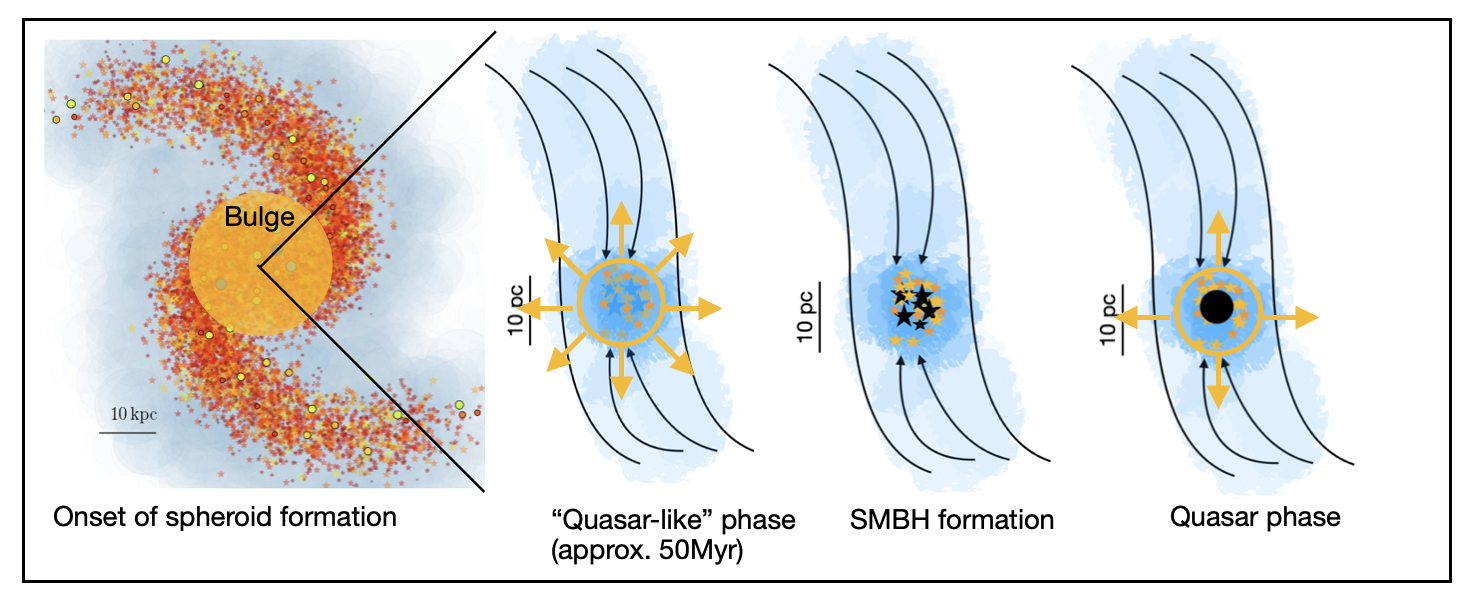}
\caption{\small{
Visualisation of the major model ingredients (see Table~\ref{tab:times}): From left to right: formation of the spheroid, the formation of the central hyper-massive star cluster (the quasar-like object), the evolution of the BH cluster left by the hyper-massive cluster through accretion of gas from the still forming spheroid to the SMBH seed, accretion onto the SMBH seed of gas from the still forming spheroid ("true" quasar phase).
}}
\label{fig:quasar_phases}
\end{figure}

It is clear that neither are all SMBHs formed at the same time nor do they all accrete in-phase such that the visibility of quasars should be spread-out over time, possibly helping to alleviate the missing quasar problem. Within the framework of the present model (see Fig.~\ref{fig:quasar_phases}), first the quasar-like hyper-massive clusters discussed above appear at very high redshift to then drop-out because they fade when the ionising stars die, for some of these to re-appear again after they formed SMBH seeds if these accrete gas sufficiently to shine as quasars. This accretion phase may last only while the spheroid forms, or later briefly again (of the order of a dynamical time, i.e. a hundred to a few hundred~Myr) when the spheroid or hosting galaxy has an encounter with another galaxy such that gas enters its central region. 

Using the model developed here we can estimate the observable number of quasars at a high redshift. This estimate also serves the purpose to visualise the redshift dependence of the two different phases which ought to exist: the first quasar-like hyper-massive star cluster phase and the subsequent accreting SMBH seed quasar phase. 

In the following we estimate the number of quasars as a function of redshift using empirical constraints to serve as a visualisation of the model rather than to provide explicit quantification which depend on structure formation in an adopted cosmological model. The estimate is very basic and we do not aim at a fully fledged quantification as this would be beyond the scope of this contribution. The calculations neither take into account flux limits nor that some quasar-like objects or quasars may be unobservable due to obscuration by intervening dust or unfavourable viewing orientation if the quasars emit anisotropically. 

Following \cite{Speagle+14} and \cite{Kroupa15} it is implicitly
assumed here that galaxies follow a common, deterministic track.  We
assume the number of galaxies in a given co-moving volume remains
constant with time. Is this a reasonable assumption?
\cite{Conselice+16} calculates the number density of galaxies observed
locally, given the number density at very high redshift, finding the
local density to be smaller than expected by a factor of about~7 and
inferring from this that galaxies merge and are destroyed in
clusters. Their calculation does not take into account that the local
400-Mpc-radius volume is a significant underdensity
(\citealt{Karachentsev12, Keenan+13}, by a factor of 3~to~5, fig.~1 in
\citealt{Kroupa15}) such that the number of galaxies merging would be
smaller.  Concerning observational evidence for the role of mergers in
galaxy formation and evolution, in "The Impossibly Early Galaxy
Problem", \cite{Steinhardt+16} note that there are several orders of
magnitude more $10^{12-13}\,M_\odot$ dark matter halos at $z=6-8$ than
expected if hierarchical assembly through mergers were to be
valid. \cite{Shankar+14} conclude, based on a sample of early-type
galaxies, that the hierarchical merging models are disfavoured
significantly and therefore that the dynamical friction time-scales
need to be longer than theoretically expected. In the same vein,
\cite{Kormendy+10} deduces that the large number of observed massive
bulgeless galaxies are in contradiction to the hierarchical formation
of such galaxies through mergers.  The observational study by
\cite{DePropris+14} of merger rates using close pairs indicates blue
galaxies to have a small merger incidence, being smaller than that
expected in standard-cosmology. \cite{DePropris+14} conclude that
their findings minimize the importance of major mergers and
interactions for galaxy evolution and they argue that most galaxy
evolution takes place via secular and internal processes. Because disk
(i.e. blue) galaxies dominate the number counts across a wide range of
redshifts \citep{Delgado+10, Tamburri+14}, we
neglect merging in the present explorative simple estimate. As a
caveat, the spheroids may form from early intense mergers
(Sec.~\ref{sec:downs}) and galaxies may be destroyed in galaxy
clusters such that more work is needed to establish the role of such
processes in galaxy evolution. The idea is thus to tag a galaxy
whether is hosts a SMBH today and compute, from the observed
present-day galaxy mass function, how many quasars we expect to see at
a given redshift.

The life-time of the first quasar-like (hyper-massive star cluster) phase is $\approx 50\,$Myr. A stochastic interval is added during which gas re-accretes onto this cluster (now of BHs), and the quasar phase (i.e. the accreting SMBH-seed) lasts for the remaining time until the hosting spheroid has assembled.

\subsection{The algorithm}

By assuming the stellar mass density of the Universe, $\rho_{\rm stars}=0.0027 \times 1.46 \times 10^{11}\,M_\odot$/Mpc$^3$ (\citealt{KarachentsevTelikova18, FukugitaPeebles04}, see also \citealt{Panter+04}), the observationally constrained Schechter function 
(\citealt{Conselice+16}, their eq.~1 with parameters from the fist row of their table~1)
can be normalised such that the number density of galaxies in the local Universe is known. The normalisation leads to an equivalent $\eta_{\rm o} = 1.13 \times 10^9$ galaxies per Gpc$^3$ with a stellar mass between $10^6\,M_\odot$ 
and $10^{13}\,M_\odot$, being in good agreement with the above number in view of the underdensity by a factor of about 3~to~5 in the local region with a radius of about 400~Mpc \citep{Karachentsev12, Keenan+13}. 

The Schechter function is randomly sampled with $N_{\rm gal}=10^8$ galaxies, i.e, we obtain $N_{\rm gal}$ values of present-day masses, $M_{\rm pgal, i}\ge 10^6\,M_\odot$. 
Using the Schechter function in this problem is not without precedent (e.g. \citealt{PacucciLoeb19}). Each galaxy is assumed to have a bulge/spheroid mass $M_{\rm sph, i}=X_1\,M_{\rm pgal,i}/2$ (leaned on the results by \citealt{Kormendy+10}; $0\le X_1 \le 1$ being a random deviate) and it is assumed that stellar-evolution mass loss does not play a role in this exploratory modelling (Fig.~\ref{fig:ipgal} shows this to be a reasonable assumption). The requirement that a spheroid is needed for the quasar phase is based on the identification of their role for quasar activity \citep{Kormendy+11} and is consistent with the model here in that the accretion onto the SMBH seed formed from the very early central cluster of BHs is fuelled by the formation of the spheroid. 

The time of formation of a $\mathrm{spheroid_i}$ (i.e. the time or age of the Universe when the spheroid begins to form) with total stellar mass (once it has assembled all its stars), $M_{\rm igal,i}$, is (all times are in~Gyr)
\begin{equation}
    t_{\rm form,i} = \tau_{\rm Hubble} - T_{\rm 1,i},
\label{eq:tformi}
\end{equation}
with
\begin{equation}
  T_{\rm 1, i}  = T_{\rm 2,i} + 0.5\,\Delta\tau_{\rm i},
\label{eq:tform1}
\end{equation}
where
\begin{equation}
T_{\rm 2,i}= a + b \, {\rm log}_{10} {M_{\rm igal,i} \over M_\odot},
\label{eq:tform2}
\end{equation}
with $a=0.125 + X_2\, (0.427 - 0.125)$ for random number deviate $0\le X_2 \le 1$, $b= a\, c_1 + c_2$,  $c_1 = (0.053-0.071)/(0.427-0.125)$ and $c_2= 0.053 - c1 \, 0.427$, which together comprise a linear interpolation between the formation time in high- and low-density environments as given by eq.~3 in \cite{Thomas05}. The above formation time combines the average age of the stellar population (eq.~3 in \citealt{Thomas05}) with the duration (i.e. down-sizing time scale) of the star formation in the spheroid (Eq.~\ref{eq:downs}, Fig.~\ref{fig:SFR}) and $\tau_{\rm Hubble}=13.8\,$Gyr is the age of the Universe \citep{PlanckVI}.  Thus, more massive spheroids form earlier and quicker, and the same occurs in low-density environments but delayed. It is thus assume here that the galaxies form evenly distributed between high- and low-density regions.  The relation between age ($t$) and redshift ($z$)  is calculated for the standard LCDM model \citep{Planck+18, PlanckVI}.

We conservatively assume every spheroid$_{\rm i}$ with a stellar mass $M_{\rm igal,i} > 10^{9.6}\,M_\odot$ (Fig.~\ref{fig:SMBHseed}, Fig.~\ref{fig:AppSMBHseed}) forms a hyper-massive cluster which appears quasar-like for~50~Myr. Following these 50~Myr, the time $t_{\rm infl,i}/{\rm Myr} = 100 - 99\,X_3$ is added during which the in-fall of gas into the BH cluster occurs. $0\le X_3 \le 1$ being a random deviate and $t_{\rm infl,i}$ describes that gas in-fall may be very rapid ($1\,$Myr) or delayed by a considerable amount ($100\,$Myr being a conservative value, Sec.~\ref{sec:nonmono}). The time, $t_{\rm compr,i}$, needed for the BH cluster to be squeezed to the SMBH seed by the gas in-fall, is then added before the quasar (i.e. the accreting-SMBH seed) phase begins.  This time is expressed here as $t_{\rm compr,i}/{\rm Myr} = 500\,X_4$, with $X_4$  being another random deviate. This allows for the possibility that the coalescence of the BH cluster to a SMBH seed may be instantaneous or may take up to $t_{\rm rel,i}=500\,$Myr (Sec.~\ref{sec:sm}, Fig.~\ref{fig:solutions_m1_1}). Note that in this simple modelling we do not take into account the dependence of $t_{\rm rel,i}$ on $M_{\rm BH,0}$ evident in Fig.~\ref{fig:solutions_m1_1}. 
The accretion phase of the now existing SMBH seed which formed in the cluster according to Sec.~\ref{sec:sm} is assumed to last for maximally the remaining downsizing time (Eq.~\ref{eq:downs}), i.e., for the time 
\begin{equation}
\Delta t_{\rm quasar,i} = \Delta \tau_{\rm i} - \left( 50\,{\rm Myr}  + t_{\rm infl,i} + t_{\rm compr,i} \right).
\label{eq:tquasar}
\end{equation}
If $\Delta t_{\rm quasar,i} \le 0\,$Myr then no visible quasar phase is recorded. Table~\ref{tab:times} and Fig.~\ref{fig:quasar_phases} summarise these different evolutionary phases. 

\begin{table}
    \centering
    \begin{tabular}{c|c}
      time/Myr  & physical process \\
      \hline
      $t=t_{\rm form,i}$ & galaxy~i begins to form \\
      $t_{\rm form, i} \le t \le t_{\rm form,i}+50 = t_{{\rm 1,i}}$  & quasar-like hyper-massive cluster \\
      $t_{\rm 1,i}< t \le t_{\rm 1,i}+ t_{\rm infl,i}= t_{\rm 2,i}$  &  infall of gas into BH cluster \\
     $t_{\rm 2,i} < t \le t_{\rm 2,i}+ t_{\rm compr,i} = t_{\rm 3,i}$ & BH cluster collapses to SMBH seed \\
        $t_{\rm 3,i}  < t \le t_{\rm form, i}+ \Delta \tau_{\rm i}$ & quasar phase only if $t_{\rm 3,i} \le t_{\rm form, i}+\Delta \tau_{\rm i}$\\
        $t> t_{\rm form, i}+\Delta \tau_{\rm i}$ & SMBH in spheroid, possible AGN phases
    \end{tabular}
    \caption{\small{Sequence of time epochs. Subscript~i refers to galaxy~i in the calculation. The final phase constitutes the SMBH in the centre of the spheroid which can become an active galactic nucleus (AGN) if the spheroid accretes gas. The quasar phases are visualised in Fig.~\ref{fig:quasar_phases} (the physical relation between AGN, Seyferts and quasars is reviewed by \citealt{HeckmanBest14}).
    }}
    \label{tab:times}
\end{table}

\subsection{Results}
\label{sec:quasar_results}

\begin{figure}
\includegraphics[scale=0.9]{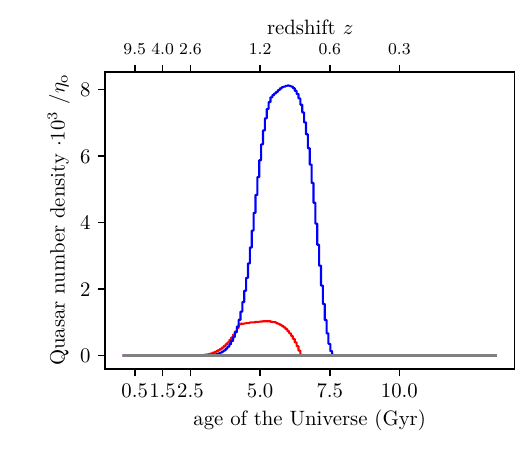}
\includegraphics[scale=0.9]{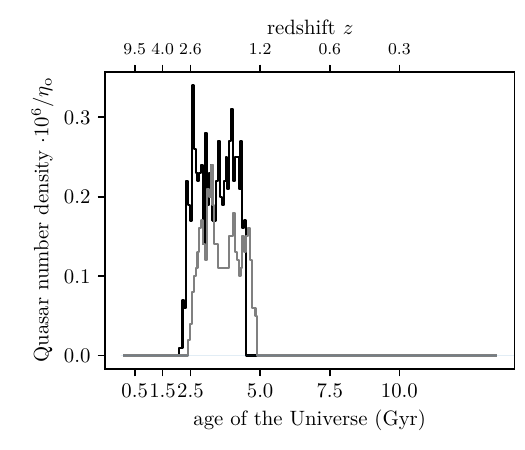}
\caption{\small{
The number of quasars in a co-moving Gpc$^3$ volume in units of the present-day number of galaxies, $\eta_{\rm o}$, as a function of the age of the Universe. If quasars were to live for ever and their number would not change, then neither the quasar number density nor $\eta_{\rm o}$ would change in this plot.
Top panel: The red histogram is for all quasar-like hyper-massive clusters while the blue histogram shows all accreting SMBH seeds (i.e. "true" quasars) which form in these. Note that the quasar-like hyper-massive clusters appear before the accreting SMBH seeds, which however are the last ones to shine. At $t> 4\,$Gyr the blue curve lies above the red one because the life-times are different, the accreting SMBH seeds being quasars for much longer and thus their numbers add-up in each time interval.
The grey histograms below the blue and red ones are shown in the bottom panel. Bottom panel: only the quasars which are in the most massive spheroids with $M_{\rm igal,i} > 10^{11.5}\,M_\odot$. These are likely to have the largest accretion rates and are thus likely to be the brightest quasars. The black histogram is for the quasar-like hyper-massive clusters, while the grey histogram is for the accreting SMBH seeds. The hyper-massive clusters are more abundant in the present model because the accreting SMBH seed phase is limited by the short $\Delta \tau_{\rm i}$ values of the massive spheroids. Thus, many hyper-massive clusters leave SMBH seeds, which however do not appear as quasars ($\Delta t_{\rm quasar, i}\le 0$, Eq.~\ref{eq:tquasar})
or do so only for a short time. For example, in the upper panel at an age $t\approx 5\,$Gyr, the model implies there to be about $8 \times 10^6$ accreting SMBH seeds per co-moving Gpc$^3$ which should appear like quasars (blue curve). In the lower panel, at 
$z \approx 2.6$, there would be $\approx 300$ quasar-like hyper-massive clusters per co-moving Gpc$^3$ 
(black histogram) while there would be about~100 accreting SMBH seeds ("true" quasars) per co-moving Gpc$^3$ 
(grey histogram). Note that the quasars in the most massive spheroids (lower panel) appear before the vast number of quasars in all spheroids (upper panel), and that, with the formation times assumed here (Eq.~\ref{eq:tform2}), no quasars ought to be observable at $z > 3$. At lower redshifts, AGN and quasars are most likely triggered through galaxy encounters and are not modelled here.
}}
\label{fig:numbdens1}
\end{figure}

\begin{figure}
\includegraphics[scale=0.9]{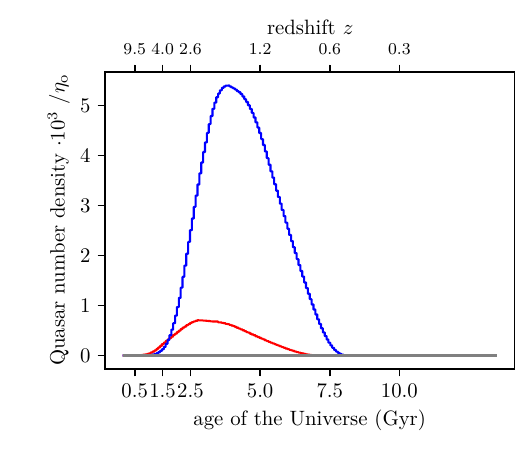}
\includegraphics[scale=0.9]{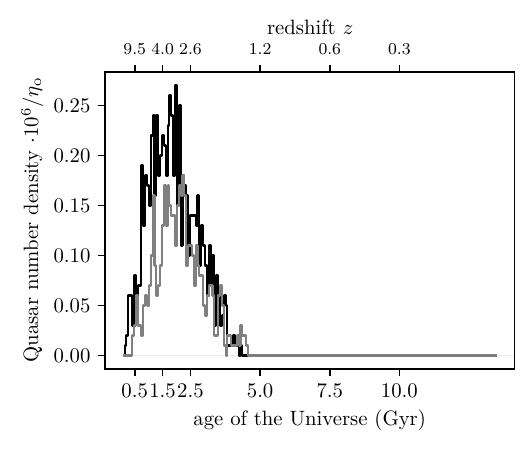}
\caption{\small{As Fig.~\ref{fig:numbdens1}, but here it is assumed that the spheroid formation time, $T_{\rm 2, i}$ (Eq.~\ref{eq:tform2}), is reduced such that all galaxies form $1.5\,$Gyr earlier, and 
that this new time $T_{\rm 2,i}$ is randomised with a uniform distribution over the interval $\pm30$~per cent (leaving $\Delta \tau_{\rm i}$ unchanged). This adds a reasonable degree of variation in the formation time without affecting the down-sizing time scale, since it is expected that spheroids which end up with the same stellar mass would not all form at exactly the same time even if forming in the same density environment since perturbations are likely to play a role (values of $t_{\rm form, i}\le 0.1\,$Gyr are transformed to a random value in the range $0.1 \le t_{\rm form, i}/{\rm Gyr} < 1$, but this affects only a negligible number of cases). 
According to this model, there ought to be a peak of $\approx5.5\times 10^6$ quasars per co-moving Gpc$^3$ at $z\approx2$ (blue curve in the upper panel), while the hyper-massive clusters would contribute about 10~per cent to the number as quasar-like objects (red curve in the upper panel).  At a redshift of about $z=5$ there would be about $100$ quasars and about $200$ quasar-like hyper-massive clusters per co-moving Gpc$^3$ (grey and black histograms, respectively, lower panel). 
}}
\label{fig:numbdens2}
\end{figure}

The co-moving number density of quasar-like objects and of quasars per~Gpc$^3$ 
is shown in Fig.~\ref{fig:numbdens1} and~\ref{fig:numbdens2}.

Fig.~\ref{fig:numbdens1} implies that the model based on the spheroid
formation times of \cite{Thomas05} leads to the too late appearance of
quasars. Quasars are observed at $z > 5$ (e.g. \citealt{Jiang+2016}, Sec.~\ref{sec:CaseinPoint})
in contrast to the present model which couples the onset of the
formation of spheroids with the formation of hyper-massive
clusters. One possibility would be to increase the downsizing time
(Eq.~\ref{eq:downs}). According to stellar population synthesis it
would be about twice longer \citep{delaRosa+11,McDermid+15}. This
approach however leads to the quasar phase lasting longer and thus to
a larger number of quasars per unit time in the redshift range when
the SMBH seeds are accreting from their forming spheroids.

\begin{figure}
\includegraphics[scale=0.9]{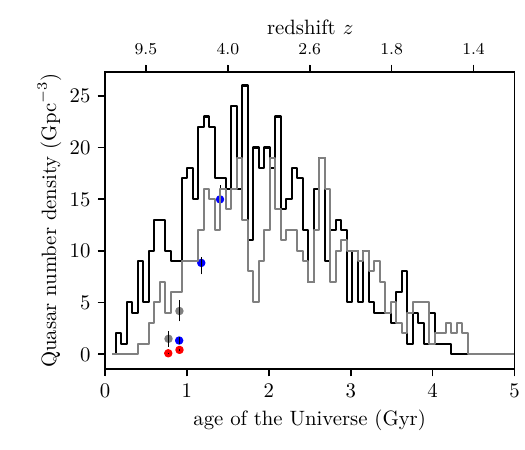}
\caption{\small{As Fig.~\ref{fig:numbdens2}, lower panel, but here showing the expected number of quasars per co-moving Gpc$^3$ (rather than the normalised value) and assuming only 10~per cent of all quasar-like hyper-massive clusters and 10~per cent of all quasars can be detected because of dust obscuration and unfavourable viewing directions. These model numbers are compared with the numbers per co-moving Gpc$^3$ gleaned from a survey of $z\approx 6$ quasars \citep{Jiang+2016} shown by the seven symbols. The reader is referred to their fig.~10 for a detailed description of the symbol types. Notable is also that a newer and more sensitive survey reports about~380 quasars per co-moving Gp$^3$ at $z\approx 5$ (\citealt{Shin+20}, their table~4), which is comparable to the model numbers shown in the bottom panel of Fig.~\ref{fig:numbdens2}.
}}
\label{fig:obs_quasars}
\end{figure}

Another possibility is to shift the spheroid formation times closer to the birth of the Universe and to increase the dispersion of these times, while keeping $\Delta \tau_{\rm i}$ unchanged in order to not enlargen the quasar population. Fig.~\ref{fig:numbdens2} indicates, in comparison to Fig.~\ref{fig:numbdens1}, how assumptions on the formation time of the spheroids affect the quasar counts.
Thus, by moving the formation times 1.5$\,$Gyr earlier to the birth of the Universe (and in addition introducing a 30~per cent spread of these formation times), quasars would be readily observed at the highest redshifts (Fig.~\ref{fig:numbdens2}). According to this model, at the very highest redshifts (i.e. at ages near $0.5\,$Gyr), the observed quasars are hyper-massive clusters (the quasar-like objects), and at later cosmological epochs (ages $\simgreat 6\,$Gyr) all quasars are accreting SMBHs. In the intermediate times both types of quasars co-exist albeit the hyper-massive cluster type contributes only about 10~per cent to the numbers. 
In this model, the quasar numbers per co-moving~Gpc$^3$ reach a maximum around $z\approx 2$, with the brightest quasars appearing at $z>9.5$. At these earliest cosmological ages ($t \simless 1.5\,$Gyr), hyper-massive clusters slightly outnumber accreting SMBH seeds, with the very first "quasars" being, in this model, hyper-massive clusters.  This model also implies that the number of quasars per co-moving Gpc$^3$ of both types decreases with increasing redshift $z > 3$. Especially the change of the number of the brightest quasars per co-moving Gpc$^3$ with increasing $z$ for $z>5$ (lower panel in Fig.~\ref{fig:numbdens2})
is reminiscent of the observed dependency \citep{Jiang+2016}. This is demonstrated in Fig.~\ref{fig:obs_quasars}.

The following two problems 
thus emerge within the present model and would need to be addressed in future studies: 

\begin{itemize}
    \item The very early formation times that are suggested by the observed very-high red-shift quasars appear to be in tension with the formation times deduced from the observational analysis of spheroids by \cite{Thomas05} which would not allow quasars to exist at $z>5$. More generally, the nearby massive-spheroid data and the high redshift observations in terms of the SFRs and timings (Sec.~\ref{sec:modelSFRs}) appear to be in conflict.\footnote{One issue relevant for this problems is the redshift-age relation which may differ in different cosmologies (e.g. \citealt{Melia09, Balakrishna+19, Merritt17}).}

\item Another issue raised here is that of the "missing quasars":
As noted above, the number of SMBHs in the local Universe should lead to $10^6 - 10^7$~quasars per co-moving Gpc$^3$ existing at intermediate  redshifts ($1 \simless z \simless 3$, Fig.~\ref{fig:numbdens2} for the more realistic spheroid formation times). Such an expected large number is difficult to verify observationally. For example, the spectroscopic survey by \cite{Eftekharzadeh+15} reports about 3000~quasars per co-moving Gpc$^3$ in the redshift range $2.2-2.8$, while 3--4 million are expected (blue curve in Fig.~\ref{fig:numbdens2}). \cite{Eftekharzadeh+15} discuss the daunting observational problems in this work, and a combination of missing quasars due to dust obscuration and unfavourable viewing direction as well as flux limits may bring the model numbers into the observed range. But the very early cosmological era during which quasars are expected to appear may lead to countable predictions. The number of brightest quasars per co-moving Gpc$^3$, expected to be hosted in the most massive forming spheroids ($M_{\rm igal,i}>10^{11.5}\,M_\odot$), is indeed much smaller
as the model calculations depicted in the lower panel of Fig.~\ref{fig:numbdens2} (about 300 quasars in total at $z\approx 5$) indicate.  The model presented here thus underlines this issue of missing quasars at intermediate redshifts, despite taking into account that the quasars would not form at the same time but in dependence of the formation time of their hosting galaxy and that they would come in two phases with interruption in detectability. Detecting quasars at a high redshift is very difficult so it is not clear yet whether a missing-quasar problem exists. Indeed, assuming only 10~per cent of the existing quasar-like objects and quasars can be detected leads to model numbers in agreement with those observed (Fig.~\ref{fig:obs_quasars}). 

The model as shown in Figs~\ref{fig:numbdens2} and~\ref{fig:obs_quasars} thus lead to about ten times more quasars being expected than are observed at $z>5$. If the large number density of quasars expected as based on the Local Cosmological Volume galaxy numbers and the model studied here are not found at a high redshift, then this would imply that the SMBHs hosted by the majority of galaxies with stellar mass $\simgreat 10^{9.6}\,M_\odot$ would need to be either existing at a high redshift but not be accreting (thus supporting a primordial origin), or that they might need to form later but largely invisibly. Both possibilities appear to pose difficult physical hurdles.
\end{itemize}

%On the other hand, a
%different cosmological model may allow for an older age of the
%Universe by a few hundred~Myr and thus more time for SMBHs to form
%\citep{YuWang2014,MM15}, possibly according to the physical model studied
%here.

\subsection{A case in point: the $1.5\times 10^9\,M_\odot$ SMBH in the luminous quasar P{\={o}}niu{\={a}}'ena at $z=7.5$}
\label{sec:CaseinPoint}

How does this model fare in comparison with the discovery by \cite{Yang+20} of the $z=7.5$ ($t \approx 700\,$Myr after birth of the Universe) quasar J1007+2115 which is interpreted to be powered by an $\approx 1.5 \times 10^9\,M_\odot$ heavy SMBH? 
The quasar is observed to be in a host with $80 \simless SFR_{\rm [CII]}/(M_\odot/{\rm yr}) \simless 520$ (based on the [CII] emission line) and $SFR_{\rm TIR}  \approx 700\, M_\odot/{\rm yr}$ (based on total infrared luminosity assuming $T_{\rm dust}=47\,$K). Assuming the IMF is top-heavy and that the correction factor is $\kappa=0.5$ (Eq.~\ref{eq:kappa} below), 
$40\simless SFR_{\rm true}/(M_\odot/{\rm yr}) \simless 350$.

According to the model this quasar may be interpreted non-exclusively as follows:
\begin{enumerate}
\item
It may be a hyper-massive star-burst cluster formed at $z=7.5$ (within about 10~Myr of the observation). It may, for example, constitute the first cluster with $M_{\rm ecl, max} \approx 3\times 10^9\,M_\odot$ formed 
at the start of assembly of the spheroid which will become an elliptical galaxy with a mass of about $10^{12}\,M_\odot$ which follows from the $M_{\rm ecl, max}(SFR)$ relation (Eq.~\ref{eq:Mdeltat}) and assuming the standard downsizing time $\Delta \tau =0.34\,$Gyr (Eq.~\ref{eq:downs}). The bolometric luminosity of this cluster (assuming it has a top-heavy IMF, Eq.~\ref{eq:IMF}) would be $L_{\rm bol}\approx 3\times  10^{13}\,L_{{\rm bol}, \odot}$ (fig.~3 in \citealt{Jerab17}) which compares favourably with the observed value. The cluster would
drive a massive outflow which may have the observed line widths (Sec.~\ref{sec:BHcontent}). The SFR would correspond to 
$SFR_{\rm true} \approx 300\, M_\odot/{\rm yr}$ for a formation time-scale of $10\,$Myr assuming only this cluster is forming. 

\item 
Alternatively, it may already be an accreting SMBH which formed as a SMBH-seed from such a cluster. In this case, the SMBH-seed might comprise the conservative 5~per cent of $M_{\rm BH, 0}$, amounting to $\approx 4.8 \times 10^7\,M_\odot$ (Fig.~\ref{fig:UCDBH}) and it would need to accrete at the Eddington-limited constant super-Eddington rate for about 172~Myr (Eq.~\ref{eq:factor}) to reach $1.5\times 10^9\,M_\odot$. This interpretation would necessitate the hosting cluster with $M_{\rm ecl, max}\approx 3 \times 10^9\,M_\odot$ to have formed sufficiently prior to the age it is observed at ($\approx 700\,$Myr) to allow (a)~all massive stars to transform into BHs (about $50\,$Myr), (b)~to allow gas-inflow from the forming spheroid to shrink the sub-cluster of BHs to the SMBH-seed ($\approx 100-200\,$Myr, Fig.~\ref{fig:solutions_m1_1}) and (c) to allow the SMBH-seed to grow to the deduced $1.5 \times 10^9\,M_\odot$ ($\approx 172\,$Myr). The onset of the formation of the spheroid would thus have to have been at an age of $280 \simless t_{\rm form}/{\rm Myr} \simless 380\,$Myr after the birth of the Universe (cf. Fig.~\ref{fig:obs_quasars} and~\ref{fig:SFHs}). The $SFR_{\rm true}$ value at the moment this object is observed at would correspond to a snapshot of the SFH of the forming spheroid (nominally at $3000\,M_\odot$/yr according to the box SFH model in Fig.~\ref{fig:SFHs}) and may be suppressed momentarily through the action of the quasar (cf. \citealt{Ploeckinger+19}). 
\end{enumerate}
These possibilities are non-exclusive because different combinations between $M_{\rm ecl,max}, t_{\rm form}$ and time for the SMBH-seed to reach $1.5\times10^9\,M_\odot$ (as long as it is $\simless \Delta \tau$) are possible in this model. 

The discovery of P{\={o}}niu{\={a}}'ena thus appears to be consistent with the present model but clearly much more work will be needed to test the above possibilities. For example, concerning the first hypothesis above, it needs to be checked if a hyper-massive cluster can drive an outflow which does resemble the observations. It will also be important to check if the $\simless 100\,$kpc (Sec.~\ref{sec:downs}) surroundings of this putative first hyper-massive cluster contains a gas cloud weighing a few times $10^{12}\,M_\odot$ from which the spheroid will form.  Concerning the second hypothesis, observations might verify if a major galaxy is in the process of assembling around the quasar. Because about $333-433\,$Myr have passed since the putative $M_{\rm ecl,max}\approx 4\times 10^9\,M_\odot$ hyper-massive starburst cluster began to form (at $t_{\rm form}\approx 280\,$ to $380\,$Myr), the spheroid, which nominally takes $\Delta \tau = 0.34\,$Gyr to form,  should already largely be present with a possibly already decreasing SFH.

%============================================
\section{Discussion}
\label{sec:disc}

The above calculations are simplified because the 
physically realistic situation of an astrophysically
and stellar-dynamically evolving super-star cluster formed at
extremely low metallicity and at the centre of a violently forming and
evolving spheroid including gas in- and out-flows and stellar mergers
cannot, for the time being, be computationally assessed in full
rigour.  In the study
of this problem performed here, the well-documented empirical constraints on freshly
formed stellar populations (which are the basic axioms of 
the IGIMF theory, Sec.~\ref{sec:gwIMF}) and 
the global evolution of a cluster of BHs, which emerges from 
the stellar population, are combined.
  
The evolution of the BH sub-cluster is primarily dictated by balanced
evolution and by gas drag, as discussed above. The physically
realistic situation is however very complex as the BHs accrete, they
exert radiation which heats the gas and also the rate of cooling of
the gas in the BH sub-cluster is governed by the opacity. In
Sec.~\ref{sec:BHaccr}
some of these processes are briefly touched upon in view of their
possible influence on the overall evolution of the BH sub-cluster and
may be helpful in future extensions of this work. In particular
  the possibility that accreting BHs may be accelerated rather than
  suffer orbital decay is raised. The mutually related problems of whether spheroids
  form as rapdly as in this model and if
  the observed SFRs at a high redshift are as high as associated with
  the formation of spheroids in the present model are addressed in
  Sec.~\ref{sec:SFRs}. 
 In Sec.~\ref{sec:other} additional caveats are attended to.

\subsection{Gas accretion onto the BHs}
\label{sec:BHaccr}

As the BHs orbit through the in-falling gas they accrete from it (for
a comprehensive treatment see \citealt{JuanKingRaine02}).  The
Bondi-Hoyle-Lyttelton accretion rate onto a BH of mass $\mstar$ moving
with velocity $v \approx \sigma$ through an ambient medium of (mass)
density $\rho_{\rm g}$ and sound speed $c_{\rm s}$ is
\begin{equation}
\dot{m}_{\rm BH}=4\,\pi \, \rho_{\rm g} \, {G^2\, \mstar^2 \over 
\left(v^2 +    c_{\rm s}^2 \right)^{3/2}} \; .
\label{eq:BHaccr}
\end{equation}
Assuming supersonic motion which is appropriate for the case of
extremely dense star clusters, $v\gg c_{\rm s}$, 
\begin{equation}
\dot{m}_{\rm BH}=4\,\pi \, \rho_{\rm g} \, {G^2\, \mstar^2 \over v^3} \; .
\label{eq:BHaccr2}
\end{equation}
The accretion onto the moving BH leads to loss of momentum,
$\dot{p} \approx -\dot{\mstar}\, v$, leading to dissipation of kinetic
energy, $\dot{E}_{\rm BH} \approx v\,\dot{p}$, which implies an equation
for the shrinkage of the BH sub-cluster similar to Eq.~\ref{eq:edisstot}
since both effects (dynamical friction and accretion) are due to the
same physical process.

With accretion-induced feedback, the accretion rate is proportional to
mass (as opposed to Eq.~\ref{eq:BHaccr}) because the feedback
luminosity, which opposes infall, is proportional to the mass of the
BH.  Eddington-limited BH growth, a limit where the in-falling matter
radiates its rest-mass energy which balances the infall, is thus
significantly slower than growth through Bondi-Hoyle accretion, which
is why taking $\mstar = \,$constant (Sec.~\ref{sec:sm}) is a
reasonable assumption.

Thus, if we assume that the accretion rate is at the Eddington limit, then
\begin{equation}
\dot{m}_{\rm BH} = \mstar / \tedd,
\end{equation}
with
\begin{equation}
\tedd \equiv \frac{c \, \thompson}{4\pi \, G \, \Mp} \approx 452\,\myr\;,
\end{equation}
where $c$ is the speed of light, $\thompson$ is the Thompson
scattering cross section and $\Mp$ is the mass of the proton. In this case the BHs double their mass every
$\approx 300\,\myr$. 
%This means that the shrinkage time
%(Eq.~\ref{eq:ts5}) will be decreased by less than a factor
%of two if it is comparable to $\tedd$.

To give an explicit example, assume that the BHs accrete at a constant
fraction of the Eddington limit, i.e., 
\begin{equation}
\dot{m}_{\rm BH} = \eddc \mstar / \tedd
\label{eq:eddlimit}
\end{equation}
with $\eddc = (1-\epsilon_{\rm r})/\epsilon_{\rm r} < 1$, the radiative efficiency
$\epsilon_{\rm r}$ being the fraction of the rest-mass energy of the in-falling matter being electromagnetically radiated. Thus, for
Eddington-limited accretion at the Eddington rate, $\epsilon_{\rm r}=0.5$, while a standard
assumption used throughout the literature is accretion to occur
Eddington-limited but with a super-Eddington rate with $\epsilon_{\rm r}=0.1$ \citep{Salpeter64,
  YooME04,BD19}). This means that the mass-growth of the black hole is regulated by the feedback from the accretion flow, while the mass accretion onto the black hole is at the super-Eddington rate as given by the customary assumption $\epsilon_{\rm r}=0.1$.
The mass of a BH then grows with accretion
time $t_{\rm accr} = t-t_{\rm start}$ as
\begin{equation}
\mstar(t) = m_{\rm BH,0} \exp\left( \frac{\eddc \, t_{\rm accr}}{\tedd} \right) \;,
\label{eq:mt}
\end{equation}
where $m_{\rm BH,0}$ is the initial mass of the BH at time $t_{\rm start}$. 
Over the formation
time-scale of the spheroid (up to a few hundred~Myr, Eq.~\ref{eq:downs}), the mass growth
of the BH is relatively small and can thus be neglected without
significantly affecting the results on the shrinkage of the BH
cluster.

The realistic situation is complicated because the dormant BH begins
to accrete according to Eq.~\ref{eq:BHaccr}. The accreted gas forms an
accretion disk about the BH feeding the BH and thus increasing
$\mstar$ in a regulated manner and more likely similar to a rate given
by Eq.~\ref{eq:mt}. The interplay between the accretion onto the disk
and the accretion onto the BH defines the mass growth of the BH plus
accretion disk system, the outflow in the form of relativistic jets
and the actual mass gain of the BH.

Assume that the BHs moving through the gaseous medium will
accrete at the Eddington-limited rate, i.e., with the accretion rate given by
Eq.~\ref{eq:eddlimit}.
%\begin{equation}
%\mdot = 4\pi\rho \frac{G^2 \mstar^2}{(v^2 + \cs^2)^{3/2}}\;,
%\end{equation}
%where $\cs$ stands for the spped of sound. For $\cs \ll v$, we may consider a simplified form,
%\begin{equation}
%\mdot = 4\pi\rho \frac{G^2 \mstar^2}{v^3}\;.
%\label{eq:BHL_mdot}
%\end{equation}
Assuming further that half of the rest mass energy of the accreted
matter turns into radiation\footnote{While this assumption is a
  reasonable first order estimate for standard disc-like accretion
  flows, a completely different geometry of the accretion flow
  considered here may lead to a different relation between the
  accretion rate and luminosity. The estimate used here, however,
  represents a rather safe upper limit.}, $\epsilon_{\rm r}=0.5$, the luminosity of individual
BHs will be $L_{BH, acc} \approx \frac{1}{2} \, \dot{m}_{\rm BH} \, c^2$. The
total accretion luminosity of the cluster of BHs is thus
\begin{equation}
\lacc \approx \frac{1}{2} \, N \, \dot{m}_{\rm BH} \, c^2.
\label{eq:clusteraccrlum}
\end{equation}
If $\lacc$ exceeds the formal Eddington luminosity, $L_{\rm Edd}$, of the whole
cluster of mass $M_{\rm cl}=N\,m_{\rm BH} + M_{\rm g}$, the gas cloud
would be blown out and its density will drop. Consequently, both
energy dissipation of the BH sub-cluster and its luminosity will be
decreased or even stopped, until the gas inflow re-establishes.  The
ratio of $\lacc$ and $\ledd$ is estimated next. Assuming the gas
consists of ionised hydrogen,
\begin{equation}
L_{\rm Edd} = {  
4\,\pi\,G\,m_{\rm P}\, c \, M_{\rm cl}   \over \sigma_{\rm T}
 },
\label{eq:Ledd}
\end{equation}
i.e
\begin{equation}
    {L_{\rm Edd} \over  L_\odot} = 3.4 \times 10^5 \, {M_{\rm Edd} \over M_\odot},
\label{eq:Ledd1}    
\end{equation}
where $M_{\rm Edd}=M_{\rm cl}$ is the luminosity-inferred Eddington mass of this accreting object with, in this case, a mass of $M_{\rm cl}$.

When $L_{\rm acc}/L_{\rm Edd} > 1$, $\rho_{\rm g}$ would decrease
within $R$ on a time-scale that may be estimated by $10^4\,$K gas
leaving the cluster radius $R$ with the sound speed (about
$10\,$km/s), thus on a time-scale of
$\tau_{\rm out} \approx R/(10\,{\rm km/s})$.  If this blow-out
time-scale is much longer than the crossing time of BHs through the
cluster, $\tau_{\rm out} \gg t_{\rm cross}$, then the BH sub-cluster would
expand to a new $R_{\rm new}=R/(1-f_{\rm out})$ (eq.~25
in \citealt{Kroupa08}), where $f_{\rm out}=M_{\rm g}/N\,m_{\rm BH}$ is
the fraction of mass blown out, assuming all gas goes and ignoring the
rest of the stellar cluster in which the BH sub-cluster is embedded. It is
unclear though if this evolution is viable, because the in-falling gas
from beyond $R$ as a result of the continued formation of the spheroid
during the first few hundred~Myr is likely to limit the decrease of
$\rho_{\rm g}$ within $R$. Consequently, we ignore the possible drop
of $\rho_{\rm g}$ during the assembly of the spheroid.
 
Given the long mass-doubling time and the likely interruptions of the
accretion activity, the conservative case is $\dot{\mstar}=0$ which is why this work assumes $M_{\rm BH}(t) = {\rm constant} = M_{\rm BH,0}$. In this
case the shrinkage time of $R$ calculated in Sec.~\ref{sec:sm} is
likely to overestimate the time-scale to SMBH-seed formation.  The special
case, $\dot{m}_{\rm BH} = 0, \dot{M}_{\rm g} \propto M^2_{\rm g}$,
leads to an analytical solution to the mass increase of the gas mass
within $R$ (see Appendix~1).

One important caveat which will need to be taken into account in future modelling is 
the possibility that the gas accretion onto the BH sub-cluster will lead to the individual BHs in the sub-cluster to accelerate. \cite{KimKim09}, \cite{Gruzinov+20} and \cite{Li+20} point out that accreting BHs which move though a gaseous medium and which drive an outflow may experience, under certain conditions, acceleration rather than a deceleration. The details are highly complex, especially in view of the calculations that were performed being by necessity simplified and idealised. The BH sub-cluster envisioned in the present model experiences gas in-fall from the forming spheroid at a high rate which is most likely time variable and also highly in-homogeneous. Since the BHs accrete some of this gas, their growing mass and accretion of gas with, on average, a negligible momentum, should be shrinking the BH sub-cluster. The dissipational gas component further shrinks the sub-cluster as its mass increases through accretion.

\subsection{On the formation of massive spheroids}
\label{sec:SFRs}

The model of SMBH formation relies on the most-massive-star-cluster (Eq.~\ref{eq:WKL} or~\ref{eq:Mdeltat}) and the formation-time-scale (Eq.~\ref{eq:downs})
to correlate with the mass of the spheroid. These relations imply large SFRs and short assembly times. 
Are these consistent with other evidence gleaned from high-redshift observations and modelling of spheroids? 
This question is relevant for understanding the SMBH--spheroid mass correlation.
Observational information on the formation of spheroids is touched upon in Sec.~\ref{sec:nonmono} and Sec.~\ref{sec:obsSFRs} contains a discussion of observational constraints on SFRs at high redshift.
In Sec.~\ref{sec:modelSFRs} the possible revision of these SFRs is considered if the high-redshift 
dust temperatures are lower than assumed and if the galaxy-wide IMF is top-heavy in spheroid formation. The implications of this on the SMBH-seed are discussed in Sec.~\ref{sec:implSFRs}. 

\subsubsection{Observational constraints on the formation of spheroids}
\label{sec:nonmono}

This model rests on the bulk of a spheroid assembling rapidly such that the centre-most very massive star cluster correlates with the rest of the mass of the spheroid through the observed SFR--most-massive-young cluster relation. This also yields the observed metal and alpha-element abundances of spheroids. If it were to be found that spheroids form from previously uncorrelated smaller proto-galaxies, then this would 
largely invalidate the model.

Concerning evidence on the assembly of spheroids obtained independently of elemental abundances:
A massive, nearly non-star-forming galaxy with a
velocity dispersion of about 270~km/s has been found to 
be present at a redshift $z=4.01$ \citep{Tanaka+2019}\footnote{The redshift--age correspondence can be found e.g. in Fig.~\ref{fig:SFHs}}.  \cite{Glazebrook17} report the discovery of a quiescetent galaxy with stellar mass $\approx 1.7\times 10^{11}\,M_\odot$ at $z=3.17$.
Concerning more-local spheroids, in their spatially-resolved stellar populations analysis of a sample of 45~elliptical galaxies, 
\cite{Martin+18} find that the bulk of stars are old and typically formed $\simgreat 10\,$Gyr ago. 
%These authors also write that in addition to the already existing dominant
%old stellar component, younger populations can form later from either the metal-rich, recycled gas expelled by supernovae and evolved stars, or from more pristine gas accreted from the intergalactic medium. 
%quotation:
%"regarding the age of the stellar populations within our sample, the bulk of stars are old, typically formed $\simgreat 10\,$Gyr ago. On top of the dominant old stellar component, younger populations can form later due to the either metal-rich, recycled gas expelled by supernovae and evolved stars, or due to accretion of more pristine gas from the intergalactic medium". 
%We refer the interested reader to their sec.~6.1 for an in-depth discussion of how this may fit-in with the expectations from the dark-matter based hierarchical galaxy formation theory. 
\cite{SR+19} conclude, based on their study of 28663 galaxies, residual star formation to be ubiquitous in massive early-type galaxies, amounting to an average mass fraction of 0.5~\% in young stars in the last 2 Gyr of their evolution. They also conclude that this fraction decreases with increasing galaxy stellar mass, being consistent with downsizing. 
%They also report synthetic galaxies from state-of-the-art cosmological numerical simulations to substantially overproduce both intermediate and young stellar populations,  
%quitation:
%"We find that residual star formation is ubiquitous in massive early-type galaxies, measuring average mass fractions of 0.5~\% in young stars in the last 2 Gyr of their evolution. This fraction shows a decreasing trend with galaxy stellar mass, consistent with a downsizing scenario. We also find that synthetic galaxies from state-of-the-art cosmological numerical simulations substantially overproduce both intermediate and young stellar populations."  
%the simulated early-type galaxies showing residual star formation even at low redshift. 
\cite{Lim+20} discovered the sustained formation of globular clusters
around the central giant elliptical galaxy of the Perseus cluster in
large-scale filamentary structures. The analysis by \cite{Vazdekis+16}
of a representative set of spheroids of varying mass implies a small
fraction of the stellar body to be young with the bulk being very old.
\cite{Seidel+15} find, that in all the galactic bulges they studied,
at least 50 per cent of the stellar mass already existed 12$\,$Gyr
ago, being more than currently predicted by simulations. These results
are consistent with the finding by \cite{Delgado+10} that the fraction
of early-type galaxies amongst galaxies with a baryonic mass
$>1.5\times 10^{10}\,M_\odot$ remains constant at 3--4~per cent over
the past~$6\,$Gyr, a similar result being reported by \cite{Tamburri+14} but
for $0.6 \le z \le 2.5$ . This is consistent with the conclusion reached by
\cite{Fernandez+14}, as based on their survey of isolated galaxies,
that the colours of the red bulges and the low bulge-to-total ratios
for AMIGA isolated galaxies are consistent with an early formation
epoch and not much subsequent growth. The major observational survey
and stellar population synthesis by \cite{delaRosa+11} and
\cite{McDermid+15} finds massive early type galaxies to have formed
their stars faster and earlier than less-massive
ones. \cite{McDermid+15} write that all of today’s spheroids share an
early period of intense star formation, making the systems compact,
metal-rich, and alpha-enhanced. It should be noted though that their
down-sizing times are systematically longer by about a factor of two
than those quantified by Eq.~\ref{eq:downs}.

Intimately linked to the formation process of spheroids are their morphological properties -- while star formation
observations indicate that the bulk of the massive spheroids formed
very early (perhaps monolithically), the evidence from
modelling of morphological and kinematical properties of spheroids is
not so clear: Massive elliptical galaxies show cores, are boxy and slow rotators, while less massive ones are disky,
have power-law slopes and are fast rotators (e.g. \citealt{Bender+89,
  Bender+92}). \cite{Krajnovic+20} provide a detailed discussion
  of the rich morphology and kinematical complexity of spheroids.
  These properties need to be accounted for in any formation theory.
  The properties range from being explainable via massive
  dissipation-less mergers and via gas accretion and gas-rich mergers \citep{Naab+14}. 
\cite{Trakhtenbrot+18} acknowledge that many of their high
redshift sources hosting quasars may be interpreted to be merging
  galaxies, and point out that in some cases the ordered gas rotation
  signatures do not support this, implying other than merging
  processes to be feeding the star-formation activity.  Early-type galaxies show a significant size evolution with
    redshift from $z\approx 2$ to $z=0$ (\citealt{Balakrishna+19} and
    references therein). Using Nbody simulations of
        minor and major dry mergers, \cite{Nipoti+09} conclude that the observed magnitude of the size
        evolution cannot be explained through dissipation-less mergers.  By using a
        high-resolution cosmological simulation, \cite{Naab+09} deduce, on the other hand, 
        that compact high-redshift spheroids can evolve into the
        observed sizes and concentration of present-day spheroids by
        undergoing minor mergers such that minor mergers may
        be the main driver for the late evolution of sizes and
        densities of the observed spheroids. A review of the current standing of the merger theory can be found in \cite{NO17}.
        Dissipation-less (dry) monolithic formation models of galaxies explored by \cite{Nipoti+07b} in 
        Milgromian gravitation \citep{Milgrom83, BK84, FamaeyMcGaugh12, Merritt20} are an alternative approach.
       The challenge facing all models is to explain
      the combination of high metallicity and high alpha-element
      abundances (Sec.~\ref{sec:downs}). 

In summary, the evidence thus seems to suggest that the bulk (at least about 50~per cent) of the present-day spheroids did form early and rapidly and possibly monolithically,  but that a part of the stellar body may have joined later. If the time-scale of the formation of the bulk of the spheroids is longer by a factor of two than the downsizing time deduced from metallicity constraints, as suggested by the stellar-population synthesis work, then the implied SFRs would be correspondingly lower (by a factor of 2) than used in the nominal modelling here which assumes all the mass of the spheroid to form in the time $\Delta \tau$ (Eq.~\ref{eq:downs}), leaving more time for hierarchical merging to play a role. 
%In Sec.~\ref{sec:ndens} we adopted the empirical results of \cite{Thomas05} concerning the times of formation of the spheroids to estimate the expected number density of quasars as a function of redshift in the present model. {\bf For completeness, we also note that the dissipation-less monolithic formation models of galaxies explored by \cite{Nipoti+07b} in Milgromian gravitation \citep{Milgrom83, BK84, FamaeyMcGaugh12, Merritt20} suggest that the spheroids do obtain some of the correct morphological properties. In Milgromian gravitation mergers are rare.
%such that spheroids will be forming mostly through monolithic collapse, larger density contrasts of the post-Big-Bang gas clouds collapsing earlier and faster in this case, being qualitatively consistent with the observed downsizing discussed in this section.

\subsubsection{Observed SFRs}
\label{sec:obsSFRs}

First the SFH of a typical model is discussed to then consider how observed SFRs at high redshifts compare.

\begin{figure}
\includegraphics[scale=0.9]{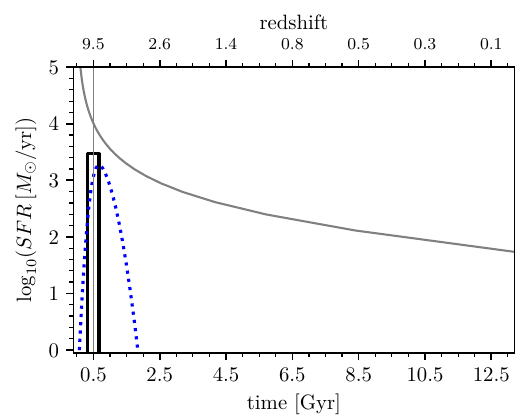}
\caption{\small 
The idealised box-shaped SFH adopted in this model is shown as the black solid line. It assumes the SFR jumps to $SFR=3\times10^3\,M_\odot$/yr implied by Eq.~\ref{eq:downs} for the downsizing time $\Delta \tau = 0.34\,$Gyr. The midpoint of the SFH lies at $0.5\,$Gyr corresponding to $t_{\rm form} = 0.33\,$Gyr (cf Fig.~\ref{fig:numbdens2}). A more realistic SFH can be approximated by a 
skew normal distribution
as shown by the dotted blue line. This example has 
the skew parameter set to~2.5 (0 corresponding to a Gaussian), 
the sigma value is $\Delta \tau$ and the location is~0.5~Gyr.
 The gray solid line corresponds to the growth rate of the baryonic component of a galaxy in the standard dark-matter-based cosmological model (Eq.~\ref{eq:dekel}).  Except for the latter, in the two SFH models the integral under each curve yields the mass formed in stars, $M_{\rm igal}=10^{12}\,M_\odot$. In both models, the 
 first hyper-massive star-burst cluster at the centre of the spheroid with a mass of $M_{\rm ecl,max}=3 \times 10^{9}\,M_\odot$ (Eq.~\ref{eq:Mdeltat}) would form during the first few~Myr. Note that the box and skew-normal models do not emerge from a self-consistent cosmological structure formation calculation such that the here applied value of $t_{\rm form}$ is not a prediction but an assumption.
}
\label{fig:SFHs}
\end{figure}

The model presented here assumes an idealised box-shaped
SFH (Fig.~\ref{fig:SFHs}). A more detailed description of monolithic formation would include star formation starting in the highest-density peak very low-metallicity gas near the centre of the future spheroid to then pick-up as the gas collapses and forms the spheroid. 
%The computations of galaxy formation in Milgromian dynamics 
%by \cite{Wittenburg+20} show this behaviour but form early-type rotationally-supported galaxies. 
Such a putative SFH is approximated in Fig.~\ref{fig:SFHs} as a skew normal distribution. 
In standard cosmology, the accretion of gas onto a galaxy can be parametrised as \citep{Dekel+09}
\begin{equation}
{SFR \over M_\odot/{\rm yr}} \approx 6.6\, 
\left( { M_{\rm DMH} \over 10^{12}\,M_\odot } \right)^{1.15}
\left(1+z\right)^{2.25}\, f_{0.165} \; ,
\label{eq:dekel}
\end{equation}
where $f_{0.165}$ is the baryon fraction in the dark matter halos in units of the cosmological value, $f_{\rm b}=0.165$, 
and $M_{\rm DMH}$ is the mass of the dark matter halo. 
It is assumed here, for the purpose of the argument, that a dark matter halo with $M_{\rm DMH}=6\times 10^{12}\,M_\odot$ (such that $M_{\rm igal}=10^{12}\,M_\odot$ for consistency with $f_{\rm 0.165}=1$, \citealt{Dekel+09}) formed at very high redshift and accretes gas such that the accretion rate equals the SFR. The resulting SFH is shown in Fig.~\ref{fig:SFHs}. Given the stellar population synthesis constraints from recent observations, the galaxy would need to shut-off star-formation after a few~100~Myr. 

Thus, star-formation begins in the highest-density gas which also cools
fastest. If the formation of stars at
very low metallicity and in the first highest-density peaks occurs mostly in extremely massive star-burst clusters, i.e.
if the initial mass function of embedded clusters, $\xi_{\rm ecl}(M_{\rm ecl})$,
is strongly top-heavy ($\beta\le 2$ in Eq.~\ref{eq:ECMF}) under these conditions, a hyper-massive star-burst cluster weighing
$10^8 \simless M_{\rm ecl}/M_\odot \simless 10^9$ would need
$20 \simless SFR_{\rm true}/(M_\odot/{\rm yr}) \simless 1000$ to form over a time of $1-5\,$Myr. Metal-enrichment proceeds
on the time-scale of a few dozen~Myr and will be especially rapid if the IMF in this cluster is top-heavy \citep{Murray09}, such that observing a metal-enriched high redshift star-burst is not necessarily in contradiction with it beginning to form from very low-metallicity gas.  

In terms of observed SFRs at high redshifts, 
\cite{Finkelstein+13} find a galaxy at a redshift of $z \approx 7.5$
(about $700\,$Myr after the Big Bang assuming the detected emission line to be Ly$\alpha$; the alternative redshift being $z = 1.78$ if the line is [O II]) with a 68~per cent confidence range $320 < SFR_{\rm obs}/(M_\odot/{\rm yr}) < 1040$ and a stellar mass $0.9\times 10^9 - 1.2 \times 10^9\,M_\odot$. 
\cite{Trakhtenbrot20} provides
a review of the properties of high redshift quasars and ALMA observations of their
hosts show observed SFRs surpassing $1000\,M_\odot/$yr at $z>6$ in some cases.
\cite{Nguyen+20} depict, in their fig.~9 and for the twelve quasars at $z\approx 4.8$
observed with ALMA, host-galaxies. These have $1000 \simless SFR_{\rm obs}/(M_\odot\,{\rm Myr}^{-1}) < 4000$.
\cite{Nguyen+20} assume a dust temperature, $T_{\rm dust}=60-70\,$K, for the objects with the highest SFRs in their sample.  \cite{Trakhtenbrot+18} point out, as an interesting case in point, that the system~J1341 ($z\approx 4.8$) has an observationally deduced $SFR_{\rm obs} \approx 3000\,M_\odot$/yr ($T_{\rm dust}=47\,$K) and evidence 
of rotation-dominated gas and no companion galaxies, being quite comparable to the galaxy-formation simulations by \cite{Wittenburg+20}.  
\cite{Forrest+20} report a $z=3.493$ quiescent galaxy with
a stellar mass of  $3.1^{+0.1}_{-0.2} \times 10^{11}\,M_\odot$.  The authors infer $SFR_{\rm obs} > 1000\,M_\odot$/yr for about $0.5\,$Gyr beginning at $z \approx 7.2$ for it to have buildup its mass and 
suggest it to be a descendant of massive dusty star-forming galaxies at $5 < z < 7$ recently observed with ALMA.
\cite{Scoville+17} report ALMA observations of a sample of 708 galaxies at $z = 0.3 - 4.5$ finding cases with 
$5000 < SFR_{\rm obs}/(M_\odot/{\rm yr})<10^4$ at $z\approx 3$ (their fig.~6; they assume $T_{\rm dust}=25\,$K).
\cite{Fan+19} report a dust-obscured quasar at $z\approx 2.9$ in a galaxy with an inferred molecular gas mass of $8.4 \times 10^{10}\,M_\odot$ and $SFR_{\rm obs} \approx 3000-7000\,M_\odot$/yr using various methods and for a derived $T_{\rm dust}=78.1\,$K.
\cite{Riechers+20} report three dusty galaxies at $z>5$ with deduced 
$SFR_{\rm obs} = 870 \pm 100\,M_\odot$/yr ($T_{\rm dust} \approx 35 \,$K), $1030^{+190}_{-150}\,M_\odot$/yr ($T_{\rm dust} \approx 50 \,$K)
and $SFR_{\rm obs} = 2500 \pm 700\,M_\odot$/yr ($T_{\rm dust} \approx 92\,$K), each with a gas mass of a few~$10^{10}\,M_\odot$, finding that their results suggest an $\;\approx 6-55$ times higher space density of such distant dusty systems within the first billion years after the Big Bang than thought until now. They also discuss that higher dust temperatures are expected at higher redshift. 

In summary, the high redshift observations indicate rather high SFRs, consistent with the present model as can be deduced by comparing the SFRs quoted above with Fig.~\ref{fig:SFHs}, but cooler dust temperatures may lower these somewhat.

\subsubsection{Revised SFRs?}
\label{sec:modelSFRs}

If the present model were relevant for the observed population of SMBHs then the model SFRs ought to be consistent with the SFRs deduced from high-red shift observations. The previous Sec.~\ref{sec:obsSFRs} suggests this to be the case, but
the observed SFRs rely on a number of assumptions and in particular on the assumed value of $T_{\rm dust}$ and on the IMF. 

In their fig.~9, \cite{Decarli+18} quantify how the assumed dust temperature influences the observationally derived SFRs. If $T_{\rm dust}$ in the star-forming system is $25\,$K instead of $50\,$K, then $SFR_{\rm 25\,K} \approx \kappa_{\rm dust} \times SFR_{\rm 50\,K}$ with the correction factor $\kappa_{\rm dust} \approx 0.1$, such that the values reported above, $1000 \simless SFR_{\rm 50\,K}/(M_\odot/{\rm yr}) < 10^4$, would become 
$100 \simless SFR_{\rm 25\,K}/(M_\odot/{\rm yr}) < 10^3$. However, some of the above mentioned results are obtained by independently fitting for $T_{\rm dust}$ with the argument that $T_{\rm dust}$ is expected to be higher at higher $z$. 

At the same time, if the gwIMF/IGIMF is top-heavy (Sec.~\ref{sec:gwIMF}), then the
true SFRs are much smaller in truth than the observed SFRs inferred using the invariant canonical IMF.
This is the case because the photons which an observer detects to measure the SFR are emitted mostly by the ionising stars while a top-heavy gwIMF changes the ratio between the mass in low-mass stars and high-mass stars towards smaller values \citep{YJK17, Jerab18}.
The IGIMF models
 indicate that canonical observed values of $SFR_{\rm obs}\approx 5000-10000\,M_\odot$/yr may in actuality be $SFR_{\rm true} \approx 200-1000\,M_\odot$/yr largely independently of the metallicity (fig.~7 in \citealt{Jerab18}).
It is to be noted though that the correction factor used here is valid only if the SFR tracer is H$\alpha$ emission. For example, using the far-UV flux would increase the correction factor closer to a value of~1  \citep{Pflamm09b}. The impact of a top-heavy gwIMF on the observationally-deduced SFRs needs to be investigated in the future and is an important point to consider.

Furthermore, combining the maximum dust-temperature reduction factor with the IGIMF correction factor would imply $SFR_{\rm true}\approx 10^{-2}\,SFR_{\rm obs}$. Thus, a local spheroid with $M_{\rm igal}=10^{12}\,M_\odot$ would have had a physical counterpart at high redshift with $SFR_{\rm true}\approx 30\,M_\odot$/yr and would need to form over a time span $\Delta \tau_{\rm true} \approx 3 \times 10^{10}$yr. This is clearly ruled out by the observed bulk ages of the local spheroids (Sec.~\ref{sec:nonmono}). Presumably, the realistic case is that the dust temperatures adopted by the observational studies mentioned in Sec.~\ref{sec:obsSFRs} are approximately correct and that the IGIMF correction factor lies in the range $0.1-1.0$ (for $SFR_{\rm obs} \simgreat 100 M_{\odot}/{\rm yr}$, with the canonical gwIMF being used to infer $SFR_{\rm obs}$, \citealt{Jerab18}, and since the high-redshift SFR tracer is not the H$\alpha$ flux) such that 
\begin{equation}
SFR_{\rm true} = \kappa \, SFR_{\rm obs}
\label{eq:kappa}
\end{equation}
with $\kappa \approx 0.5$ for the combined IGIMF and dust-temperature correction factor.

In summary, a tension has emerged between the high SFRs necessary to form the
locally-observed massive spheroids within a~Gyr  and the here possibly implied smaller SFRs (if the dust temperature at very high redshift is in fact much smaller than usually assumed, and if the observed SFRs are overestimated due to the assumption of a canonical gwIMF).
Given that local observations are typically more reliably interpreted than high redshift ones, it may well be that the local constraints (Sec.~\ref{sec:nonmono}) might inform us about the observational biases acting in high-redshift observations. A plausible resolution of this tension might be that the downsizing times are somewhat longer together with a possible bias that typically the most-massive cluster would be seen in formation (which has a much lower SFR than the whole spheroid) in combination with the rarity of the most massive spheroids.

\subsubsection{Implications of revised SFRs?}
\label{sec:implSFRs}

Would the model developed here be invalidated if the observed SFRs at high redshift are in fact much lower (Sec.~\ref{sec:modelSFRs}) than implied by the downsizing time (Eq.~\ref{eq:downs})? 

The observed emission from high redshift star-forming objects stems mostly from ionising stars.  The mass, $M_{\rm max, m*>8}$, in stars more massive than $8\,M_\odot$ in the hyper-massive cluster with mass $M_{\rm ecl, max}$, corresponds to an absolute luminosity (in some relevant wavelength range), $L_{\rm max, m*>8}$, from all such stars combined. $M_{\rm max, m*>8}$ associated with an observed object is thus fixed through the received photons once the luminosity distance is known. The observationally deduced $L_{\rm max, m>8}$ is transformed into a SFR by usually assuming an invariant (e.g. the canonical) gwIMF (Sec.~\ref{sec:gwIMF}) which essentially associates the luminosity, and thus mass of the ionising stellar component, with the total stellar mass including all low-mass stars. If the gwIMF is, however, top-heavy, then the same $M_{\rm max, m*>8}$ corresponds to a smaller total stellar mass including low mass stars. 

A case in point is the measurement of the SFR using the H$\alpha$ flux, i.e. if $L_{\rm max, m*>8}=L_{{\rm H}\alpha}$. \cite{Jerab18} quantifies the galaxy-wide $L_{{\rm H}\alpha}$ assuming a canonical gwIMF and the gwIMF calculated using the IGIMF theory (Sec.~\ref{sec:gwIMF}). Using $L_{{\rm H}\alpha}$ leads to the largest sensitivity on the gwIMF shape since it is the most direct luminosity measure of the ionising-star massive-stellar content. The dependency of the true SFR on the gwIMF is quantified in fig.7 in \cite{Jerab18} from which follows that for an observed value assuming the canonical gwIMF of $SFR_{{\rm H}\alpha}\simgreat 100\,M_\odot$/yr the correction factor is $0.1-0.05$, i.e. $0.05 \simless SFR_{\rm true} / SFR_{{\rm H}\alpha} \simless 0.1$ 
with a weak dependence on metallicity. 

\begin{figure}
\includegraphics[scale=0.9]{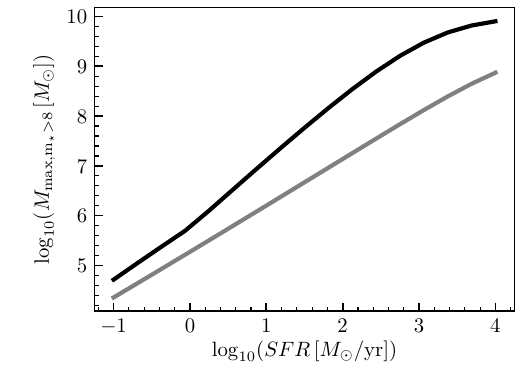}
\caption{\small 
The mass, $M_{\rm max, m*>8}$, in stars more massive than $8\,M_\odot$ in the first-formed most-massive cluster (of mass $M_{\rm ecl,max}$ as obtained by solving Eq.~\ref{eq:Mdeltat}) formed in the model with a box-shaped SFH as a function of the SFR of the spheroid ($SFR=M_{\rm igal}/\Delta \tau$). 
This mass, $M_{\rm max, m>8}$ yields the mass in BHs (Fig.~\ref{fig:UCDBH}) which, according to the present model, may coalesce to the SMBH. The gray line is the relation valid for the canonical IMF (corresponding to the dashed red line in the bottom-right panel of Fig.~\ref{fig:SMBHseed}; note that the values here are larger because all stars with $m>8\,M_\odot$ are counted). The  black line is the relation assuming the 
IMF is top-heavy (Eq.~\ref{eq:IMF}; corresponding to the dashed red line in the lower-left panel of Fig.~\ref{fig:SMBHseed}; note that the values here are larger because all stars with $m>8\,M_\odot$ are counted). For example, an observer might deduce a high redshift star-forming spheroid to have $SFR_{\rm obs}=5000\,M_\odot$/yr which would correspond to $M_{\rm max, m*>8} \approx 3\times 10^8\,M_\odot$ assuming a canonical IMF (Sec.~\ref{sec:canIMF}) and $\beta=2.0$. If, however, the IMF and thus the gwIMF is top-heavy then the true SFR is smaller, $SFR_{\rm true}\approx 250\,M_\odot$/yr \citep{Jerab18}. This is the case because the same number of ionising photons emitted by the massive stars is observed, but it now corresponds to a smaller total stellar mass being formed. The mass in ionising stars corresponding to this smaller SFR is thus unchanged \citep{Chruslinska+2020}, as is the mass in the BH sub-clusters which forms in these clusters.
}
\label{fig:correctedSFRs}
\end{figure}

This implies that $SFR_{\rm obs}=5000\,M_\odot$/yr would correspond to $SFR_{\rm true}\approx 250\,M_\odot$/yr while $M_{\rm max, m*>8}$ remains the same. Fig.~\ref{fig:correctedSFRs} depicts this situation.  The implication is that the observed SFRs may in fact be lower by a factor of perhaps ten, but the mass in ionising stars corresponding to these SFRs remains unchanged, as explicitly pointed out by \cite{Chruslinska+2020}. 

Thus, given the top-heavy gwIMF, the implied lower true SFRs transform to the same mass of the BH cluster as used in this model such that the model remains unaffected apart from requiring longer $\Delta \tau$ to form $M_{\rm igal}$. In other words, Fig.~\ref{fig:SFR} will change accordingly, while Fig.~\ref{fig:ipgal}, Fig.~\ref{fig:SMBHseed} and~\ref{fig:AppSMBHseed} will not change. The number density of high redshift quasars (Sec.~\ref{sec:ndens}) would need to be recomputed taking into account the new $\Delta \tau$. This is not done here because the calculation presented there is very rough in any case and more detailed galaxy-formation models would need to be incorporated.

\subsection{Other caveats}
\label{sec:other}

Further caveats and uncertainties, all of which can be included in future research programs, are touched upon in this section.

\subsubsection{The IMF}

The initial BH content of a cluster depends on the IMF, with the
invariant canonical IMF (Sec.~\ref{sec:canIMF}) having fewer and the
variable IMF (Eq.~\ref{eq:IMF}) allowing for many more. But the
invariant IMF is inconsistent with a broad range of observational
constraints and basic star-formation theory robustly expects the IMF
to vary with physical conditions (Sec.~\ref{sec:stpop}). 

How does this variation look like in view of the many existing
observational constraints and in view of the universality of the
canonical IMF in Milky Way star-forming clouds?  The coefficients used
in Eq.~\ref{eq:IMF} are originally obtained from GC data. The
inferred initial conditions of the GCs are based on the
semi-analytical model of \cite{Marks10} which quantifies the expansion
of a young GC due to residual gas expulsion. The explicit constraints
on the IMF variation (Eq.~\ref{eq:IMF}) are obtained by \cite{Marks12}
who, by that process, account for the lack of low-mass stars in those
GCs which have a low concentration \citep{deMarchi07}, as well as
taking the high dynamical mass-to-light ratios \citep{Dabring09}, the
overabundance of low-mass X-ray binaries \citep{Dabring12} in some
UCDs and the Milky Way data into account. The need to evacuate a
certain fraction of low-mass stars from a GC with an observed
concentration, allows the massive-star IMF to be constrained because
the massive stars provide the energy to expel the residual gas thereby
expanding the mass-segregated young GC.  But neither \cite{Marks10}
nor \cite{Marks12} take into account the impact on the long-term
evolution of a GC by the implied large population of BHs
(e.g. \citealt{BH13b, Chatterjee+17, Giersz19, Wang20}). For
$\alpha_3 <1.5$, BHs would dominate the total mass of a GC. This can
result in the fast dissolution of the luminous component of the GC and
the formation of a dark cluster \citep{BK11}. This would imply that
the observed morphology of those GCs, which, according to the analysis
by \cite{Marks12} ought to have been born with a very top-heavy IMF,
can probably not be matched to their observed versions. A case in
point is 47~Tuc \citep{H-B+20}. However, none of the existing
modelling of the long-term evolution of GCs takes into account the
evolving potential of the hosting galaxy (e.g. \citealt{Speagle+14,
  delaRosa+16,Mancini+19}), and also other astrophysical processes not
yet included in the first~Gyr of GC evolution will affect the
present-day appearance of GCs. 

Thus, considering the sum of
observational evidence (Sec.~\ref{sec:stpop}), the systematic
variation of the IMF with density and metallicity appears to be a
robust general trend. Indeed, if one were to construct an IMF
dependency on metallicity and density on the embedded-star-cluster
scale in view of the observational constraints from the Milky Way and
other galaxies (Sec.~\ref{sec:varIMF} and~\ref{sec:gwIMF}), 
this dependency would need to be similar to
Eq.~\ref{eq:IMF}.  The ultra-low metallicity IMF may even be much more
top-heavy than assumed here. For instance, in the Galactic centre, the
observed population of very young stars appears to follow a very
top-heavy IMF with low-mass stars apparently largely missing
(Sec.~\ref{sec:varIMF}) despite being a metal-rich population. This
may indicate a more extreme variation of the IMF than encompassed here
through Eq.~\ref{eq:IMF}, and would imply even larger BH sub-cluster
masses than in the models computed here.  

Thus, given the above
caveats, the specific formulation via Eq.~\ref{eq:IMF} is strongly
suggestive but not definitive. In view of this we have also included
models with an invariant canonical IMF.

\subsubsection{The low-mass stars and other simplifications}

In finding the collapsing solutions, we ignored the low mass
stars. These are likely to help (speed-up) the shrinkage of the BH
sub-cluster due to energy equipartition. We applied, in
  Sec.~\ref{sec:heat}, the balanced-evolution theory as formulated via
  eq.~1 in \cite{Antonini19} to the case of a cluster of equal-mass
  BHs. By doing so we neglected the rest of the stellar cluster,
  assumed the BH sub-cluster to be self-gravitating and thermally
  isolated from its surroundings. Clearly these are significant
  simplifications, but the complexity of the realistic situation which
  involves the dynamical evolution and the heating rate through BH
  binaries being modulated by the gas infall warrants an argument in
  support of these simplifications to ensure technical
  feasibility. The stellar component is likely to enhance the mass
  growth of the central massive BH since it will also be shrunk onto
  the BH cluster through the in-falling gas. The BHs will have
different masses in the BH sub-cluster. This speeds-up core collapse
and merging rates (e.g. \citealt{Khalisi+07}). We have ignored the
likely rotation of the BH sub-cluster, but rotating systems collapse
faster, the collapse being accelerated even more if the BHs have a
mass spectrum \citep{Kim+04}. Spin-spin interactions during BH mergers
have not been considered here, but simulations of BH clusters in which
the BHs have no, aligned or randomly oriented spins do not lead to
measurable differences in the merging behaviour \citep{bre13}.

\subsubsection{Further growth of the SMBH seeds}
\label{sec:furthergrowth}

The SMBH-seed will additionally grow through gas accretion in the time
$\Delta \tau$ during which the spheroid forms.  The gas in-flow and
out-flows will be modulated by the accretion activity of the
SMBH-seed(s), being dependent also on the progressive metal enrichment
of the gas as the spheroid forms (e.g. \citealt{YJK19}).  The
accretion of gas onto the BHs and the likely feedback effect are
discussed in Secs~\ref{sec:BHaccr}.  Ultimately, it will be necessary
to perform N-body or equivalent simulations of such systems, but these
currently remain out of computational reach (e.g. \citealt{Lee93,
  Lee95, Spu99, Kupi06, bre13, Giersz15, Rodriguez+16, Wang16}).  Once
the bulk of the spheroid is assembled, their central SMBHs may grow
slowly through the disruption of stars (e.g. \citealt{Brockamp+11}),
and this will need to be also included in future more comprehensive
modelling of the SMBH--spheroid correlation.

\subsubsection{Multi-generation clusters and their merging}
\label{sec:cavMultiCl}

As the spheroid assembles, many massive clusters will form near its
centre (Eq.~\ref{eq:Ngen}). These will have their own BH sub-clusters and the masses of
these will depend on how quickly the IMF changes to a more canonical
form as the metallicity of the gas from which they form
increases. Whether or not the BH sub-cluster in any one of these
evolves sufficiently rapidly into the pressureless state to allow
collapse to a SMBH-seed depends on whether the hosting cluster can
accrete a sufficient amount of gas for a sufficiently long time from
the forming spheroid.  Thus, more than one SMBH-seed is likely to
co-exist near the centre of the forming spheroid, but BH sub-clusters
which have not collapsed should also exist. These would be
identifiable as ``dark clusters'' \citep{BK11}.  Gas inflow and
dynamical friction on field stars \citep{Bekki10} brings the most
massive clusters to merge near the centre of the spheroid, enhancing
the growth of the central SMBH-seed which forms from the first
ultra-metal-poor star-burst cluster. Observational evidence for this
process, albeit in dwarf galaxies, has been documented
\citep{Fahrion+2020}. The detailed physical evolutionary processes
operating in the interactions and mergers of multiple massive star
clusters and their BH components leading to the formation of the
present-day nuclear regions in spheroids and the continued build-up of
the nuclear cluster in late-type galaxies are complex and need special
attention (e.g. \citealt{Arca+15, Arca+16, Arca+19}, for a review of
nuclear star clusters see \citealt{Neumayer+20}).

\subsubsection{Multiple SMBH seeds}
\label{sec:multipleSeeds}

Mergers between SMBH-seeds formed in the individual cluster discussed
in Sec.~\ref{sec:cavMultiCl} are likely to contribute to the growth of
the SMBHs and will add a stochastic element to the final mass
correlation between SMBH and spheroid mass (Fig.~\ref{fig:SMBHseed}). 
Such mergers may lead to recoils from gravitational wave
emission and this would imply displaced SMBHs from the centres of
their hosting spheroids \citep{Lena+14}. Dynamical friction of the
SMBHs on the stellar population of the spheroid will bring the SMBH
back to the centre within the mass-segregation time scale of about
$10^9\,$yr (Eq.~\ref{eq:t_msegr}).
The observed absence of a significant population of
displaced SMBHs, which \cite{Lena+14} use to constrain the merging
rate of galaxies, thus appears to be broadly consistent with the model
developed here.
%, given that galaxy--galaxy mergers are expectedly rare
%in a non-dark-matter-based cosmological model \citep{Kroupa15}. 

Less-massive SMBHs, IMBHs,  or dark clusters would not suffer sufficient dynamical friction to merge with the central cluster though such that the inner regions of spheroids should, according to the present model, contain many such objects. The IRS13E object near the
Galactic centre source Sgr~A* may be such an example (\citealt{BK11,
  Tsuboi+17}, but see \citealt{ReidBrunthaler20} for astrometric
limits on such companions of Sgr~A*). 
Another example may be the candidate 12-yr-period binary-SMBH in the active galactic nucleus or blazar OJ287 \citep{Britzen+18, Laine+20}.

\subsubsection{IMBHs}
\label{sec:IMBHs}

As can be seen from Fig.~\ref{fig:solutions_m1_1}, the process of BH
cluster collapse presented here is not limited to the formation of
SMBHs. Indeed, assuming~5~per cent of $M_{\rm BH,0}$ merge and in the
case of a low to moderate gas fraction ($\eta \simless 1$), the
formation of IMBH seeds is possible within a wide range of initial
parameters which may be in-line with the hierarchical scenario of the
formation of massive galaxies.  Challenges to be addressed in this
context are that the model presented here relies on massive spheroids
harbouring a SMBH to have formed rapidly, because smaller progenitors
would violate the $M_{\rm ecl,max}-SFR$ relation
(Eq.~\ref{eq:Mdeltat}), that the observed metallicity and
alpha-element abundances of the spheroids need to be fulfilled, and
that the IMBHs would subsequently have to merge rapidly by some
mechanism to form the SMBHs observed at very high redshifts. For
example, \cite{WirthBekki20} study models in which pre-existing IMBHs
assemble at the centre of a UCD through dynamical friction where they
form a sub-cluster in which mergers lead to the build-up of an SMBH.

The model presented here though {\em allows} the direct formation of the SMBH-seeds on a very short time-scale.

%==========================================
\section{Conclusion}
\label{sec:concs}

The problem of the observed rapid appearance of quasars in the very
young Universe and of the mass correlation between the SMBH and its spheroid are
considered. Here we concentrate on a conservative
approach resting on the properties of observed stellar populations and
known stellar-dynamical processes. The model
relies on the mass in massive stars in the first-formed star-burst cluster(s) near the centre of the later spheroid to correlate with the global SFR at assembly time of the bulk of the spheroid as given by observational data in the nearby Universe. The sequence of star-burst clusters assembled at the centre of the forming spheroid provides the seed-SMBH(s) and the formation of the spheroid leads to gas inflow which shrinks and merges the star-burst cluster's stellar-mass BHs and leads to additional mass growth of the seed(s). 
It is this physical correlation between the central gas density and the proto-spheroid gas mass which ultimately leads to the observed SMBH-mass--spheroid-mass correlation.

By applying the IGIMF theory, i.e. the observationally independently derived relations describing stellar
populations in individual star-formation events on a molecular
cloud-core scale (i.e. embedded clusters) and their distribution by
mass in a star-forming galaxy, to assess the expected content of
stellar-mass BHs in the very early Universe, this work suggests that
SMBH-seeds with masses up to a few times $10^6\,M_\odot$ might emerge 
within a few
hundred~Myr of the first star formation (Figs~\ref{fig:SMBHseed} and~\ref{fig:AppSMBHseed}).  This is a conservative estimate relying 
on only 5~per cent of the BH content of the first massive cluster 
at the centre of the forming spheroid to merge to the SMBH seed. 

The existence of quasars within $\approx500\,$Myr 
after the birth of the Universe and the existence of massive SMBHs at 
lower redshift is explained as a two-stage process (see Fig.~\ref{fig:quasar_phases} in 
Sec.~\ref{sec:ndens} for a visualisation using explorative models):

During the first phase, an extremely massive ($>10^8\,M_\odot$) 
star-burst cluster forms as the first
ultra-low-metallicity population with a top-heavy IMF 
\citep{Jerab18, Yan+2019, YJK19} at the centre of the forming spheroid. This 
cluster might appear, in terms of emission line widths, spatial resolution
and luminosity, similar to a quasar (\citealt{Jerab17}, Sec.~\ref{sec:BHcontent}), but this needs to be ascertained using radiation-hydrodynamic simulations. 
This phase lasts a few dozen Myr during which the massive stars explode as 
supernovae driving an energetic and metal-rich cluster wind. 
Once the last core-collapse supernova explodes.
leaving the BH sub-cluster, such an object fades, and this is where
the SMBH formation ensues as per the model presented here if the BH
sub-cluster accretes gas from the surrounding environment. 

During the second phase, which begins after the first few dozen~Myr 
when the cluster
luminosity has decreased significantly \citep{Jerab17, Ploeckinger+19}, 
gas will fall onto the central region from the still violently forming spheroid (Sec.~\ref{sec:model}). 
The key mechanism for the appearance of the actual SMBH-seeds 
is accretion of gas from the surrounding forming
spheroid into the central region. The gas 
accretion leads to contraction of the BH containing
cluster (modelled here as a BH sub-cluster, ignoring the stellar component) 
to a relativistic state (Eq.~\ref{eq:sigmarel}) within a
few hundred Myr from which it collapses through energy loss by
radiating gravitational waves during BH--BH encounters. The BH
sub-cluster has an incompressible phase-space distribution function
and would, if left on its own, be in a state of balanced evolution 
(with the rest of the stellar cluster),
but the gas spatially shrinks it to the point where its equation of
state becomes pressure-less, allowing the relativistic collapse to a
SMBH-seed. Post-Newtonian N-body simulations show such a BH sub-cluster to
core-collapse within 10--15 median two body relaxation times leading
to runaway merging between the BHs and coalescence of the core of the
BH sub-cluster to a SMBH-seed \citep{Lee95} which may amount to 5~per
cent of the BH sub-cluster mass \citep{Lee93, Kupi06}. Due to the
small physical size, core collapse occurs within a dozen~Myr once the
BH sub-cluster reaches this relativistic state.  The SMBH-seed mass
may in reality reach much larger values if the compression through the gas forces
faster shrinkage, faster BH--BH coalescences and if it would squeeze the rest of the stellar cluster into the BHs. The key physical mechanism
provided by the in-falling gas from the forming spheroid is thus to
force the BH sub-cluster out of its balanced evolution and into the
state where core collapse occurs within a dozen~Myr.

The first and second phase together last as long as the spheroid
continues to form on the downsizing time-scale (Eq.~\ref{eq:downs}).
As a case in point, the model developed here is applied in
Sec.~\ref{sec:CaseinPoint} to the $1.5\times 10^9\,M_\odot$ SMBH in
the luminous quasar P{\={o}}niu{\={a}}'ena at $z=7.5$ discovered by
\cite{Yang+20}.  Since spheroids remain largely dormant once they have
formed (Sec.~\ref{sec:nonmono}), the SMBH--spheroid correlation will
not change significantly with redshift once spheroid assembly
quenches, although some slow further growth will likely take place
(Sec.~\ref{sec:furthergrowth}). Indeed, \cite{Suh+20} find a lack of
evolution of the SMBH--spheroid relation out to $z\approx 2.5$.

While possibly being relevant for the 
rapid emergence of SMBHs in the
Universe, the present model also at the same time relaxes the need to interpret extremely high redshift quasars as already formed accreting
very massive SMBHs.
This is explicitly shown in 
Sec.~\ref{sec:ndens} with a calculation based on the formation times of galaxies adopted from \cite{Thomas05} and the different model phases of the quasars. In that section also the possible issue of missing quasars is discussed.

The model studied here might be relevant for explaining why evidence for IMBHs exists only at the centres of galaxies, them not having been 
found in star clusters
\citep{Baumgardt+19}: If a BH sub-cluster is embedded only in a star
cluster and is locked in a state of balanced evolution thereby slowly
ejecting its BH content through BH--BH binary encounters, it will
core-collapse on a time-scale too long to form a substantial IMBH
(e.g. \citealt{Giersz15}). The same BH sub-cluster which suffers gas
infall from the forming and evolving galaxy will, on the other hand,
be compressed sufficiently for runaway merging. The results obtained
here (Fig.~\ref{fig:solutions_m1_1}) suggest that only when the BH
sub-cluster has a mass $M_{\rm BH,0}>10^4\,M_\odot$ will the relativistic state
be reached. More research on this question is clearly desired.

According to the present model, SMBHs are expected to occur from the formation of spheroids, i.e., a disk galaxy with a classical bulge should have an SMBH, while a 
disk galaxy with the same rotation speed but without a classical bulge may not host an SMBH, or at best, only a compact central BH sub-cluster.

Noteworthy is that the SMBH--spheroid mass
correlation (Figs~\ref{fig:SMBHseed} and~\ref{fig:AppSMBHseed}) comes out naturally in terms of
its slope, without needing to adjust any parameters to achieve this
purpose, the model being consistent with observational data on
pc-scale stellar systems (Fig.~\ref{fig:UCDML}) and with galaxy-scale
stellar populations, as it results from applying the IGIMF theory (Sec.~\ref{sec:gwIMF}). A correlation emerges
also if the IMF remains canonical, but in a weaker form
(Figs~\ref{fig:SMBHseed} and~\ref{fig:AppSMBHseed}). But the IMF variation with metallicity and
density as inferred from independent data by 
\cite{Dabring09, Dabring12, Marks12} implies, surprisingly, a
near-perfect correlation in excellent agreement with the observed
one. Further Eddington-limited mass-growth of the thus formed SMBH-seeds 
will depend on further infall of gas as modulated
by the formation of the spheroid and may lead to extremely massive
SMBHs within an additional few hundred~Myr period as allowed by the
down-sizing time (Eq.~\ref{eq:downs}). Interestingly, taking the
conservative estimate that if only the 5~per
cent SMBH-seed masses grow for about~345~Myr (Sec.~\ref{sec:corrs}) at the super-Eddington 
rate ($\epsilon_{\rm r}=0.1$, Sec.~\ref{sec:BHaccr}),
then they reach the observed correlation (Eq.~\ref{eq:factor}).
That is, this model achieves a most remarkable consistency of time-scales in that the need to grow the SMBH-seeds per spheroid to the observed SMBH masses comes out to be within the downsizing time. {\it This is not a trivial correspondence, because the result might have been that the growth time would have turned out to be much longer than the downsizing time, which would have violated the model.}

Note also that if all of the mass formed into stars in all the $N_{\rm gen}$ central most-massive clusters would be combined into a black hole, then the resulting SMBH--spheroid correlation is impressively consistent with the data (dotted line in Figs~\ref{fig:SMBHseed} and~\ref{fig:AppSMBHseed}).

%This suggests a
%possible revision of the downsizing relation in the sense that
%$\Delta \tau_{\rm YJK} \approx t_{\rm rel}+ \tau_{\rm Edd} \approx
%1\,$Gyr is the overall formation time-for spheroids of all masses, as
%obtained independently from chemical evolution analysis (Yan,
%Jerabkova \& Kroupa, in preparation).

The here presented model is expected to lead to a dispersion of the SMBH
masses at a given present-day spheroid mass because the SMBH-seeds are
expected to not be identical, given the stochasticity of mergers
between central clusters and of the gas infall feeding SMBH-seed
growth. According to the present model, the correlation of SMBH--spheroid
ceases near a spheroid mass of $10^{9}\,M_\odot$ (for $\beta=2.0$, Fig.~\ref{fig:SMBHseed}) and 
$10^{9.6}\,M_\odot$ (for $\beta=2.4$, Fig.~\ref{fig:AppSMBHseed}) such that at lower
masses only the central (i.e. nuclear) star cluster with its central
component of BHs takes over the correlation towards smaller masses due
to Eq.~\ref{eq:Mdeltat}. This appears to be consistent with
this same observed phenomenon \citep{WehnerHarris06, Capuzzo+17}.

Since the environment of the central cluster 
need not necessarily become a massive spheroid, this
object may, under certain circumstances, 
later appear as an isolated UCD with a SMBH in it as
observed (Sec.~\ref{sec:introd}). This may also be relevant to the
finding of a quasar at a redshift of 3.84 which is inferred to be a
SMBH weighing $\approx 2.5\times 10^{10}\,M_\odot$ embedded in a
host with stellar mass $\simless 6.3\times 10^{10}\,M_\odot$
\citep{Schramm+19}.  This may happen in the present context if
mergers of forming spheroids eject some of their very massive
clusters with associated material. An implication of this scenario
would be that some such ejected UCDs may have been cutoff from the gas
accretion from their forming spheroid such that their BH sub-cluster
may not have reached the relativistic limit. The SMBH detected in such
UCDs would then constitute a compact (pc-scale) sub-cluster of BHs
which is in the state of balanced evolution, rather than an actual
SMBH. At the moment this is merely a speculation, and detailed
computations would be needed to test if this scenario for the
formation of SMBH-bearing UCDs is viable. 

The model studied here has two implications: (i) The central region
of spheroids should contain many dark clusters and SMBH-seeds
(including IMBHs) as left over relics (Sec.~\ref{sec:multipleSeeds}).
(ii) When the BH sub-clusters
enter the relativistic phase, they will radiate gravitational waves at
an increasing rate from the many BH--BH mergers, including the final
runaway collapse to the SMBH-seed. This emission ought to be
detectable with the appropriate gravitational wave observatories, and
must not be wrongly interpreted as a sign of the hierarchical assembly
of early galaxies in their speculative dark matter halos. It will be
useful to calculate the signal expected from SMBH assembly according
to the model presented here.

The here presented results rest on, by computational necessity, important simplifications (Sec.~\ref{sec:disc}). 
Sec.~\ref{sec:SFRs} dissects the current observational evidence for and against the high SFRs that ought to be evident at high redshift when the massive spheroids were forming, and how the evidence compares with the theory developed here. 
%The community will need a longer sustained research effort before 
%a consensus about the cosmological-formation of early type galaxies which bear massive SMBHs can be reached. The possibility that accreting BHs may, under certain conditions, accelerate rather than suffer dynamical friction, leading to longer BH--BH merging times, needs to be taken into account in future modelling (Sec.~\ref{sec:cavAccelDecl}).
Thus, while encouraging, much more research will be needed to assess
the importance of gas accretion onto the BH sub-cluster, its formation
and evolution in a realistic spheroid formation context. We note that
the model presented here would not work (i) if either the bulk of a
spheroid does not form monolithically and thus with the high SFRs
assumed, or (ii) the first star-formation events at extremely low
metallicity and modest SFRs do not produce hyper-massive star-burst
clusters with a top-heavy IMF before the bulk of the spheroid
assembles later. It is quite likely that a combination of processes
play a role. For example, a hyper-massive star (formed either as a
population~III object or from runaway stellar mergers in the central
hyper-massive star cluster, Sec.~\ref{sec:introd}) may directly
collapse to a SMBH-seed which subsequently accretes such that the
outcome (the SMBH-mass--spheroid-mass correlation) emerges also just
as in the present model, although it is not clear if this scenario can
reach the large SMBH-seed masses as in the present model such that the
further growth were consistent with the downsizing time.  It is also
imaginable that the squeezing of a cluster of BHs into the
relativistically unstable regime studied here through gas drag in
addition to the coalescence of the whole stellar body of the nuclear
cluster (generations) by gas drag, leads to an, in effect,
hyper-massive first star, similarly as discussed in the
Introduction. This hypothetical possibility would be physically
related to the postulated Thorne-Zytkow objects \citep{ThorneZytkow75,
  ThorneZytkow1977}.  Observationally, it will be important to
constrain the number density of quasars at the highest reachable
redshifts and to assess the masses the most massive SMBHs may reach as
is the goal of the Gargantua search strategy \citep{Brockamp+16}.

\section*{Acknowledgements}
An early version of this research programm was begun in Concepcion,
Chile, at the Modelling and Observing Dense Stellar Systems 15 (MODEST15) 
conference. Pavel Kroupa and Long Wang thank the organisers of
that memorable event (notably Mike Fellhauer), 
Sverre Aarseth, Paulina Assmann, Joerg Dabringhausen and Rainer
Spurzem for detailed discussions on the N-body dynamics of BH clusters.
We thank Moritz Haslbauer for useful discussions on galaxy evolution 
and Zhiqiang Yan and Alexandre Vazdekis for important discussions on the formation of elliptical galaxies.
Pavel Kroupa and Ladislav \v{S}ubr acknowledge support from the Grant Agency of the Czech Republic
under grant number 20-21855S.
Tereza Jerabkova acknowledges support through a two-year European Southern Observatory (ESO) Studentship in
  Garching, the Erasmus+ programme of the European Union under
grant number 2017-1-CZ01- KA203-035562 on La Palma, and the European Space Agency fellowship programme.  Long Wang acknowledges
support through an Alexander von Humboldt Fellowship while 
being hosted by the Stellar Populations and Dynamics Research (SPODYR) group at the
University of Bonn.  Long Wang also acknowledges the present support through
a Japan Society for the Promotion of Science International Research Fellowship. Some of this work was performed
at ESO in Garching and Pavel Kroupa thanks ESO for a science-visiting position in 2019.
Pavel Kroupa and Long Wang thank the non-professorial staff of the Argelander-Institute for Astronomy for technical support.
This work benefited from the
International Space Science Institute (ISSI/ISSI-BJ) in Bern and
Beijing, thanks to the funding of the team ``Chemical abundances in the
ISM: the litmus test of stellar IMF variations in galaxies across
cosmic time'' (Donatella Romano and Zhi-Yu Zhang). Pavel Kroupa is grateful to the Isaac Newton Group staff on La Palma for hospitality where a part of this manuscript was written just prior to the Corona Crisis and acknowledges the Erasmus+ programme of the European Union under
grant number 2017-1-CZ01- KA203-035562 for enabling this visit while on a contract with the Czech Academy of Sciences through Ond\v{r}ejov Observatory.
Havelska Kuchine (``knedlikarna'') in Prague was an essential location throughout this
work -- Knedlikum zdar! 
%%%%%%%%%%%%%%%%%%%%%%%%%%%%%%%%%%%%%%%%%%%%%%%%%%

\section*{Data Availability}
The results are based on calculated models as described within this
manuscript. 

%%%%%%%%%%%%%%%%%%%% REFERENCES %%%%%%%%%%%%%%%%%%

\bibliographystyle{mnras}
\interlinepenalty=10000
\bibliography{literature}

%%%%%%%%%%%%%%%%%%%%%%%%%%%%%%%%%%%%%%%%%%%%%%%%%%
%=================== Appendix===========================
\newpage

\noindent{\bf \Large Appendix~1: Special case of Sec.~\ref{sec:BHaccr}}\\

The special case, $\dot{m}_{\rm BH} = 0, \dot{M}_{\rm g} \propto M^2_{\rm g}$, leads to an analytical
solution to the mass increase of the gas mass within $R$: Assuming the gas mass within $R$ to
grow through Bondi-Hoyle accretion from the surrounding forming spheroid.  This is essentially
accelerating gas infall from the forming host galaxy and is given by Eq.~\ref{eq:BHaccr} with velocity
$v=0$ since the central cluster is approximately at rest relative to the forming spheroid, and
$\rho_{\rm g}$ being the average gas density of the protogalaxy, i.e.  
$\rho_{\rm g}=M_{\rm gal}/V_{\rm gal}$. Here, $M_{\rm gal}$ is the initial mass of the post-Big-Bang 
gas cloud which will form the spheroid with a stellar mass of $M_{\rm igal}=M_{\rm gal}/3$ for a star-formation efficiency of~$1/3$, and $V_{\rm gal}$ is the volume of this gas cloud. This volume 
can be taken to be a sphere of radius of $R_{\rm gal}\approx 10\,$kpc.

The analytical solution for the gas mass, $M_g$, as a function of time
is derived.  We have,
\begin{equation}
    \dot{M}_g = \frac{\dd{M_g}}{\dd{t}} = 4\pi \rho_g
    \frac{G^2(M_g+Nm_{\ff{BH}})^2}{c_s^3}\,,
\label{eq:gasBHaccr}
\end{equation}
where
\begin{equation}
    \rho_g = \frac{M_{\ff{gal}}}{4/3 \pi R_{\ff{gal}^3}}\,,
\end{equation}
is given, as well as $c_s$. Eq.~\ref{eq:gasBHaccr} can be rewritten as
\begin{equation}
    \frac{\dd{M_g}}{\dd{t}} = K (M_g+Nm_{\ff{BH}})^2\,
\end{equation}
where $K=4\,\pi\,\rho_{\rm g}\,G^2/c_s^3$ 
combines all the given constants. We can then separate the variables and 
integrate, obtaining 
\begin{equation}
    \int_{M_{g}^{0}}^{M_g} \frac{\dd{M_g}}{(M_g+Nm_{\ff{BH}})^2} = \int_{0}^{t} K\cdot t\,,
\end{equation}
which results in
\begin{equation}
    \frac{1}{Nm_{\ff{BH}}+M_g^0}-\frac{1}{Nm_{\ff{BH}}+M_g} = K \cdot t \, .
\end{equation}

Thus $M_g$ can be expressed as a function of time as,
\begin{equation}
M_g(t) = \frac{1}{\frac{1}{Nm_{\ff{BH}}+M_g^0} - K \cdot t} - Nm_{\ff{BH}}\, .
\label{eq:BH_sol}
\end{equation}

\newpage
\newpage

\noindent{\bf \Large Appendix~2: SMBH--spheroid-mass relations }\\

\begin{figure*}
\includegraphics[scale=0.9]{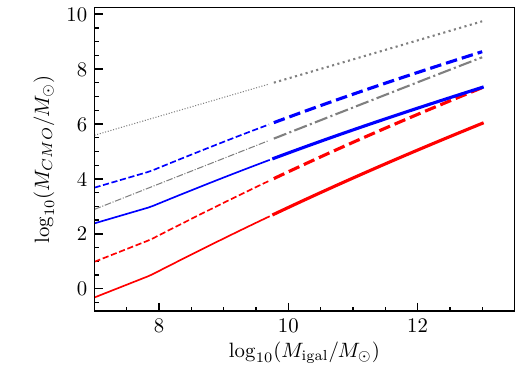}
\includegraphics[scale=0.9]{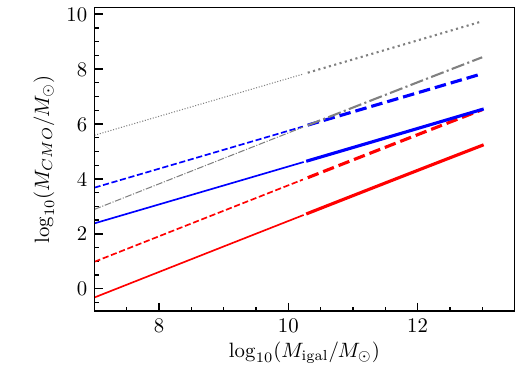}
\includegraphics[scale=0.9]{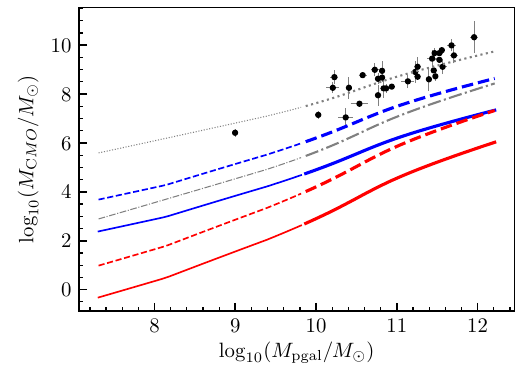}
\includegraphics[scale=0.9]{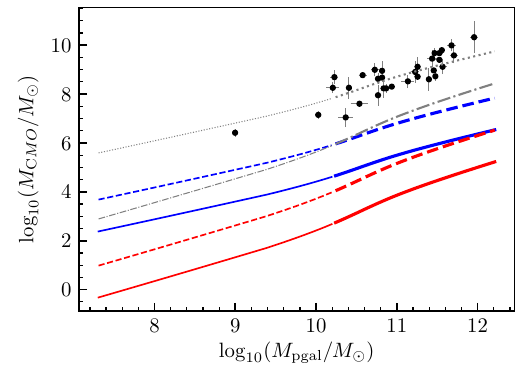}
\caption{\small As Fig.~\ref{fig:SMBHseed} but for $\beta=2.4$. Note the smaller SMBH-seed masses at a given 
value of present-day spheroid mass, $M_{\rm pgal}$. }
\label{fig:AppSMBHseed}
\end{figure*}

Additional Fig.~\ref{fig:AppSMBHseed} to show the effect of using a bottom-heavy mass function of embedded clusters with Salpeter power-law index $\beta=2.4$.

% Don't change these lines
\bsp	% typesetting comment
\label{lastpage}
\end{document}